\documentclass{article}
\renewcommand{\thesection}{\Roman{section}} 
\renewcommand{\thesubsection}{\thesection.\Alph{subsection}}

\usepackage[utf8]{inputenc}
\usepackage[english]{babel}
\usepackage{amsmath}
\usepackage{graphicx}
\usepackage{ulem}
\usepackage{geometry}
\usepackage{afterpage}
\usepackage{caption}
\usepackage{subcaption}
\usepackage{color}
\usepackage{float}
\usepackage[sort,numbers]{natbib}
\usepackage{placeins}
\newcommand{\pD}[2]{\frac{\partial #1}{\partial #2}}

\newcommand{\D}[2]{\frac{d #1}{d #2}}

\newcommand{\eval}[1]{\bigg{|}_{#1}}

\newcommand{\intcell}{\int^{x_R(k,t)}_{x_L(k,t)}\!dx\,}

\newcommand{\cv}{C_\mathrm{v}}
\newcommand{\lpsi}{\overline{\psi}}
\newcommand{\lphi}{\overline{\phi}}
\newcommand{\lT}{\overline{T}}
\newcommand{\Le}{\overline{e}}
\newcommand{\lcvnaught}{\overline{\mathrm{C}}_\mathrm{v0}}

\newcommand{\wgb}[1]{\textcolor{black}{#1}}
\newcommand{\wgbt}[1]{\textcolor{black}{#1}}

\begin{document}
\title{Benchmark solutions for radiative transfer with a moving mesh and exact uncollided source treatments}
\author{William Bennett\\
Ryan G. McClarren}
\date{\today}
\maketitle

\begin{abstract}
The set of benchmark solutions used in the thermal radiative transfer community suffer some coverage gaps, in particular nonlinear, non\wgb{-}equilibrium problems. 
Also, there are no non\wgb{-}equilibrium, optically thick benchmarks. These shortcomings motivated the \wgb{development} of a numerical method \wgb{free from the requirement} of linearity and easily able to converge on smooth optically thick problems, a moving mesh Discontinuous Galerkin (DG) framework that utilizes an uncollided source treatment. Having already proven this method on time dependent scattering transport problems, we present here solutions to non-equilibrium thermal radiative transfer problems for familiar linearized systems \wgb{together with} more physical nonlinear systems in both optically thin and thick regimes, including both the full transport and the $S_2$/$P_1$ solution. Geometric convergence is observed for smooth sources at all times and some nonsmooth sources at late times when there is local equilibrium\wgb{.} \wgb{Also,} accurate solutions are achieved for step sources when the solution is not smooth.
\end{abstract}

\section{Introduction}
The Stefan-Boltzmann law \cite{stefan1879uber}, which describes the relationship between radiation emitted from a material and its temperature as proportional to temperature to the fourth power, is \wgb{partly responsible} for the obdurate nonlinearity in high-energy density radiative heat transfer models. \wgb{Also, the opacity of material of interest can have a nonlinear dependence on temperature.} For \wgb{these} reason\wgb{s}, the extant analytic benchmarks with space and time dependence in this field are predicated on assumptions of linearity or equilibrium. 
There are solutions that assume the radiation energy and internal energy of the material instantly equilibrate, inducing a Marshak Wave \cite{marshak, petschek1960penetration,bennett2021self}, and solutions for non-equilibrium problems that  linearize the system in $T^4$ by invoking a form for the material heat capacity that is proportional to temperature cubed, an innovation of Pomraning \cite{pomraning1979non}\wgb{, and specify a constant opacity}. This technique of defining a heat capacity to linearize the system has been used to produce an abundance of solutions, including  transport treatments for the $P_1$ equations \cite{mcclarren2008analytic}, full transport solutions with one speed \cite{ganapol1983non, SU19971035}, and non-grey problems \cite{SU1999279}. In the diffusion limit, benchmarks solutions have been provided for one temperature \cite{bingjing1996benchmark}, three temperature \cite{mcclarren2011solutions}, and a non-grey treatment \cite{McClarren2022semi}.  

While these solutions are invaluable to code developers for verification, it is necessary to point out that there are certain drawbacks to using linear problems to verify codes whose purpose it is to solve nonlinear systems. While ideally the numerical code in question would solve the fully nonlinear equations and implement a special equation of state when running these verification problems, there is nothing to prohibit the curators of these codes from simply solving a linearized system when the benchmark is being run. Also, solutions to linear systems can be scaled to match benchmarks, unlike the unforgiving solutions to nonlinear systems, and the solution to the linearized equations equilibrates more quickly than a nonlinear problem as the temperature increases as a result of the special equation of state. Solving a linear problem does not completely verify the functionality of a radiative transfer code. 

Although nonlinearity is an impediment for analytic methods, it is not necessarily a source of difficulty for spectral methods. This was the impetus for our development of a moving mesh, uncollided source treatment Discontinuous Galerkin (DG) method for solving transport problems \cite{movingmesh}. The time dependent cell edges, which we call a moving mesh, and the uncollided source treatment were added on to the DG implementation because the transport equation with its finite wavespeeds admits discontinuities that inhibit DG methods from attaining their higher order convergence potential. The moving mesh and uncollided source can present a smoother problem for the method to solve: the moving mesh by matching edges to moving wavefront discontinuities and the uncollided source by analytically resolving the most structured part of the solution.

As documented in \cite{movingmesh}, we have already conducted extensive tests with this method on time-dependent transport problems, which allowed for a detailed analysis of the efficacy of the moving mesh and uncollided mesh for different source types. For example, for finite width, nonsmooth sources that induce a nonsmooth solution that is smoothed over time, the method proved the most beneficial when compared to a standard DG implementation, but displayed only algebraic error convergence, not the optimal geometric convergence that DG methods are capable of. \wgb{However,} \wgb{for} smooth Gaussian sources we we able to achieve spectral convergence, but the importance of the moving mesh was diminished. 

With an understanding of the effectiveness of this method on linear systems, we  apply it to nonlinear radiative transfer problems and obtain results with accuracy comparable to an analytic solution, which is the stated intent of this work. \wgb{The nonlinear problems we consider are close to the aforementioned linearized benchmarks, but with a more physical constant specific heat. While we could specify a temperature dependent opacity and provide a more physical benchmark, doing so would disallow some of the orthogonality simplifications in the DG derivation. Temperature dependent opacity will be left for a future work.} Before attempting fully nonlinear problems however, we first apply our method the existing linear radiative transfer problems. This will allow us to test our method on problems with known solutions and uncover deficiencies in a more forgiving arena. For nonlinear problems, we can still gauge the precision of our solution by inspecting magnitude of the expansion coefficients and the accuracy by checking against existing numerical $S_n$ solvers. 

We selected the Su and Olson transport benchmark \cite{SU19971035} as an ideal verification solution for our method. Unfortunately, the results are not given to enough digits to fully demonstrate the effectiveness of our scheme and recalculation of these results is non-trivial. Therefore, we rely on integration of a $P_1$ version of this benchmark \cite{mcclarren2008analytic} to create solutions which are not necessarily as physically accurate as the full transport solution, but can be \wgb{evaluated with additional precision}. 

There are no existing radiative transfer transport verification solutions for optically thick problems outside of the equilibrium diffusion limit. By optically thick, we mean that the source width or the support of an initial condition is orders of magnitude larger than a mean free path. Conversely, optically thin problems have source widths comparable to a mean free path. For a transport code to have sufficient coverage of verification problems, converging to a diffusion benchmark of a thick problem while not resolving a mean free path is a good test. If the code resolves a mean free path, however, it will converge to the diffusion problem plus a transport correction. It is for the purpose of verifying this transport correction that we include transport solutions and the $S_2$/$P_1$ solutions for optically thick problems.



The remaining sections of the paper are organized as follows. Section \ref{sec:equations} contains an introduction to our model equations, nondimensionalization, and derivation of the uncollided source. Our DG implementation is laid out briefly in Section \ref{sec:DG}, but a more detailed derivation is left to \cite{movingmesh}. Section \ref{sec:s2_bench} is devoted to the calculation of $S_2$ benchmarks and the corresponding uncollided solutions used in verifying our method. Following this is a description of how the convergence of the error in the results section is calculated (Section \ref{sec:converge}.  then our results (Sections \ref{sec:thin_results} and \ref{sec:thick_results}). The results sections  also contain specific details of the methods used in each problem and discussion of the solution characteristics.

\section{Equations}\label{sec:equations}
We study non-equilbrium time dependent radiative heat transfer in an infinite, purely absorbing, \wgb{constant opacity}, stationary medium with an internal radiation source. The radiation transport and material balance equations for this system are,
\begin{equation}\label{eq:transport_master}
    \left(\frac{1}{v}\pD{}{\tau} + \mu \pD{}{z} + \sigma_a\right)\psi(z,\tau,\mu) = \sigma_a\left(\frac{1}{2}av T(z,\tau)^4 \right)  + \frac{1}{2}S(z,\tau),
\end{equation}
\begin{equation}\label{eq:material_master}
     \pD{}{\tau}e(z,\tau)  = \sigma_a\left(\phi(z,\tau) -av T(z,\tau)^4\right),
\end{equation}
where the general form of the equation of state is,
\begin{equation}\label{eq:state_general}
    e = \int_0^T \!dT'\,\cv(T').
\end{equation}

The variables in these equations are, $\psi$, the angular flux or intensity, $\phi = \int_{-1}^{1}\!d\mu'\psi(x,t,\mu')$ 
the scalar flux, $T$, the temperature, and $e$, the material energy density.
$\psi$ and $\phi$ have units of energy per area per time ([GJ$\cdot$cm$^{-2}$ns$^{-1}$]) and $e$ has units of energy density ([GJ$\cdot$cm$^{-3}$]). $S$ is a source term with units of energy density per time. $\mu \in [-1,1]$  is the cosine of the particle direction with respect to the $z$ axis. $v$ is the particle velocity, which is the speed of light in a vacuum for our application, $v=29.998$ cm ns$^{-1}$.  The radiation constant is $a = 4\sigma_\mathrm{SB}/v=0.0137225$ GJ cm$^{-3}$ keV$^{-4}$, where $\sigma_\mathrm{SB}$ is the Stefan-Boltzman\wgb{n} constant. The absorption cross section, $\sigma_a$, is in units of inverse length.   

We seek a non-dimensionalization for these equations that is compatible with the non-dimensionalization given the in Su-Olson benchmark \cite{SU1999279} and that may be used in optically thick problems without enlarging the non-dimensionalized length to accommodate the larger opacity,
\begin{equation}
    x = l \sigma_a z \qquad t = l v \sigma_a \tau.
\end{equation}
$l$ is a dimensionless scaling variable that is set to one for thin problems and a small number to offset the greater $\sigma_a$ in optically thick problems. 

Each equation is transformed into the new variables and divided by $a v\sigma_aT_H^4$, where $T_H$ is the reference temperature, called the hohlraum temperature in previous work, 
\begin{equation}\label{eq:transport_nondim}
    \left(l\pD{}{t} + \mu l \pD{}{x} + 1 \right)\lpsi(x,t,\mu) = c_a\left(\frac{1}{2} \lT(x,t)^4 \right)  + \frac{1}{2}Q(x,t),
\end{equation}
\begin{equation}\label{eq:material_nondim}
     l\pD{}{t}\Le(x,t)  = c_a\left(\lphi(x,t) - \lT(x,t)^4\right).
\end{equation}
Our non-dimensional dependent variables are now,
\begin{equation}\label{eq:non-dim-variables}
    \lpsi(x,t) = \frac{\psi(x,t)}{a v T_H^4} \qquad \lphi(x,t) = \frac{\phi(x,t)}{a v T_H^4} \qquad \lT(x,t) = \frac{T(x,t)}{T_H} \qquad \Le(x,t) = \frac{e(x,t)}{a T_H^4}, 
\end{equation}
the non-dimensional source is,
\begin{equation}
    Q(x,t) = \frac{S(x,t)}{\sigma_a a v T_H^4 }
\end{equation}
and the absorption ratio is defined,
\begin{equation}
    c_a \wgb{=\frac{\sigma_a}{\sigma_t}}= 1.
\end{equation}
In this work, we consider two functional forms for $\cv$. To solve the Su-Olson benchmark problem, we use the familiar form,
\begin{equation}\label{eq:pomrainium}
    \cv = \alpha T^3.
\end{equation}
which renders Eq.~\eqref{eq:material_nondim} linear in $T^4$. With the conventional choice of  $\alpha = 4a$, now 
\begin{equation}\label{eq:eos_su}
    \Le_{\mathrm{SU}} = \lT^4,
\end{equation}
where the subscript ``SU" indicates that this is the equation of state for the linear Su-Olson problem.

While it is important for our investigation to solve these linear problems the novel aspect of this paper is results for nonlinear problems. For these, we choose a more physical, constant specific heat, $\cv = \mathrm{C}_\mathrm{v0},$ with units of of energy density per temperature. This choice renders $e = \mathrm{C}_\mathrm{v0} T.$ To find the relationship between the nondimensional variables with this equation of state, we define $\overline{\mathrm{C}_\mathrm{v0}} =\frac{\mathrm{C}_\mathrm{v0}}{  a T_H^3}.$ Now we can write
\begin{equation}\label{eq:eos:nonlin}
    \Le_{\mathrm{N}} = \lcvnaught  \overline{T},
\end{equation}
where the subscript ``N'' indicates that this is our equation of state for the nonlinear problems. 

\subsection{Uncollided solutions}
In time dependent transport trials, we found that the deployment of an an uncollided source treatment, where using the solution to the equation, 
\begin{equation}\label{eq:transport_uncol}
    \left(l \pD{}{t} + \mu l  \pD{}{x} + 1 \right)\lpsi_u(x,t,\mu) =   \frac{1}{2}Q(x,t),
\end{equation}
as a source term to solve for the collided flux is a significant boon for accuracy when the solution is not smooth. The Green's \wgb{function} solution to Eq.~\eqref{eq:transport_uncol} with $l = 1$ was provided by \cite{ganapol}. This solution is integrated for different source configurations in \cite{bennett2022benchmarks}, including a square and a Gaussian source. Using these solutions, we can say that $\lpsi_u$ is known and can be integrated analytically to find $\lphi_u$. For problems where $l\neq 1$, a simple scaling is required. For optically thick problems when $l \ll 1$, the uncollided solution is not as useful since it has decayed to zero by the pertinent evaluation times.

To solve for the remaining collided portion of the flux, we have the system, 
\begin{equation}\label{eq:transport_nondim_uncol_source}
    \left(l \pD{}{t}  + \mu l  \pD{}{x} + 1 \right)\lpsi_c(x,t,\mu) = c_a\left(\frac{1}{2} \lT(x,t)^4 \right)  
\end{equation}
\begin{equation}\label{eq:material_nondim_uncol_source}
     l\pD{}{t}\Le(x,t)  = c_a\left(\lphi_c(x,t) + \lphi_u(x,t) - \lT(x,t)^4\right). 
\end{equation}
In linear transport applications, it is possible to decompose the flux infinitely, not just into uncolllided and collided flux, but uncollided, first collided, second collided, etc. Even though the radiative transfer equations are nonlinear, we are able to use this linear solution technique since the uncollided flux has no interaction with the material. However, we cannot not further decompose the flux as we could in a linear transport problem. 

Answers obtained with an ``uncollided source" treatment refer to solutions \wgb{of the collided equations} \wgb{(}Eqs.~\eqref{eq:transport_nondim_uncol_source} and \eqref{eq:material_nondim_uncol_source}\wgb{)} \wgb{where the uncollided solution from Eq.~\eqref{eq:transport_uncol} is evaluated at the final time and added to the collided portion}. A ``standard source" treatment refers to Eqs.~\eqref{eq:transport_nondim} and \eqref{eq:material_nondim}. The most useful source treatment used in a specific problem is determined by the behavior of the uncollided flux during the solution time window. Tests run in \cite{movingmesh} showed that integrating the uncollided source could require more computation time than a standard source. This is because the uncollided source is a complex function of space and more difficult to integrate with quadrature than the standard source. As a rule, problems at times where the uncollided flux has not decayed enough to be a negligible portion of the flux are good candidates for an uncollided source treatment. In these problems, \cite{movingmesh} showed \wgb{an} increase in accuracy and rate of convergence. At times where the uncollided solution has decayed, the uncollided source treatment is not as helpful. 

While the problems investigated in this paper are in purely absorbing media, the coupling between the material energy density and the radiation energy density acts as a scatterer in that it can smooth discontinuities over time. For this reason, we expect that the insights derived from solving purely scattering transport problems with uncollided source treatments will extend to these purely absorbing radiative transfer problems.

\section{Moving Mesh DG spatial discretization}\label{sec:DG}
Similar to the procedure in \cite{movingmesh}, we define a DG spatial discretization with a moving mesh to solve equations of the form~\eqref{eq:transport_nondim} and~\eqref{eq:material_nondim}. We leave some of the details of the derivation to \cite{movingmesh}. To solve for the integral over $\mu$ to find the scalar flux, we discretize in angle via the method of discrete ordinates, where the solid angle $\mu \in [-1,1]$ is discretized by choosing the points with a Gauss-Lobatto quadrature rule \cite{1960ratr.book.....C} for our full transport solution or, in the case of the $S_2$ solution, a Gauss-Legendre rule. With the corresponding weights from our chosen quadrature, we can define the scalar flux as a weighted sum,
\begin{equation}\label{eq:sn_phi}
    \lphi \approx \sum^N_{n'=1} w_{n'} \lpsi^{n'},
\end{equation}
where $w_n$ are the weights and $\psi^{n}$ is the scalar flux evaluated at a given angle. This choice makes Eq.~\eqref{eq:transport_nondim} and~\eqref{eq:material_nondim},

\begin{equation}\label{eq:transport_nondim_sn}
    \left(l\pD{}{t} + \mu_n l   \pD{}{x} + 1 \right)\lpsi^n(x,t) = c_a\left(\frac{1}{2} \lT(x,t)^4 \right)  + \frac{1}{2}Q(x,t) \:\:\: \mathrm{for}\: n=1\dots N,
\end{equation}
\begin{equation}\label{eq:material_nondim_sn}
     \pD{}{t}\Le(x,t)  = c_a\left(\sum^N_{n'=1} w_{n'} \psi^{n'} - \lT(x,t)^4\right).
\end{equation}

To discretize the spatial domain, we define $K$ non-overlapping cells with time dependent edges $x_L(k,t)$ and $x_R(k,t)$. To allow simplifications to the coming weak form of the equations, we define a mapping variable $x'(k,t)$ that maps $x$ to [-1,1] inside a cell,
\begin{equation}
    x'(k,t) \equiv \frac{x_L(k,t)+x_R(k,t)-2x}{x_L(k,t)-x_R(k,t)}, \qquad k=1\dots K.
\end{equation}
Now we define an orthonormalized Legendre polynomial basis function in $x'$ for each cell, 
\begin{equation}
    B_{i,k}(x') = \frac{\sqrt{2i +1}}{\sqrt{x_R(k,t)-x_L(k,t)}}P_i(x').
\end{equation}
Therefore, the weak solution of the angular flux in a cell for a given angle is
\begin{equation}\label{eq:solution_psi}
    \lpsi^n(x,t) \approx \sum_{j=0}^M B_{j,k}(x') \,u^n_{k,j}.
\end{equation}
where $u$ is an entry in our three dimensional solution matrix. 
Likewise, the solution for the energy density in a given cell is,
\begin{equation}\label{eq:solution_e}
    \Le(x,t) \approx \sum_{j=0}^M B_{j,k}(x') \,u^{N+1}_{k,j},
\end{equation}

The standard DG procedure for finding the weak form of the equations involves multiplying each equation by a basis function, integrating over a cell, invoking integration by parts to shift the spatial derivative onto the basis function, and taking advantage of orthogonality to simplify the system. Our moving mesh method is similar to this, but with the added step of invoking the Reynolds Transport Theorem \cite{marsden2003vector} since our integration domain is time dependent. Leaving the general outline of this procedure to \cite{movingmesh}, we arrive at,
\textbf{\begin{equation}\label{eq:vectorEQ}
    \D{}{t}\boldsymbol{U_n} - \boldsymbol{\underline{\underline{G}}U_n} +  \left(\boldsymbol{\underline{\underline{L}}U_n}\right)^{(\mathrm{surf})} - \mu_n \boldsymbol{LU_n} + \frac{1}{l}\boldsymbol{U_n} =  \frac{c_a}{2l} \boldsymbol{H} + \frac{1}{2l}\boldsymbol{Q} \quad \mathrm{for} \: n=1\dots N,
\end{equation}}
\begin{equation}\label{eq:vectorEQ_e}
    \D{}{t}\boldsymbol{U_{N+1}} + \boldsymbol{RU}^{\mathrm{surf}} - \boldsymbol{\underline{\underline{G}}U}_{N+1} = \frac{c_a}{l}\left(\sum_{n'=1}^N
  w_n'\boldsymbol{U}_{n'}  - \boldsymbol{H}\right),
\end{equation}
where the time dependent solution vector is
 $$\boldsymbol{U}_{n,k} = [u^n_{k,0},u^n_{k,1},...,u^n_{k,M}]^T,$$  where $M+1$ is the number of basis functions. We also define
\begin{equation}
    L_{i,j} = \int_{x_L}^{x_R}\!\,dx\, B_{j,k}(x')\,\D{B_{i,k}(x')}{x},
\end{equation}
\begin{equation}
    G_{i,j} = \int^{x_R}_{x_L}\!dx\,B_{j,k}(x')\,\D{B_i(x')}{t},
\end{equation}
\begin{equation}\label{eq:S}
        Q_i = \int^{x_R}_{x_L}\!dx\,B_{i,k}(x')\,Q(x,t),
\end{equation}
\begin{equation}\label{eq:H}
    H_i =  \intcell B_i(x') \, \lT^4(x,t).
\end{equation}
The numerical flux terms, which calculate the direction of flow of the solution with an upwinding scheme based on the relative velocity of a particle with the cell edges
\begin{equation}
    \left(LU\right)^{\mathrm{surf}}_i =  
    \left(\mu_n- \D{x_R}{t}\right)B_{i,k}(x'=1)\boldsymbol{\lpsi^{n+}}-\left(\mu_l-\D{x_L}{t}\right)B_{i,k}(x'=-1)\boldsymbol{\lpsi^{n-}}.
\end{equation}

\begin{equation}
    \left(RU\right)^{\mathrm{surf}}_i =  
    \left(- \D{x_R}{t}\right)B_{i,k}(x'=1)\boldsymbol{\Le^{+}}-\left(-\D{x_L}{t}\right)B_{i,k}(x'=-1)\boldsymbol{\Le^{-}}.
\end{equation}
$\boldsymbol{\lpsi^{l+}}$ and $\boldsymbol{\lpsi^{l-}}$ are found by evaluating Eq.~\eqref{eq:solution_psi} and $\boldsymbol{\Le^{+}}$ and $\boldsymbol{\Le^{-}}$ are found by evaluating Eq.~\eqref{eq:solution_e}.

If we choose to employ an uncollided source treatment, Eqs.~\eqref{eq:vectorEQ} and~\eqref{eq:vectorEQ_e} change slightly in that the source term $\boldsymbol{Q}$ disappears from the RHS of Eq.~\eqref{eq:vectorEQ} and $\frac{c_a}{l}\boldsymbol{\lphi}$ is added to the RHS of Eq.~\eqref{eq:vectorEQ_e} where,
\begin{equation}
    \overline{\boldsymbol{\phi_u}} = \int_{x_L}^{x_R}\!\,dx\, B_{j,k}(x')\,\overline{\phi_u}(x,t).
\end{equation}
In this case, the numerical solution is for the collided flux, so it is necessary to add the uncollided flux at the final step to obtain the full solution.

The general solution procedure is as follows. First, parameters such as the number of basis functions, the number of \wgb{spatial cells}, and the $S_n$ order are set. A source is specified and depending on the source treatment, the uncollided solution or the standard source is integrated at each timestep with a standard Gaussian integrator with points equal to $2M+1$. The edges of the mesh are governed by a function designed to optimize the solution for the specific source. This function also returns velocities of the mesh edges in order to calculate the numerical flux. The temperature balance terms are found by the equation of state and integrated in the same way as the source term. The solver returns the coefficient arrays and the scalar flux and material energy are reconstructed via the expansions defined in Eqs.~\eqref{eq:solution_psi} and \eqref{eq:solution_e} and, depending on the source treatment, the uncollided flux is added onto the scalar flux.

To obtain solutions from our equations~\eqref{eq:vectorEQ} and~\eqref{eq:vectorEQ_e}, we calculate the quadrature weights with the \texttt{python} package \texttt{quadpy} \cite{quadpy} and integrate the \wgb{ordinary differential equations} \wgb{(}ODE\wgb{s}\wgb{)} in with a built in integrator from \texttt{scipy} \cite{Virtanen_2020}. Our \texttt{python} implementation can be found on Github\footnote{{www.github.com/wbennett39/moving\textunderscore mesh\textunderscore radiative\textunderscore transfer \cite{Bennett_Moving_Mesh_Radiative_2022} }}.


\section{Benchmarks and uncollided solutions for the $S_2$ radiative transfer equations}\label{sec:s2_bench}
In order to show \wgb{greater precision} in our linear problems than are given in \cite{SU19971035}, it was expedient to calculate our own analytic benchmarks. Since we already include results to each problem calculated with $S_2$, we choose to verify our solver by using a $S_2/P_1$ benchmark given by \cite{mcclarren2008analytic}. This benchmark gives the analytic expression for the scalar flux and energy density solutions to Eq.~\eqref{eq:transport_nondim_sn} and \eqref{eq:material_nondim_sn} with $\wgb{(}N=\wgb{)}\:2$, angles and Gauss-Legendre weighting, ($[\mu^1,\mu^2]=[\frac{-1}{\sqrt{3}}, \frac{1}{\sqrt{3}}]$, and $[w_1, w_2]=  [1,1]$) \wgb{where} the source is a delta function in space and time. \wgb{The characteristic wavespeed ($\frac{c}{\sqrt{3}}$) of the $S_2$/$P_1$ treatment of one speed transport problems}
\wgbt{is the result of an assumption made in the derivation of the $P_1$ approximation that the angular flux is an affine function of angle. This approximation causes a $\frac{1}{3}$ to multiply the gradient term in the current equation (the first angular moment of the angular flux) that limits the speed of information propagation. See \cite[p.~221]{duderstadt1979transport} for a thorough explanation. It is also interesting to note that if the radiation is modeled as a gas of photons with a specific intensity given by a Planckian distribution, the speed of sound in that gas is $\frac{c}{\sqrt{3}}$ \cite{leveque1998radiation}. }


The Green's function given by \cite{mcclarren2008analytic} for $\lphi = \lpsi_1+\lpsi_2$ for a delta function source at position $s$ is,
\begin{equation}
    G(x,s,t) = \frac{v}{2\sqrt{3}}e^{-t}\left(\frac{t I_1\left(\sqrt{t^2-3(x-s)^2}\right)}{\wgb{\sqrt{t^2-3(x-s)^2}}}\Theta\left(t-\sqrt{3}|x-s|\right) + I_0\left(\sqrt{t^2-3(x-s)^2}\right)\delta\left(t-\sqrt{3}|x-s|\right)\right),
\end{equation}
where $\Theta$ is a step function, $\delta$ is a Dirac delta function and $I_0$ and $I_1$ are modified Bessel functions of the first kind. The Green's function for the material energy density is,
\begin{equation}
    G_U(x,s,t) = \frac{\sqrt{3}}{2}e^{-t}\left( I_0\left(\sqrt{t^2-3(x-s)^2}\right)\Theta\left(t-\sqrt{3}|x-s|\right)\right).
\end{equation}
We choose to find solutions for a square source and a Gaussian source, to test our method on both smooth and nonsmooth problems. Therefore, the solution to the integral,
\begin{equation}\label{eq:s2_analytic_phi}
    \lphi_{ss,gs} = \int^\infty_{-\infty}\!ds\int^\infty_{0}\!dt'\, S_{ss,gs}(s,t')\, G(x,s,t-t')
\end{equation}
gives the scalar flux. The energy density is likewise obtained by
\begin{equation}\label{eq:s2_analytic_e}
    \Le_{ss,gs} = \int^\infty_{-\infty}\!ds\int^\infty_{0}\!dt'\, S_{ss,gs}(s,t')\, G_u(x,s,t-t').
\end{equation}
\wgb{Here} the subscript in the solution and the source is either ``ss" for square source or ``gs" for Gaussian. The source term is
\begin{equation}\label{eq:square_source}
    S^{ss}(x,t) = \Theta(x_0-x)\Theta(t_0-t),
\end{equation}
for the square source, or
\begin{equation}\label{eq:gaussian_source}
    S^{gs}(x,t) = \exp\left(\frac{\wgb{-}x^2}{x_0^2}\right)\Theta(t_0-t),
\end{equation}
for the Gaussian source. 

Since finding a benchmark for an optically thick problem requires evaluating these integrals at extremely late times where the integrand is not well behaved, we only calculate $S_2$ benchmarks for our thin problems. 

\subsection{$S_2$ uncollided solutions}
The uncollided solutions that we have utilized so far for the uncollided source treatment have been full transport solutions from \cite{bennett2022benchmarks}. We cannot use these to solve the $S_2$ transport equation, since the two uncollided fluxes are not equal. The full transport solutions are based on the assumption that the $S_n$ order of the ODE's sufficiently resolves the angular error and that the collided flux calculated with quadrature is a good approximation of the analytic integral over $\mu$, i.e. 
\begin{equation}\label{eq:sn_phi_assumption}
    \sum^N_{n'=1} w_{n'} \lpsi^{n'} \approx \int^1_{-1}\!d\mu'\,\psi(x,t,\mu'),
\end{equation}
so that it is acceptable to employ the uncollided scalar flux found by integrating analytically the solution for the uncollided angular flux. In the $S_2$ equations, the assumption of Eq.~\eqref{eq:sn_phi_assumption} does not hold and the uncollided scalar flux must be found by numerical quadrature of the angular flux. 

Therefore, the process for finding the uncollided scalar flux to use as a source in the $S_2$ solutions to our radiative transfer problems is to  find the Green's solution for the angular flux, integrate that solution with quadrature, and then integrate again over the given source. The uncollided solution to Eq.~\ref{eq:transport_uncol} with a delta function source $(\delta(x)\delta(t))$ is \cite{ganapol},
\begin{equation}\label{eq:uncol_delta}
    \lpsi_u(x,t) = \frac{e^{-t}}{2t}\delta\left(\mu-\frac{x}{t}\right).
\end{equation}
 To find the $S_2$ uncollided scalar flux, the integral is done by Gauss-Legendre quadrature with $N=2$ to give the uncollided scalar flux, 
\begin{equation}\label{eq:s2_scalar_uncol}
     \lphi_u^{pl}(x,t) = \frac{e^{-t}}{2t}\left(\delta\left(-\frac{1}{\sqrt{3}}-\frac{x}{t}\right) + \delta\left(\frac{1}{\sqrt{3}}-\frac{x}{t}\right) \right).
\end{equation}
To finally find the uncollided scalar flux that corresponds to the benchmark solutions calculated with Eqs.~\eqref{eq:s2_analytic_phi}, we integrate
\begin{equation}\label{eq:uncol_s2_greens}
    \lphi^{ss,gs}_u(x,t) = \int^\infty_{-\infty}\!ds\int^{\infty}_0\!dt' \, \lphi_u^{pl}(x-s,t-t') \, S^{ss, gs}(s,t'),
\end{equation}
where $S^{ss,gs}$ is given by Eq.~\eqref{eq:square_source} or Eq.~\eqref{eq:gaussian_source}. Solutions to Eq.~\eqref{eq:uncol_s2_greens} are given in Appendix \ref{sec:s2_uncol}.

\section{Error estimation methods}\label{sec:converge}
In the problems presented, two methods are used to estimate the solution accuracy. For problems with a benchmark solution, we use the root mean square error (RMSE) as our error metric. This is calculated by
\begin{equation}
    \mathrm{RMSE} = \sqrt{\sum^N_i\frac{|y_i - \hat{y_i}|^2}{N}},
\end{equation}
where $y_i$ is either the calculated scalar flux or the calculated material energy density at a given node, $\hat{y_i}$ is the corresponding benchmark solution, and $N$ is the total number of nodes in the computational solution.

For problems that demonstrate geometric spectral convergence, as $M\rightarrow \infty$, the error can be modeled as 
\begin{equation}\label{eq:spectral}
    \mathrm{ERROR} = C \,\exp(-c_1 M ),
\end{equation}
where $M$ is the highest polynomial order of the basis and $C$ and $c_1$ are constants that could depend on the number of cells used in the problem. This curve is a straight line on a logarithmic-linear scale. 

For all of the problems in the following section, we plot the average of the absolute value of the coefficients in the solution expansion to characterize the solution convergence. We define the average value of the $j^{\mathrm{th}}$ coefficient in the solution expansion,
\begin{equation}\label{eq:coeffs_phi}
    |c_j| = \frac{\sum_{k=1}^{K} |a_{\wgb{j,k}}|}{K},
\end{equation}
where $j$ corresponds to the order of the Legendre polynomial in the basis and $K$ is the number of cells.

When characterizing the error of $\lphi$, since we are interested in the residual error of scalar flux, $a_{\wgb{j,k}}$ is the weighted average using the weights from Eq.~\eqref{eq:sn_phi},
\begin{equation}
    a_{\wgb{j,k}} = \frac{\sum_{l'=1}^N w_{l'} u_{(l',k,j)}}{\sum_{l'=1}^N w_{l'}}.
\end{equation}
For the material energy density, $a_{\wgb{j,k}}$ is, 
\begin{equation}
    a_{\wgb{j,k}} = u_{(N+1, k,j)}.
\end{equation}

\section{Optically thin results}\label{sec:thin_results}
The results in this section are for problems where the source width is equal to a mean free path, meeting our definition of an optically thin problem. These problems are characterized by solutions where the uncollided solution is a significant portion of the flux and travelling wavefronts. Therefore, the problems in this section all use an uncollided source and the square sources which have travelling discontinuities employ a moving mesh. 
\subsection{Su-Olson problem with a square source}\label{subsec:su_square_thin}
We first replicate the Su-Olson problem  using the same square source
originally presented in \cite{SU19971035} with $\sigma_a=1$ cm$^{-1}$, the source width $x_0=0.5$ and the source duration $t_0=10$. The uncollided solution for this source has already been presented in \cite{bennett2022benchmarks}. For the $S_2$ treatment of this problem, the uncollided source is given by Eq.~\eqref{eq:uncol_square_s2}. The temperature is calculated by Eq.~\eqref{eq:eos_su}. 

Some modifications were made to the original mesh function invented to solve the square source transport problem in \cite{movingmesh}. In that mesh, the mesh edges inside the source never moved while the edges outside travelled outwards with the wavespeed. This was done to resolve the static discontinuities at the source edge and the travelling discontinuities at the wavefront.  In the original Su-Olson results, the source turns off at $t_0=10$ and solutions are required long afterwards ($t=31.6228, 100$). With our previous square source mesh, the edges would remain clustered around the source region long after the source has ceased to introduce nonsmoothness. This is not the optimal distribution of computational zones. 

Therefore, the mesh function used in this problem is as follows. If the mesh edges are defined as the vector,
\begin{equation}\label{eq:edge_vector}
    \boldsymbol{X}(t) = \left[x^0(t), x^1(t),...,x^{K}(t)\right],
\end{equation}
and initialized to be,
\begin{align}
    \mathrm{if}\:\frac{K}{4}\leq k \leq \frac{3K}{4}, \:\: x^k_o = \wgb{x_0} y_j, \\
    \mathrm{if}\: k < \frac{K}{4}, \:\: x^k_o = \frac{s_k(\delta x) \wgb{-} 2x_0 \wgb{-} \delta x}{2} \\
        \mathrm{if}\:\: k > \frac{3K}{4}, \:\: x^k_o = \frac{s_l(\delta x) \wgb{+} 2x_0 \wgb{+} \delta x}{2}
\end{align}
where $y_j$ are the Gauss-Lobatto evaluation points with $N$, the number of points, equal to $\frac{K}{2}+1$ numbered from $0$ and $s_m$ are the Gauss-Lobatto evaluation points for $N = \frac{K}{4} + 1$. The indices $j$ and \wgb{$l$} are equal to $k-\frac{K}{4}$ and $k- \wgb{\frac{3K}{4}}$ respectively. $\delta x$ is a small initial width, and $K$ is always an even number. This initialization assigns one third of the edges to the source and the other two thirds to cover the rest of the solution domain. Each subdomain is spanned by edges with Gauss-Lobatto spacing, which has the effect of concentrating cells near the source edges and the outgoing wavefronts, where discontinuities are most likely. 

As time progresses and the outside edges move outwards with the solution, their position is defined as,
\begin{equation}\label{eq:mesh_edges_square1}
    \mathrm{if}\:\: t \leq t_0, \:\:x^k(t) = x^k_{o}  +\frac{x^k}{x^k_{o}} \times v t,
\end{equation}
where $v$ is the wavespeed,\wgb{c} for the transport problems and $\frac{\wgb{c}}{\sqrt{3}}$ for the $S_2$ problems. The edge velocity is defined, 
\begin{equation}\label{eq:mesh_vel_square1}
    \mathrm{if}\:\: t \leq t_0, \:\:\D{x^k}{t} = \frac{x^k}{x^k_{o}} \times v t,
\end{equation}

Defining the edge positions and velocities this way preserves the relative spacing of the initialized edges, meaning that the edges are clustered at the source edges and the leading wavefronts. 

At later times when the source is off, the solution to a square source in an optically thin problem will behave more like the solution for a Gaussian source since the solution will become smoother without the source emitting uncollided particles. Information flow is no longer dominated by the wavespeed. The solution will be practically zero some distance from the origin that is much less than $v t$. For instance, in the Su-Olson problem at $t=100$ the solution is practically zero past $x = \pm 30$. 

For these reasons, when the source turns off, a constant acceleration will divert the trajectory of each edge so that at the final time, they are evenly spaced over a specified width. This width is an estimate to how far the solution will have traveled by the evaluation time. We chose a constant acceleration instead of a instantaneous velocity change because the latter induced numerical errors which resulted in failure to converge to the benchmark solutions. The acceleration for each edge is found by, 
\begin{equation}
    c^k = 2  \frac{\left(\D{x^k}{t}\eval{t_0}  (t_0 -t_{\mathrm{final}}) - x^k\eval{t_0} + x^k\eval{t_\mathrm{final}}\right)}{\left(t_0-t_\mathrm{final}\right)^2}
\end{equation}
where we find $x^k\eval{t_\mathrm{final}}$ by specifying that the final positions vector $\boldsymbol{X}(t_\mathrm{final})$ evenly spans $[-x_f/2, x_f/2]$\wgb{,} where $x_f$ is our estimate for the width of the solution domain at the evaluation time. 

\begin{figure}
    \centering
    \begin{subfigure}[b]{0.48\textwidth}
    \includegraphics[width=\textwidth]{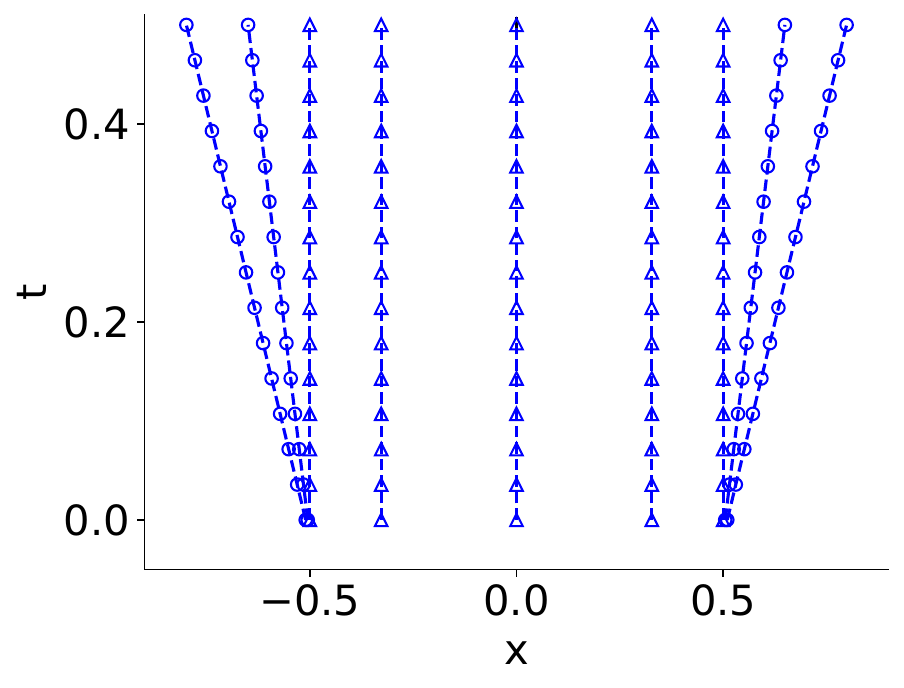}
    \caption{\wgb{t = 0.5}}
    \label{subfig:edgesearly}
    \end{subfigure}
    \begin{subfigure}[b]{0.48\textwidth}
    \includegraphics[width=\textwidth]{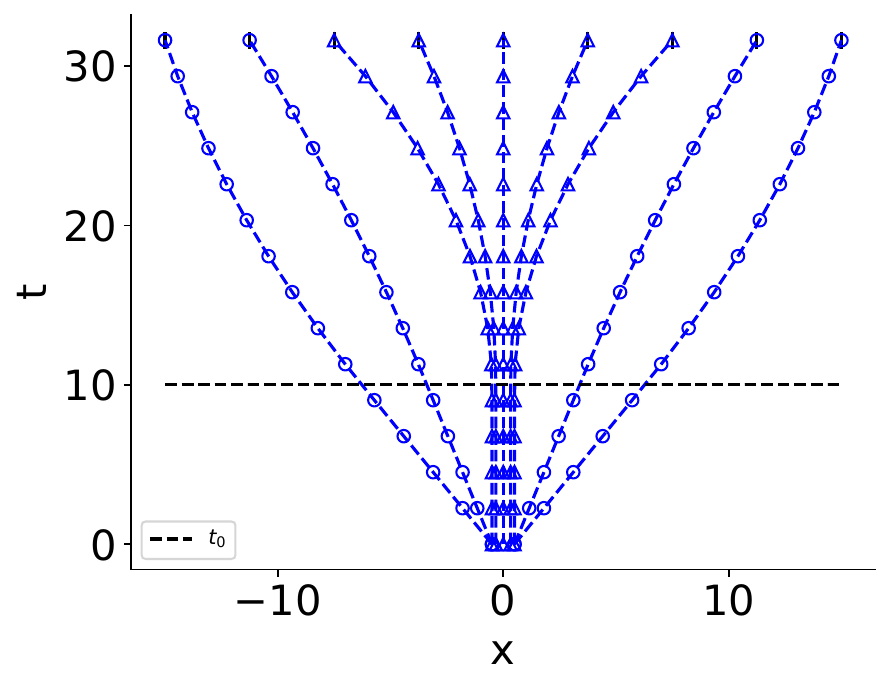}
    \caption{t = 31.6228}
    \label{subfig:edgeslate}
    \end{subfigure}
    \caption{Edge position \wgb{at early and late times} for each edge in the thin square source mesh with $8$ spaces, a wavespeed $\frac{\wgb{c}}{\sqrt{3}}$, and a final domain width $x_f = 30$.}
    \label{fig:xvst}
\end{figure}

With the acceleration defined, we calculate positions of the edges after the source has turned off with, 
\begin{equation}
    \mathrm{if} \:\: t > t_0,\:\: x^k(t) = \frac{1}{2} c^k \left(t-t_0\right)^2 + \D{x^k}{t} \eval{t_0} \left(t-t_0\right) + x^k\eval{t_0}, 
\end{equation}
and the velocities,
\begin{equation}
    \mathrm{if}\:\: t > t_0, \D{x^k}{t} =  c^k \left(t-t_0\right) + \D{x^k}{t}\eval{t_0}.
\end{equation}
We refer to this method for governing the mesh edges and velocities as the ``thin square source mesh''. \wgb{E}xample $x$ vs $t$ diagram\wgb{s} of the mesh edges \wgb{are} given in Figure \ref{fig:xvst} for clarification. \wgb{Figure \ref{subfig:edgesearly} shows the Legendre spacing of the edges inside the source and the clustering of edges outside the source around the wavefront and Figure \ref{subfig:edgeslate} shows how the edges relax into an even pattern at later times.}

\begin{figure}
    \centering
    \begin{subfigure}[b]{0.48\textwidth}
    \includegraphics[width=\textwidth]{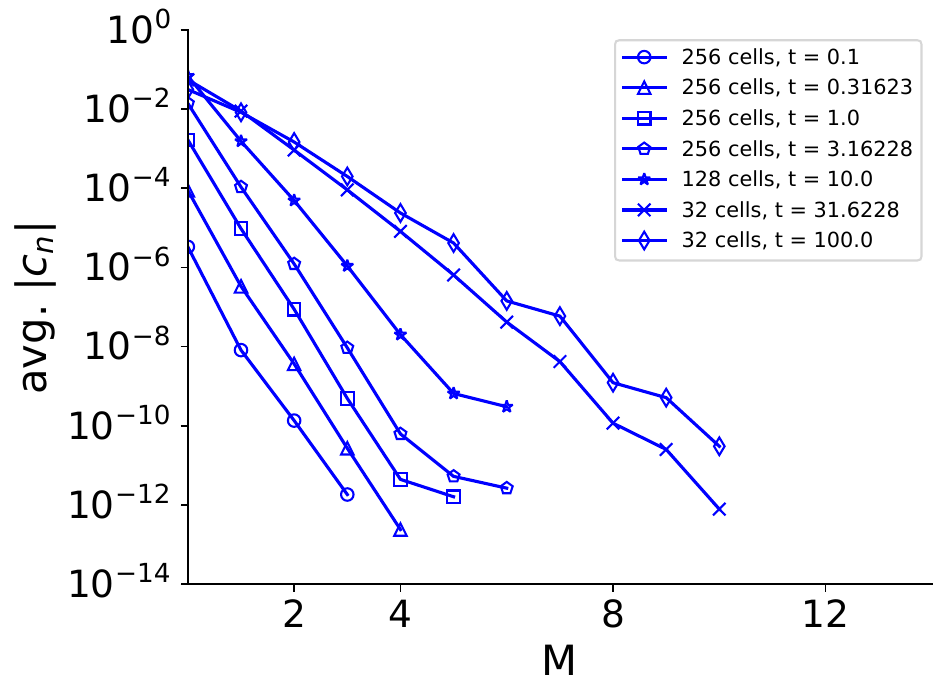}
    \caption{Radiation energy density, $\lphi$}
    \label{subfig:su_square_phi}
    \end{subfigure}
    \centering
    \begin{subfigure}[b]{0.48\textwidth}
        \includegraphics[width=\textwidth]{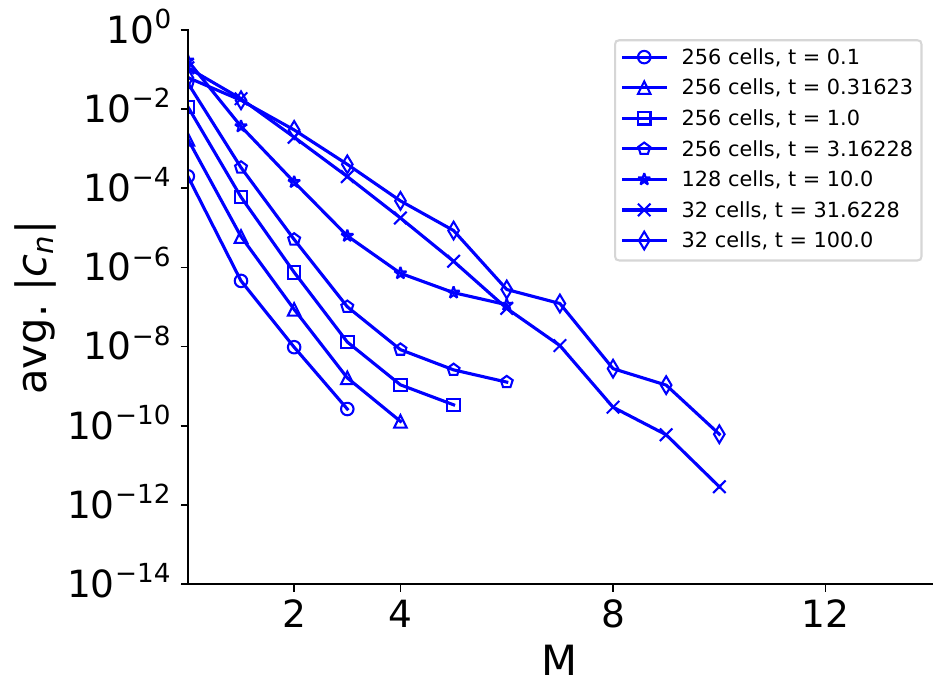}
        \caption{Material energy density, $\Le$}
        \label{subfig:su_square_e}
    \end{subfigure}
      \caption{Log-linear scaled average value of the solution expansion coefficients (found by Eqs.~\eqref{eq:coeffs_phi}) for the optically thin ($\sigma_a=1$ cm$^{-1}$) Su-Olson square source problem where $x_0=0.5$, $t_0=10$. The quadrature order for all results is $S_{256}$. All results were calculated with a moving mesh and uncollided source treatment.}
    \label{fig:su_square_convergence}
\end{figure}

\begin{figure}
    \centering
    \begin{subfigure}[b]{0.48\textwidth}
    \includegraphics[width=\textwidth]{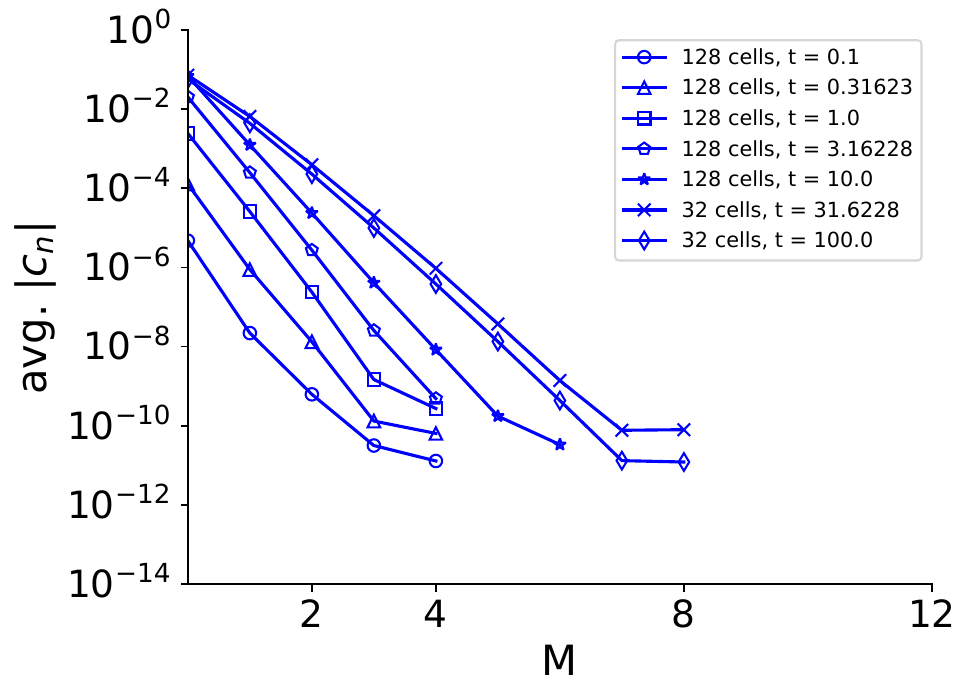}
    \caption{Radiation energy density, $\lphi$}
    \label{subfig:su_square_s2_phi}
    \end{subfigure}
    \centering
    \begin{subfigure}[b]{0.48\textwidth}
        \includegraphics[width=\textwidth]{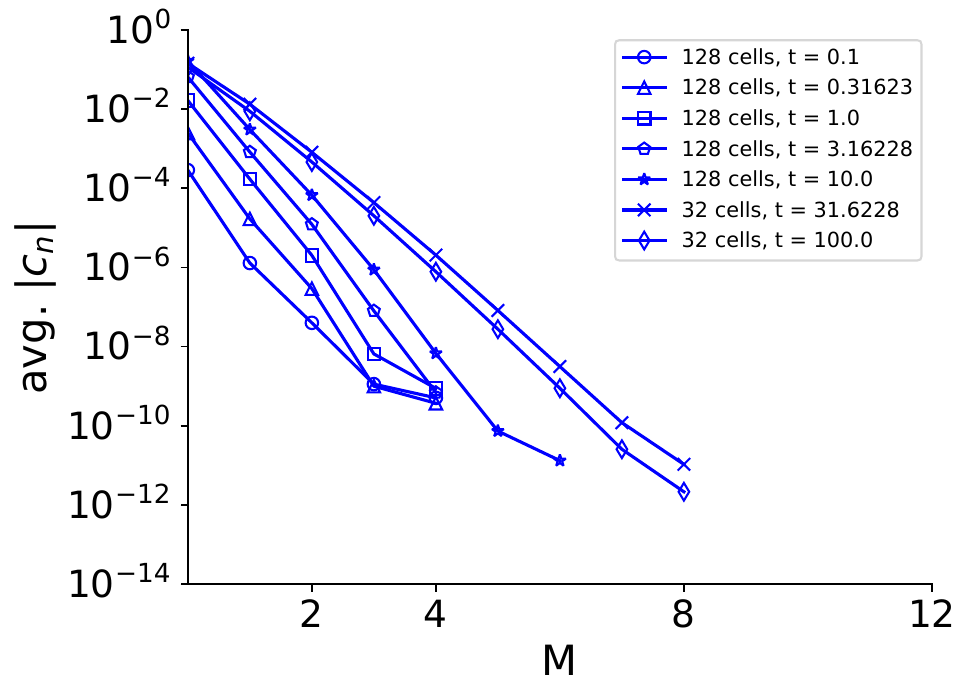}
        \caption{Material energy density, $\Le$}
        \label{subfig:su_square_s2_e}
    \end{subfigure}
      \caption{Log-linear scaled average value of the solution expansion coefficients (found by Eqs.~\eqref{eq:coeffs_phi})) for the optically thin ($\sigma_a=1$ cm$^{-1}$) $S_2$ Su-Olson square source problem where $x_0=0.5$, $t_0=10$. All results were calculated with a moving mesh and uncollided source treatment except for the $t=31.6228$ and $t=100$ cases where a standard source treatment was used.}
    \label{fig:su_square_s2_convergence}
\end{figure}

\begin{figure}
    \centering
    \begin{subfigure}[b]{0.3\textwidth}
    \includegraphics[width=\textwidth]{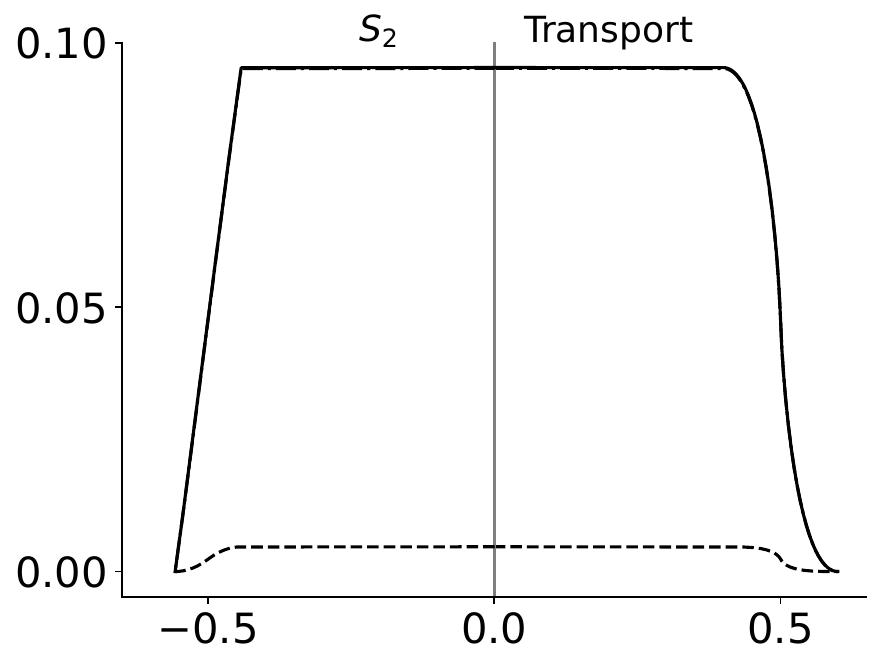}
    \caption{$t=0.1$}
    \end{subfigure}
    \centering
    \begin{subfigure}[b]{0.3\textwidth}
        \includegraphics[width=\textwidth]{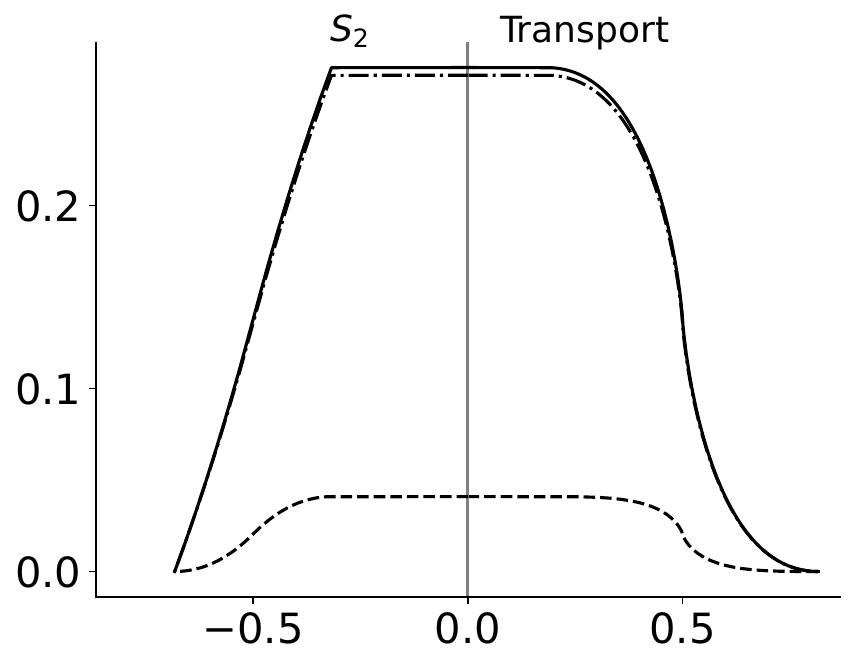}
        \caption{$t=0.31623$}
    \end{subfigure}
    \centering
    \begin{subfigure}[b]{0.3\textwidth}
        \includegraphics[width=\textwidth]{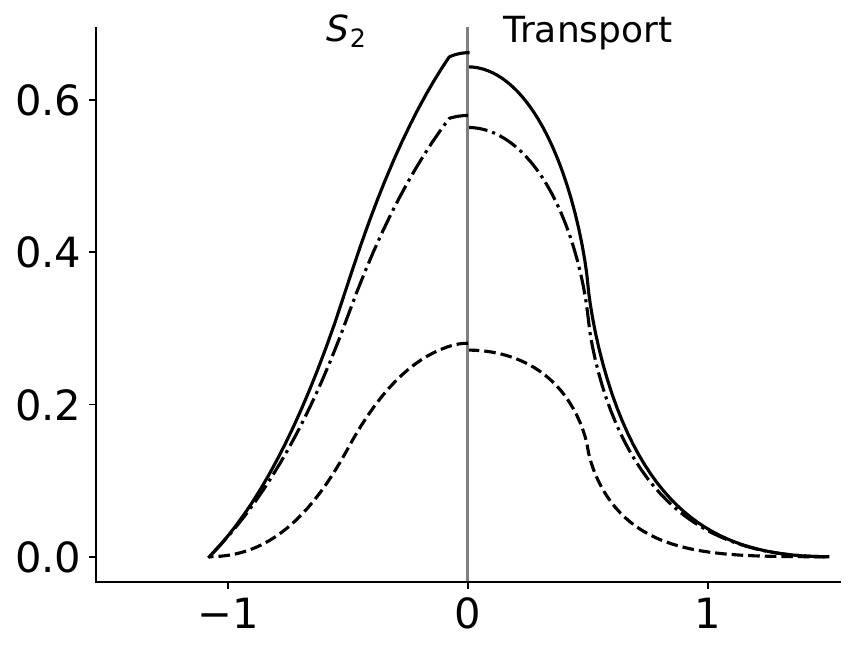}
        \caption{$t=1$}
    \end{subfigure}
    \begin{subfigure}[b]{0.3\textwidth}
        \includegraphics[width=\textwidth]{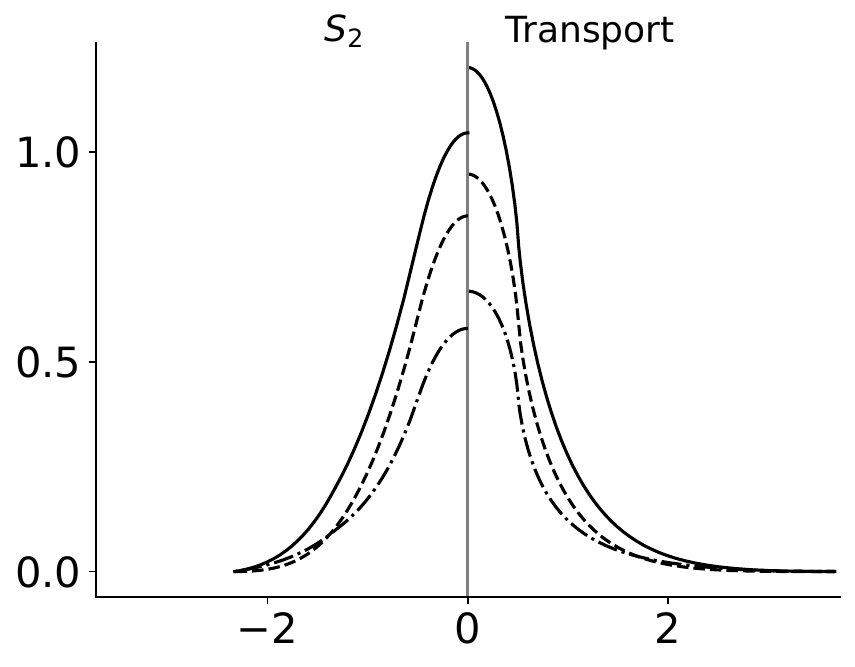}
        \caption{$t=3.16228$}
        
    \end{subfigure}
    \begin{subfigure}[b]{0.3\textwidth}
        \includegraphics[width=\textwidth]{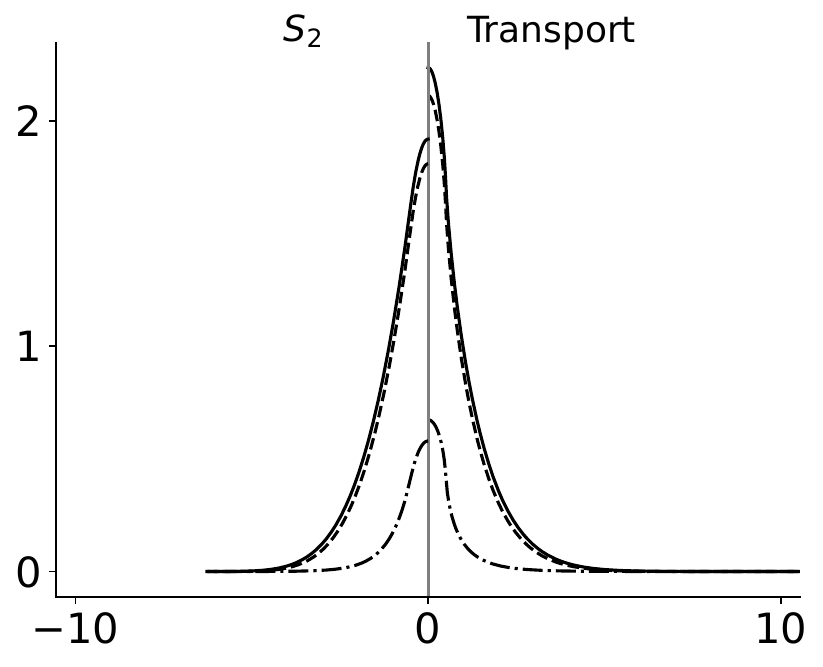}
        \caption{$t=10$}
    \end{subfigure}
    \begin{subfigure}[b]{0.3\textwidth}
        \includegraphics[width=\textwidth]{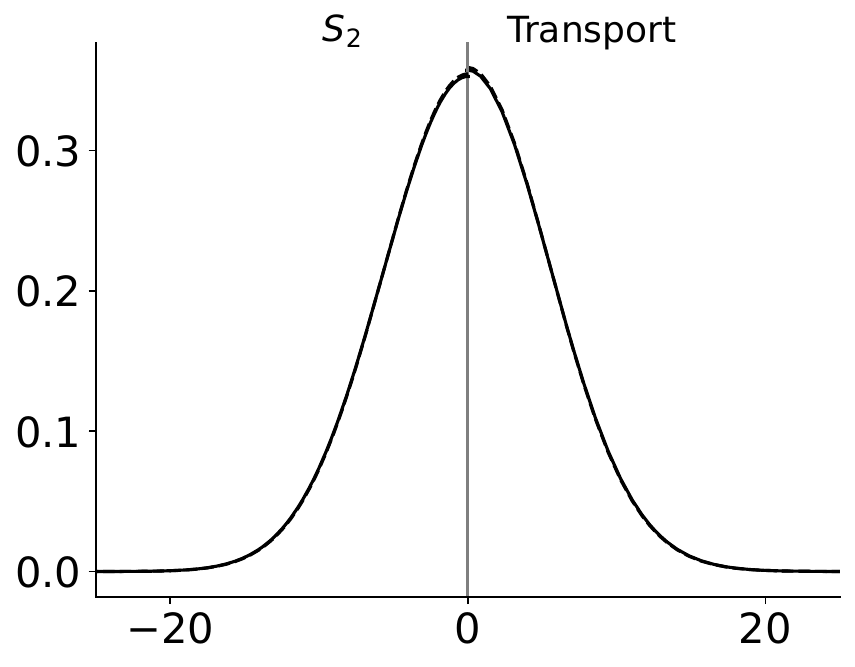}
        \caption{$t=100$}
    \end{subfigure}
      \caption{$S_2$ (left of $x=0$) and full transport (right of $x=0$) solutions for the optically thin Su-Olson square source problem with $x_0=0.5$, $t_0=10$. Solid lines are scalar flux, $\lphi$, \wgb{dash-dotted lines are the uncollided scalar flux, $\lphi_u$} and dashed are material energy density, $\Le$.}
    \label{fig:su_square_solutions} 
\end{figure}

After completing the necessary steps of defining a source, choosing a functional form for the temperature, and defining a mesh, we may present our results for this problem. Tables \ref{tab:su_square_phi} and \ref{tab:su_square_e} give our solutions, with digits that agree bolded, for the same points and evaluation times as in Table 1 and 2 of \cite{SU19971035}. The convergence results for the coefficient expansions of these results are plotted in Figures \ref{fig:su_square_convergence} and \ref{fig:su_square_s2_convergence}. The solutions at a few selected times are plotted in Figure \ref{fig:su_square_solutions}\wgb{.} For each case, a moving mesh and the uncollided source was used except in the $S_2$ solution at times greater than $t_0$\wgb{,} since the $S_2$ uncollided solution becomes sharp and difficult to resolve via quadrature. In these cases, the standard square source was integrated. 


The convergence results show that for this problem, $S_2$ is considerably smoother at early times. Twice as many spatial divisions were required at early times in the full transport solution $256$ to achieve similar levels of convergence to the $S_2$ solutions. Both cases exhibited similar behavior over time. At early times the significantly nonsmooth uncollided flux induced discontinuities in the material energy density and required far more spatial divisions to resolve, stiffening the problem and limiting the number of basis functions that could reasonably be used in the solution. After the source turned off, the solution smoothed and equilibrated locally (complete equilibrium is impossible with an infinite material). The solution became smoother and could easily be resolved with fewer \wgb{spatial cells} and more basis functions. We also note that at later times after the source has turned off, the full transport and $S_2$ solutions become more similar. 

\begin{table}[!ht]
    \centering
    \caption{Transport (top) and $S_2$ (bottom) results for the scalar flux, $\lphi$ for the thin square source Su-Olson problem with $x_0=0.5$, $t_0=10$}
    
    \begin{tabular}{|l|l|l|l|l|l|l|l|}
    \hline
        $\textbf{x/t}$ & 0.1 & 0.31623 & 1.0 & 3.16228 & 10.0 & 31.6228 & 100.0 \\ \hline
        0.01 & \textbf{0.09531}7 & \textbf{0.2752}94 & \textbf{0.643}151 & \textbf{1.200}69 & \textbf{2.235}815 & \textbf{0.690}187 & \textbf{0.357}195 \\ \hline
        0.1 & \textbf{0.09531}7 & \textbf{0.2752}94 & \textbf{0.635}943 & \textbf{1.1887}24 & \textbf{2.219}553 & \textbf{0.6897}43 & \textbf{0.3571}37 \\ \hline
        0.17783 & \textbf{0.0953}17 & \textbf{0.2752}94 & \textbf{0.619}626 & \textbf{1.16}2044 & \textbf{2.183}558 & \textbf{0.6887}73 & \textbf{0.357}011 \\ \hline
        0.31623 & \textbf{0.095}317 & \textbf{0.262}715 & \textbf{0.5618}96 & \textbf{1.071}861 & \textbf{2.064}534 & \textbf{0.685}719 & \textbf{0.3566}12 \\ \hline
        0.45 & \textbf{0.0882}4 & \textbf{0.20312}8 & \textbf{0.44711}4 & \textbf{0.9095}26 & \textbf{1.8607}58 & \textbf{0.6811}68 & \textbf{0.35}6016 \\ \hline
        0.5 & \textbf{0.04765}8 & \textbf{0.1376}47 & \textbf{0.3580}83 & \textbf{0.79902}7 & \textbf{1.731}816 & \textbf{0.6790}72 & \textbf{0.35574} \\ \hline
        0.56234 & \textbf{0.0037}62 & \textbf{0.06277}6 & \textbf{0.2537}22 & \textbf{0.666}804 & \textbf{1.5749}55 & \textbf{0.6761}6 & \textbf{0.3553}55 \\ \hline
        0.75 & - & \textbf{0.002}793 & \textbf{0.1143}15 & \textbf{0.44675}2 & \textbf{1.27398}4 & \textbf{0.6654}59 & \textbf{0.3539}29 \\ \hline
        1.0 & - & - & \textbf{0.0364}71 & \textbf{0.275}396 & \textbf{0.9878}15 & \textbf{0.6469}22 & \textbf{0.3514}09 \\ \hline
        1.33352 & - & - & \textbf{0.0028}94 & \textbf{0.1453}09 & \textbf{0.70822}1 & \textbf{0.61538}1 & \textbf{0.34697}2 \\ \hline
        1.77828 & - & - & - & \textbf{0.0596}74 & \textbf{0.45016}3 & \textbf{0.5635}09 & \textbf{0.3392}23 \\ \hline
        3.16228 & - & - & - & \textbf{0.001}155 & \textbf{0.096}453 & \textbf{0.36965}9 & \textbf{0.30346}6 \\ \hline
        5.62341 & - & - & - & - & \textbf{0.003}632 & \textbf{0.10830}5 & \textbf{0.2138}18 \\ \hline
        10.0 & - & - & - & - & - & \textbf{0.0039}14 & \textbf{0.0720}59 \\ \hline
        17.78279 & - & - & - & - & - & - & \textbf{0.00272}1 \\ \hline
        \hline
        \hline
        $\boldsymbol{x/t}$ & 0.1 & 0.31623 & 1.0 & 3.16228 & 10.0 & 31.6228 & 100.0 \\ \hline
        0.01 & \textbf{0.09531}7 & \textbf{0.2752}94 & \textbf{0.6}61668 & \textbf{1}.0451 & 1.918396 & \textbf{0.6}59852 & \textbf{0.35}2728 \\ \hline
        0.1 & \textbf{0.09531}7 & \textbf{0.2752}94 & \textbf{0.6}45629 & \textbf{1}.034885 & 1.906562 & \textbf{0.6}59502 & \textbf{0.35}2673 \\ \hline
        0.17783 & \textbf{0.0953}17 & \textbf{0.2752}94 & \textbf{0.6}04866 & \textbf{1}.012559 & 1.88071 & \textbf{0.6}58738 & \textbf{0.35}2555 \\ \hline
        0.31623 & \textbf{0.095}317 & \textbf{0.2}75294 & \textbf{0.5}15344 & \textbf{0.9}41841 & \textbf{1.7}98924 & \textbf{0.6}56328 & \textbf{0.35}2181 \\ \hline
        0.45 & \textbf{0.08}9213 & \textbf{0}.179984 & \textbf{0.4}05572 & \textbf{0}.835497 & \textbf{1}.67622 & \textbf{0.6}52732 & \textbf{0.35}1621 \\ \hline
        0.5 & \textbf{0.04765}8 & \textbf{0.1376}47 & \textbf{0.358}083 & \textbf{0.7}86079 & \textbf{1}.619313 & \textbf{0.6}51071 & \textbf{0.35}1362 \\ \hline
        0.56234 & \textbf{0}.000000 & \textbf{0.0}85419 & \textbf{0.2}99433 & \textbf{0}.722866 & \textbf{1.5}45756 & \textbf{0.6}48764 & \textbf{0.35}1001 \\ \hline
        0.75 & - & \textbf{0}.0000000 & \textbf{0.1}55119 & \textbf{0}.552606 & \textbf{1}.338003 & \textbf{0.6}40252 & \textbf{0.3}49662 \\ \hline
        1.0 & - & - & \textbf{0.0}27203 & \textbf{0}.369788 & 1.092239 & \textbf{0.6}25397 & \textbf{0.3}47295 \\ \hline
        1.33352 & - & - & \textbf{0}.000000 & \textbf{0.1}9493 & \textbf{0}.816475 & \textbf{0}.599795 & \textbf{0.34}3124 \\ \hline
        1.77828 & - & - & - & \textbf{0.05}6328 & \textbf{0}.532488 & \textbf{0.5}56761 & \textbf{0.33}5829 \\ \hline
        3.16228 & - & - & - & - & \textbf{0.09}838 & \textbf{0.3}84521 & \textbf{0.30}198 \\ \hline
        5.62341 & - & - & - & - & \textbf{0.00}028 & \textbf{0.1}16245 & \textbf{0.21}5656 \\ \hline
        10.0 & - & - & - & - & - & \textbf{0.00}2009 & \textbf{0.07}3828 \\ \hline
        17.78279 & - & - & - & - & - & - & \textbf{0.002}307 \\ \hline
        $\textbf{RMSE}$ & 2.656e-07 & 1.747e-07 & 1.642e-06 & 3.589e-07 & 2.647e-07 & 9.157e-08 & 6.128e-09 \\ \hline
    \end{tabular}
    \label{tab:su_square_phi}
\end{table}

\begin{table}[!ht]
    \centering
    \caption{Transport (top) and $S_2$ (bottom) results for the material energy density, $\Le$  for the thin square source Su-Olson problem with $x_0=0.5$, $t_0=10$}

    \begin{tabular}{|l|l|l|l|l|l|l|l|}
    \hline
        \textbf{x/t} & 0.1 & 0.31623 & 1.0 & 3.16228 & 10.0 & 31.6228 & 100.0 \\ \hline
        0.01 & \textbf{0.00468}2 & \textbf{0.04093}5 & \textbf{0.271}307 & \textbf{0.946}87 & \textbf{2.111}923 & \textbf{0.70499}1 & \textbf{0.3591}36 \\ \hline
        0.1 & \textbf{0.00468}2 & \textbf{0.04093}5 & \textbf{0.268}692 & \textbf{0.9371}54 & \textbf{2.095}970 & \textbf{0.7045}14 & \textbf{0.3590}78 \\ \hline
        0.17783 & \textbf{0.00468}2 & \textbf{0.04093}5 & \textbf{0.2626}4 & \textbf{0.915}402 & \textbf{2.060}646 & \textbf{0.7034}74 & \textbf{0.3589}49 \\ \hline
        0.31623 & \textbf{0.00468}2 & \textbf{0.0403}4 & \textbf{0.239}814 & \textbf{0.840}926 & \textbf{1.943}709 & \textbf{0.700}198 & \textbf{0.35854}4 \\ \hline
        0.45 & \textbf{0.00455}2 & \textbf{0.03314}2 & \textbf{0.18826}4 & \textbf{0.7028}83 & \textbf{1.7429}67 & \textbf{0.69532} & \textbf{0.35793}8 \\ \hline
        0.5 & \textbf{0.00234}2 & \textbf{0.02046}9 & \textbf{0.141}918 & \textbf{0.6049}35 & \textbf{1.615}402 & \textbf{0.6930}73 & \textbf{0.3576}57 \\ \hline
        0.56234 & \textbf{0.00005} & \textbf{0.00635} & \textbf{0.08838} & \textbf{0.4884}6 & \textbf{1.460}394 & \textbf{0.6899}54 & \textbf{0.3572}66 \\ \hline
        0.75 &- & \textbf{0.0000}63 & \textbf{0.03014}1 & \textbf{0.3065}58 & \textbf{1.16591}2 & \textbf{0.678}498 & \textbf{0.35581}6 \\ \hline
        1.0 &- &- & \textbf{0.00625} & \textbf{0.17519}2 & \textbf{0.8899}08 & \textbf{0.65868}5 & \textbf{0.3532}54 \\ \hline
        1.33352 &- &- & \textbf{0.0001}62 & \textbf{0.08352} & \textbf{0.62521}3 & \textbf{0.6250}66 & \textbf{0.3487}44 \\ \hline
        1.77828 &- &- &- & \textbf{0.0293}49 & \textbf{0.38688}4 & \textbf{0.5700}27 & \textbf{0.3408}7 \\ \hline
        3.16228 &- &- &- & \textbf{0.000}183 & \textbf{0.076}146 & \textbf{0.3672}69 & \textbf{0.30}4561 \\ \hline
        5.62341 &- &- &- &- & \textbf{0.002}412 & \textbf{0.1031}14 & \textbf{0.2137}68 \\ \hline
        10.0 &- &- &- &- &- & \textbf{0.00342}6 & \textbf{0.07122}6 \\ \hline
        17.78279 &- &- &- &- &- &- & \textbf{0.0026}09 \\ \hline
        \hline
        \hline
        \hline
        \textbf{x/t} & 0.1 & 0.31623 & 1.0 & 3.16228 & 10.0 & 31.6228 & 100.0 \\ \hline
        0.01 & \textbf{0.00468}2 & \textbf{0.04093}5 & \textbf{0.2}80241 & \textbf{0}.847357 & 1.808991 & \textbf{0}.672725 & \textbf{0.35}4597 \\ \hline
        0.1 & \textbf{0.00468}2 & \textbf{0.04093}5 & \textbf{0.2}73782 & \textbf{0}.837875 & 1.797308 & \textbf{0}.672354 & \textbf{0.35}4541 \\ \hline
        0.17783 & \textbf{0.00468}2 & \textbf{0.04093}5 & \textbf{0.26}1727 & \textbf{0}.817144 & 1.771785 & \textbf{0}.671544 & \textbf{0.35}4421 \\ \hline
        0.31623 & \textbf{0.00468}2 & \textbf{0.04093}6 & \textbf{0.2}25609 & \textbf{0}.751409 & \textbf{1}.691033 & \textbf{0}.668989 & \textbf{0.35}4041 \\ \hline
        0.45 & \textbf{0.004}642 & \textbf{0.03}0469 & \textbf{0.1}69249 & \textbf{0}.652361 & \textbf{1}.569863 & \textbf{0.6}65177 & \textbf{0.35}3472 \\ \hline
        0.5 & \textbf{0.00234}1 & \textbf{0.02046}7 & \textbf{0.141}916 & \textbf{0.60}6256 & \textbf{1}.513661 & \textbf{0.6}63418 & \textbf{0.35}3209 \\ \hline
        0.56234 & \textbf{0.00000}0 & \textbf{0.00}8542 & \textbf{0}.10839 & \textbf{0}.547588 & \textbf{1.4}41078 & \textbf{0.6}60972 & \textbf{0.35}2842 \\ \hline
        0.75 &- & \textbf{0.0000}00 & \textbf{0.03}8396 & \textbf{0.3}93514 & \textbf{1}.23688 & \textbf{0.6}51954 & \textbf{0.35}1482 \\ \hline
        1.0 &- &- & \textbf{0.00}1762 & \textbf{0}.236801 & \textbf{0}.997162 & \textbf{0}.636229 & \textbf{0.3}49077 \\ \hline
        1.33352 &- &- &\textbf{0.000}000 & \textbf{0}.101119 & \textbf{0}.731352 & \textbf{0.6}09161 & \textbf{0.34}4841 \\ \hline
        1.77828 &- &- &- & \textbf{0.0}19412 & \textbf{0}.462775 & \textbf{0.5}63768 & \textbf{0.3}37432 \\ \hline
        3.16228 &- &- &- & \textbf{0.000}000 & \textbf{0.07}4037 & \textbf{0.3}83609 & \textbf{0.30}3078 \\ \hline
        5.62341 &- &- &- &- & \textbf{0.00}0086 & \textbf{0.1}10507 & \textbf{0.21}5664 \\ \hline
        10.0 &- &- &- &- &- & \textbf{0.00}1626 & \textbf{0.07}2987 \\ \hline
        17.78279 &- &- &- &- &- &- & \textbf{0.002}196 \\ \hline
        \textbf{RMSE} & 1.233e-08 & 2.372e-08 & 9.65e-08 & 1.366e-07 & 4.052e-08 & 9.937e-08 & 5.936e-09 \\ \hline
    \end{tabular}
    \label{tab:su_square_e}
\end{table}

Since we claim to present benchmark quality results, we are obliged to discuss the accuracy of Tables \ref{tab:su_square_phi} and \ref{tab:su_square_e}. While \cite{SU19971035} claims to converge their solution to four digits, the observant reader will see that in some cases only 3 digits match. Given that our solutions are converged \wgb{to much greater accuracy}, we believe that our reported digits are correct.


\subsection{Constant $\cv$ problem with a square source}\label{sec:nl_square_thin}
This problem uses the same source as the problem of the last section but with a different functional form of the heat capacity. Using Eq.~\eqref{eq:eos:nonlin} our system becomes nonlinear. We choose $C_\mathrm{v0} = 0.03 \:\mathrm{GJ}\cdot\mathrm{cm}^{-3}\cdot\mathrm{keV}^{-1}$. This value was chosen to see an appreciable change in temperature during the selected time window. Now that we no longer have the convenient condition that $\Le=\lT^4$, the local equilibrium condition is not $\lphi = \Le$ as in the Su-Olson problem but $\lphi^{1/4} = \lT$. For this reason, the solution plots for this problem and all subsequent constant $\cv$ problems do not show scalar flux and material energy density but rather radiation temperature and material temperature.

Though we can no longer rely on benchmark solutions for this problem, we can be confident that our solution is converged by plotting the magnitude of the coefficients and check for systematic errors with a $S_n$ solver. Also, since the mesh method employed here is the same method described in Section \ref{subsec:su_square_thin}, we can be confident that the mesh is not introducing error.  

\begin{figure}
    \centering
    \begin{subfigure}[b]{0.3\textwidth}
    \includegraphics[width=\textwidth]{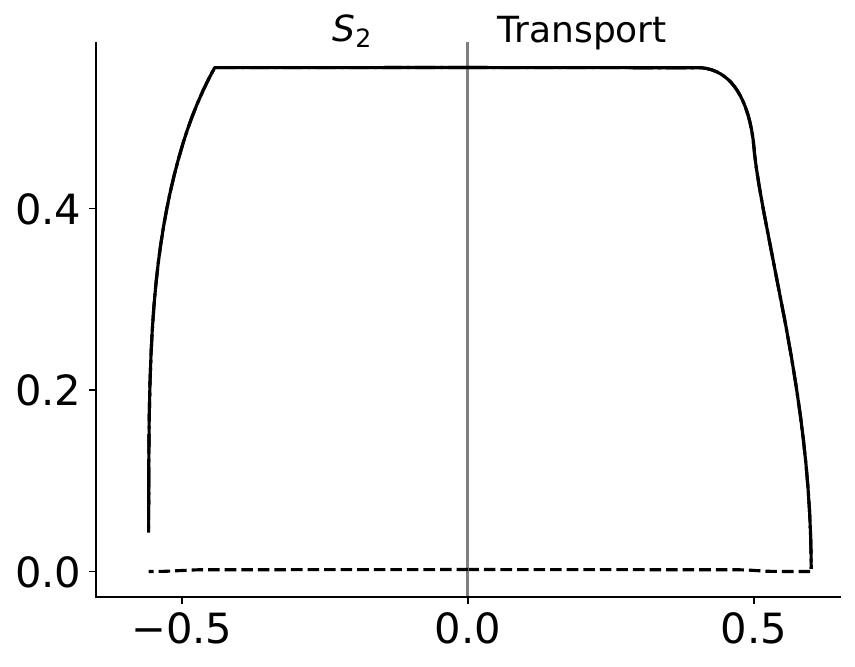}
    \caption{$t=0.1$}
    \end{subfigure}
    \centering
    \begin{subfigure}[b]{0.3\textwidth}
        \includegraphics[width=\textwidth]{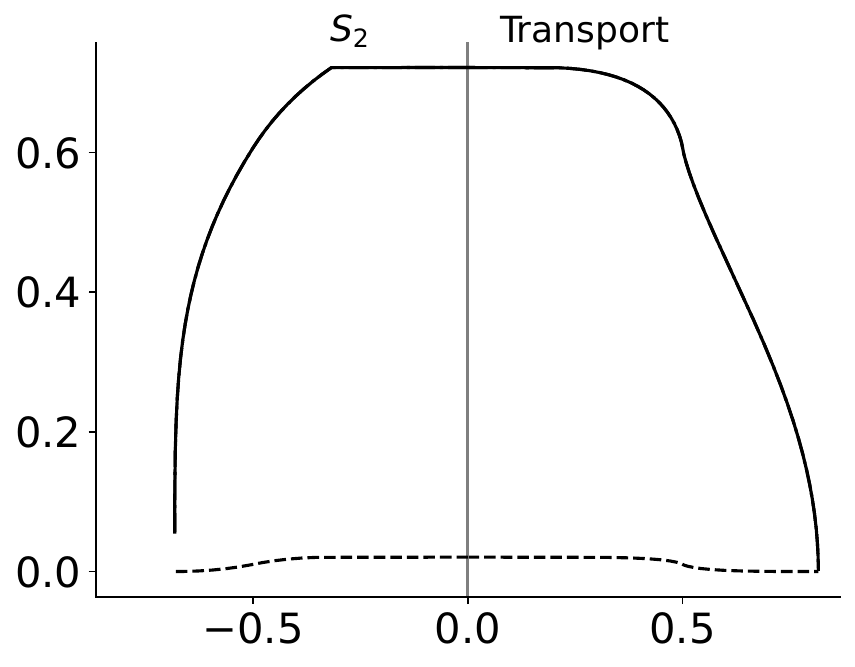}
        \caption{$t=0.31623$}
    \end{subfigure}
    \centering
    \begin{subfigure}[b]{0.3\textwidth}
        \includegraphics[width=\textwidth]{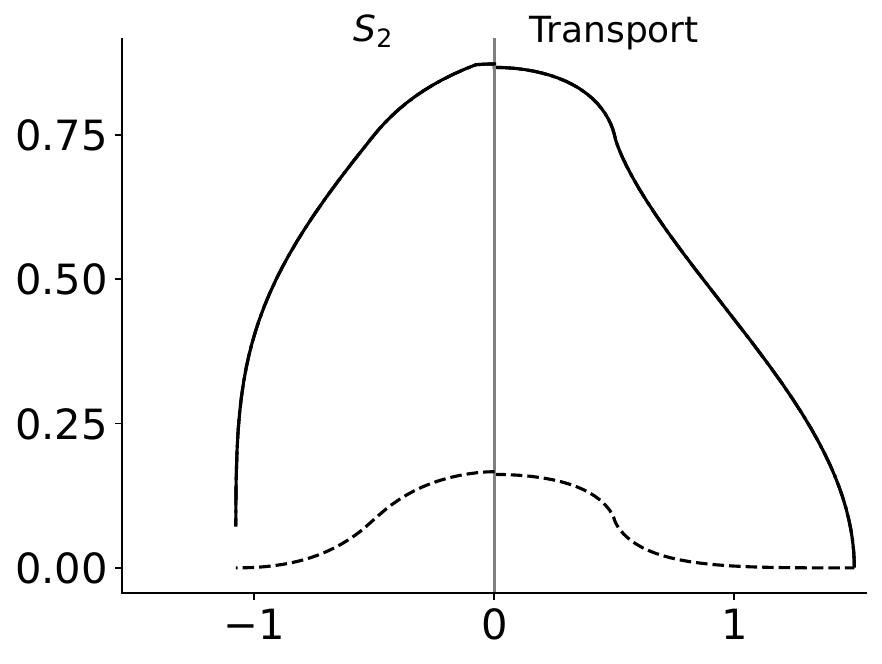}
        \caption{$t=1$}
    \end{subfigure}
    \begin{subfigure}[b]{0.3\textwidth}
        \includegraphics[width=\textwidth]{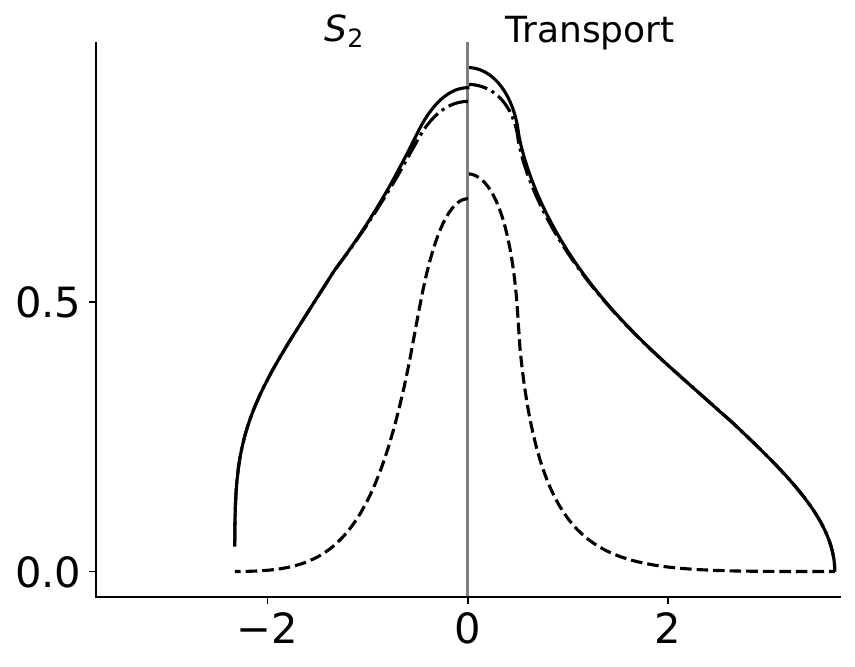}
        \caption{$t=3.16228$}
    \end{subfigure}
    \begin{subfigure}[b]{0.3\textwidth}
        \includegraphics[width=\textwidth]{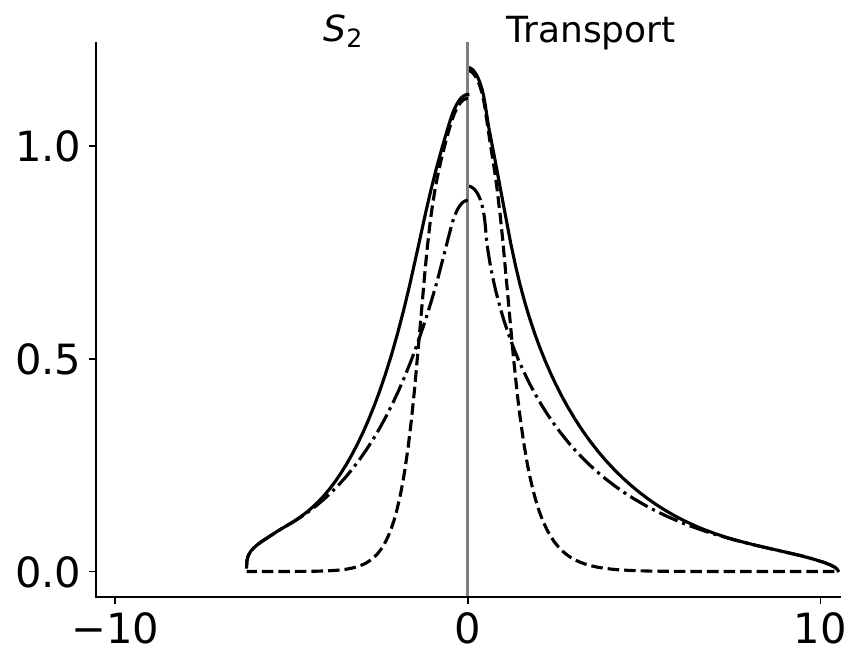}
        \caption{$t=10$}
    \end{subfigure}
    \begin{subfigure}[b]{0.3\textwidth}
        \includegraphics[width=\textwidth]{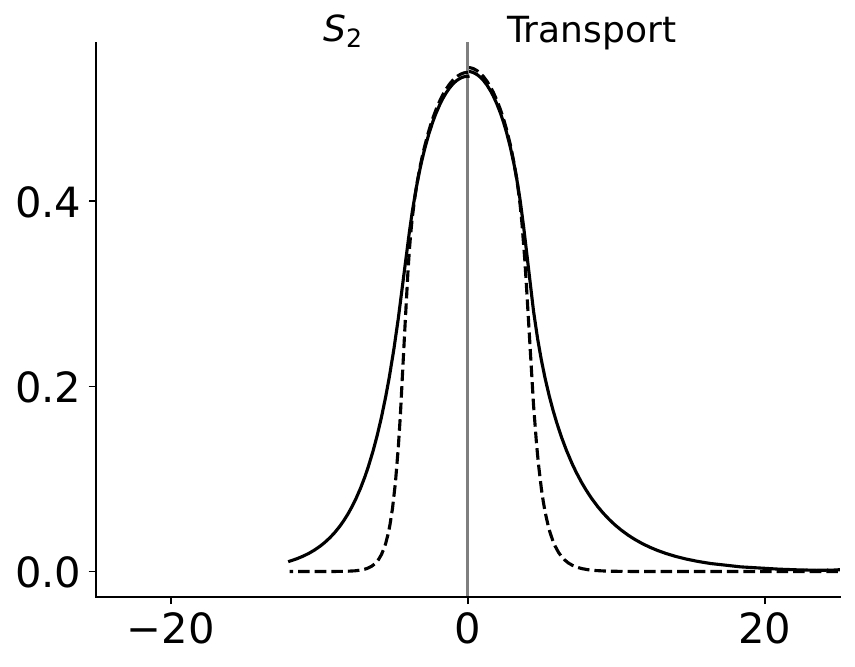}
        \caption{$t=100$}
    \end{subfigure}
      \caption{$S_2$ (left of $x=0$) and full transport (right of $x=0$) solutions for the optically thin constant $\cv$ square source problem with $x_0=0.5$, $t_0=10$. Solid lines are radiation temperature $\lphi^{1/4}$, \wgb{dash-dotted lines are the uncollided radiation temperature, $\lphi_u^{1/4}$}, and dashed are temperature, $\lT$.}
    \label{fig:nl_square_solutions} 
\end{figure}

\begin{figure}
    \centering
    \begin{subfigure}[b]{0.48\textwidth}
    \includegraphics[width=\textwidth]{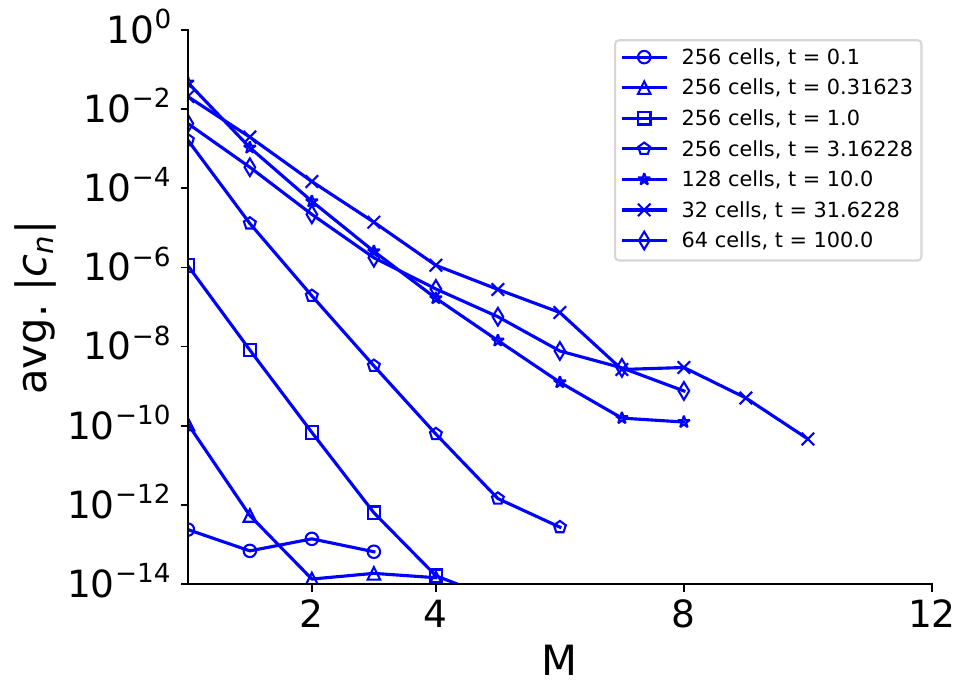}
    \caption{Radiation energy density, $\lphi$}
    \label{subfig:nl_square_phi}
    \end{subfigure}
    \centering
    \begin{subfigure}[b]{0.48\textwidth}
        \includegraphics[width=\textwidth]{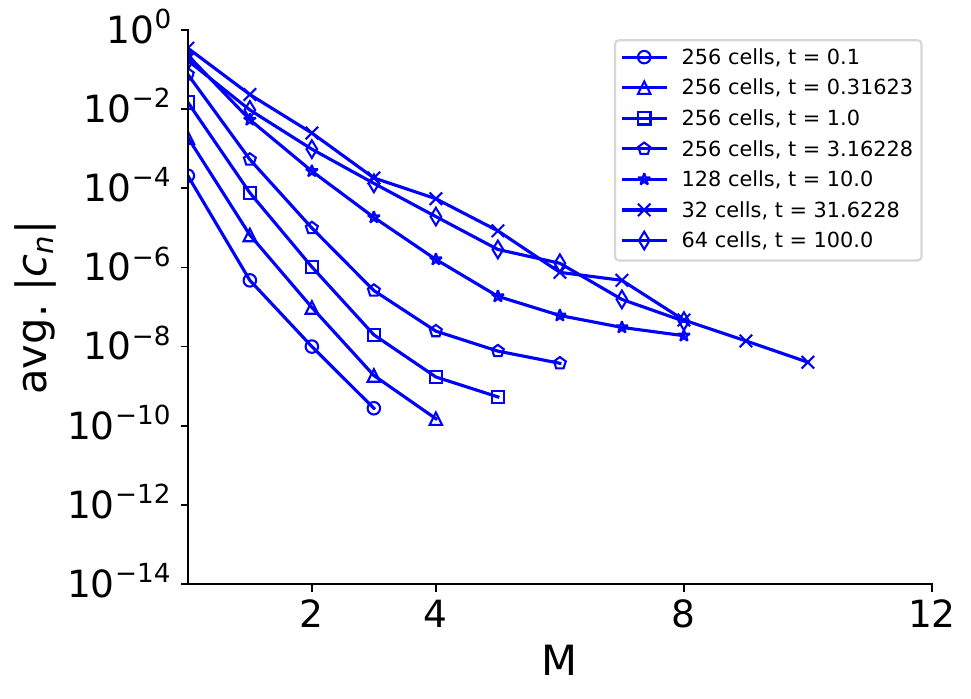}
        \caption{Material energy density, $\Le$}
        \label{subfig:nl_square_e}
    \end{subfigure}
      \caption{Log-linear scaled average value of the solution expansion coefficients (found by Eqs.~\eqref{eq:coeffs_phi}) for the optically thin ($\sigma_a=1$ cm$^{-1}$) constant $\cv$ square source problem where $x_0=0.5$, $t_0=10$. The quadrature order for all results is $S_{256}$. All results were calculated with a moving mesh and uncollided source treatment.}
    \label{fig:const_cv_square_thin_conv}
\end{figure}

\begin{figure}
    \centering
    \begin{subfigure}[b]{0.48\textwidth}
    \includegraphics[width=\textwidth]{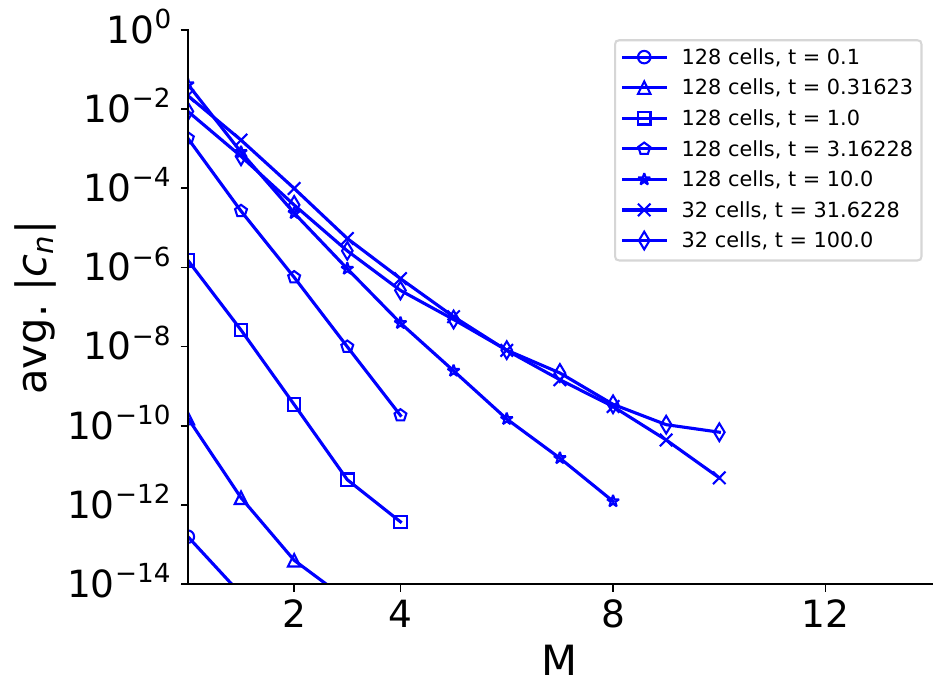}
    \caption{Radiation energy density, $\lphi$}
    \label{subfig:nl_square_s2_phi}
    \end{subfigure}
    \centering
    \begin{subfigure}[b]{0.48\textwidth}
        \includegraphics[width=\textwidth]{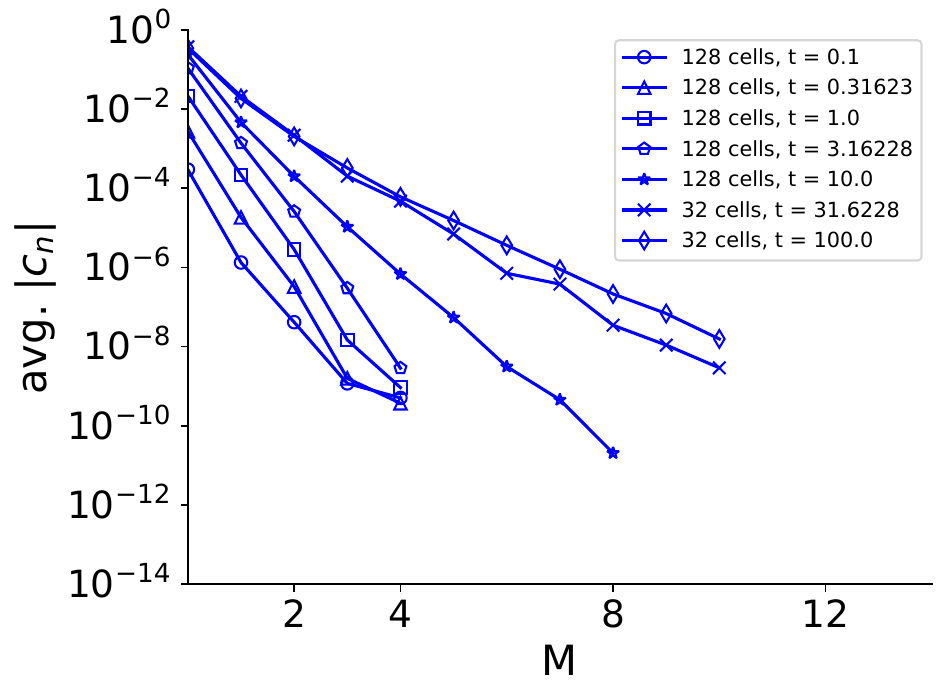}
        \caption{Material energy density, $\Le$}
        \label{subfig:nl_square_s2_e}
    \end{subfigure}
      \caption{Log-linear scaled average value of the solution expansion coefficients (found by Eqs.~\eqref{eq:coeffs_phi}) for the optically thin ($\sigma_a=1$ cm$^{-1}$) $S_2$ constant $\cv$ square source problem where $x_0=0.5$, $t_0=10$. All results were calculated with a moving mesh and uncollided source treatment except for the $t=31.6228$ and $t=100$ cases where a standard source treatment was used.}
    \label{fig:const_cv_square_thin_s2_conv}
\end{figure}

Solutions to this problem are plotted in Figure \ref{fig:nl_square_solutions}. We note that the problem is not everywhere at equilibrium by $t=100$ as the Su-Olson problem is. Also, we note that the solution does not travel as far. In the Su-Olson problem, the specific heat is very small when temperature is small and increases with the cube of the temperature. This has the effect of attracting the solution to equilibrium. This effect is not present in a constant $\cv$ case and there is less incentive for the solution to fall into local equilibrium. It is also noteworthy that at very early times ($t<1$) the scalar flux has not interacted with the material as much as in the Su-Olson problem and is mostly made up of the uncollided flux. This is apparent in Figures \ref{subfig:nl_square_phi} and \ref{subfig:nl_square_s2_phi}.

Since the solution has not fully equilibrated at later times, the solutions are less smooth compared to the Su-Olson problem. The repercussions of this can be observed by comparing the convergence results at late times for this problem in Figures \ref{fig:const_cv_square_thin_conv} and \ref{fig:const_cv_square_thin_s2_conv} to the convergence results of the Su-Olson problem in Figure \ref{fig:su_square_convergence}. Also, the convergence results show that the material energy density is generally more nonsmooth than the scalar flux. Nevertheless, we are satisfied with the convergence of these results and present them in Tables \ref{tab:const_cv_phi} and \ref{tab:const_cv_e}.

The difference between the full transport solution and our $S_2$ result is also of interest, as it provides insight into the physical characteristics of the system. We note that the two solutions only begin to look similar at later times as the solution 
equilibrates. This tells us that the solution becomes less angularly dependent and better approximated by only two angles.

\begin{table}[!ht]
    \centering
    \caption{Transport (top) and $S_2$ (bottom) results for the scalar flux, $\lphi$, for the thin square source constant $\cv$ problem with $x_0=0.5$, $t_0=10$, and $C_\mathrm{v0} = 0.03 \:\mathrm{GJ}\cdot\mathrm{cm}^{-3}\cdot\mathrm{keV}^{-1}$}

    \begin{tabular}{|l|l|l|l|l|l|l|l|}
    \hline
        $\boldsymbol{x/t}$ & 0.1 & 0.31623 & 1.0 & 3.16228 & 10.0 & 31.6228 & 100.0 \\ \hline
        0.01 & 0.095162 & 0.271108 & 0.563683 & 0.765084 & 1.96832 & 0.267247 & 0.085108 \\ \hline
        0.1 & 0.095162 & 0.271108 & 0.557609 & 0.756116 & 1.950367 & 0.266877 & 0.085054 \\ \hline
        0.17783 & 0.095162 & 0.271108 & 0.543861 & 0.736106 & 1.910675 & 0.266071 & 0.084937 \\ \hline
        0.31623 & 0.095162 & 0.258592 & 0.495115 & 0.668231 & 1.779896 & 0.263527 & 0.084565 \\ \hline
        0.45 & 0.08809 & 0.199962 & 0.396442 & 0.543721 & 1.558248 & 0.259729 & 0.084008 \\ \hline
        0.5 & 0.047581 & 0.135554 & 0.316071 & 0.453151 & 1.420865 & 0.257976 & 0.08375 \\ \hline
        0.56234 & 0.00376 & 0.061935 & 0.222261 & 0.349209 & 1.252213 & 0.255538 & 0.083392 \\ \hline
        0.75 &- & 0.002788 & 0.102348 & 0.21078 & 0.908755 & 0.246543 & 0.082061 \\ \hline
        1.0 &- &- & 0.034228 & 0.124305 & 0.562958 & 0.230831 & 0.079715 \\ \hline
        1.33352 &- &- & 0.002864 & 0.067319 & 0.27752 & 0.203718 & 0.075591 \\ \hline
        1.77828 &- &- &- & 0.031357 & 0.120054 & 0.158039 & 0.068419 \\ \hline
        3.16228 &- &- &- & 0.001057 & 0.013737 & 0.022075 & 0.036021 \\ \hline
        5.62341 &- &- &- &- & 0.000413 & 0.000814 & 0.001068 \\ \hline
        10.0 &- &- &- &- &- & 5e-06 & 5e-06 \\ \hline
        17.78279 &- &- &- &- &- &- &- \\ \hline
        \hline
        \hline
        $\boldsymbol{x/t}$ & 0.1 & 0.31623 & 1.0 & 3.16228 & 10.0 & 31.6228 & 100.0 \\ \hline
        0.01 & 0.095162 & 0.271108 & 0.579404 & 0.649946 & 1.57957 & 0.243368 & 0.081882 \\ \hline
        0.1 & 0.095162 & 0.271108 & 0.56606 & 0.641494 & 1.567037 & 0.243074 & 0.081833 \\ \hline
        0.17783 & 0.095162 & 0.271108 & 0.529958 & 0.623092 & 1.539665 & 0.242432 & 0.081725 \\ \hline
        0.31623 & 0.095162 & 0.271108 & 0.45241 & 0.565417 & 1.453129 & 0.240406 & 0.081383 \\ \hline
        0.45 & 0.08906 & 0.177033 & 0.357556 & 0.480232 & 1.323479 & 0.237381 & 0.080871 \\ \hline
        0.5 & 0.047581 & 0.135554 & 0.31607 & 0.441134 & 1.263425 & 0.235984 & 0.080634 \\ \hline
        0.56234 &- & 0.084378 & 0.264889 & 0.392645 & 1.185561 & 0.234042 & 0.080305 \\ \hline
        0.75 &- &- & 0.140336 & 0.276699 & 0.962617 & 0.226873 & 0.079082 \\ \hline
        1.0 &- &- & 0.02637 & 0.175485 & 0.692975 & 0.214346 & 0.076927 \\ \hline
        1.33352 &- &- &- & 0.097064 & 0.392628 & 0.192725 & 0.073141 \\ \hline
        1.77828 &- &- &- & 0.033465 & 0.155709 & 0.156395 & 0.066564 \\ \hline
        3.16228 &- &- &- &- & 0.007509 & 0.03175 & 0.03718 \\ \hline
        5.62341 &- &- &- &- & 4.7e-05 & 0.000413 & 0.001075 \\ \hline
        10.0 &- &- &- &- &- &- &- \\ \hline
        17.78279 &- &- &- &- &- &- &- \\ \hline
    \end{tabular}
    \label{tab:const_cv_phi}
\end{table}

\begin{table}[!ht]
    \centering
    \caption{Transport (top) and $S_2$ (bottom) results for the material energy density, $\Le$, for the thin square source constant $\cv$ problem with $x_0=0.5$, $t_0=10$, and $C_\mathrm{v0} = 0.03 \:\mathrm{GJ}\cdot\mathrm{cm}^{-3}\cdot\mathrm{keV}^{-1}$}  

    \begin{tabular}{|l|l|l|l|l|l|l|l|}
    \hline
        $\boldsymbol{x/t}$ & 0.1 & 0.31623 & 1.0 & 3.16228 & 10.0 & 31.6228 & 100.0 \\ \hline
        0.01 & 0.004837 & 0.045121 & 0.354022 & 1.613529 & 2.57461 & 1.592549 & 1.190296 \\ \hline
        0.1 & 0.004837 & 0.045121 & 0.350958 & 1.601467 & 2.568476 & 1.591998 & 1.190108 \\ \hline
        0.17783 & 0.004837 & 0.045121 & 0.343803 & 1.573757 & 2.554747 & 1.590795 & 1.189698 \\ \hline
        0.31623 & 0.004837 & 0.044507 & 0.316063 & 1.47078 & 2.507772 & 1.586979 & 1.188398 \\ \hline
        0.45 & 0.004705 & 0.036765 & 0.249325 & 1.238666 & 2.421019 & 1.581228 & 1.186445 \\ \hline
        0.5 & 0.002419 & 0.022562 & 0.183937 & 1.025219 & 2.361647 & 1.578549 & 1.185538 \\ \hline
        0.56234 & 5.1e-05 & 0.006779 & 0.108887 & 0.759317 & 2.280932 & 1.5748 & 1.184271 \\ \hline
        0.75 &- & 6.4e-05 & 0.034842 & 0.416175 & 2.069946 & 1.56071 & 1.179537 \\ \hline
        1.0 &- &- & 0.006872 & 0.214491 & 1.68516 & 1.535052 & 1.171036 \\ \hline
        1.33352 &- &- & 0.000168 & 0.094966 & 1.028758 & 1.487096 & 1.155611 \\ \hline
        1.77828 &- &- &- & 0.032116 & 0.471906 & 1.391456 & 1.127131 \\ \hline
        3.16228 &- &- &- & 0.000196 & 0.049604 & 0.471468 & 0.954827 \\ \hline
        5.62341 &- &- &- &- & 0.001163 & 0.019493 & 0.082189 \\ \hline
        10.0 &- &- &- &- &- & 0.000113 & 0.000487 \\ \hline
        17.78279 &- &- &- &- &- &- &- \\ \hline
        \hline
        \hline
        $\boldsymbol{x/t}$ & 0.1 & 0.31623 & 1.0 & 3.16228 & 10.0 & 31.6228 & 100.0 \\ \hline
        0.01 & 0.004837 & 0.045121 & 0.364106 & 1.512533 & 2.431781 & 1.555683 & 1.178842 \\ \hline
        0.1 & 0.004837 & 0.045121 & 0.357116 & 1.497067 & 2.426675 & 1.555215 & 1.178665 \\ \hline
        0.17783 & 0.004837 & 0.045121 & 0.343553 & 1.462337 & 2.415402 & 1.554193 & 1.178277 \\ \hline
        0.31623 & 0.004837 & 0.045121 & 0.299658 & 1.343527 & 2.378606 & 1.550952 & 1.177048 \\ \hline
        0.45 & 0.004796 & 0.033907 & 0.223636 & 1.138096 & 2.319758 & 1.546073 & 1.175203 \\ \hline
        0.5 & 0.002418 & 0.02256 & 0.183934 & 1.031348 & 2.290735 & 1.543804 & 1.174346 \\ \hline
        0.56234 &- & 0.00909 & 0.135476 & 0.89324 & 2.251138 & 1.54063 & 1.173149 \\ \hline
        0.75 &- &- & 0.043524 & 0.56287 & 2.120765 & 1.528729 & 1.16868 \\ \hline
        1.0 &- &- & 0.001805 & 0.289458 & 1.885442 & 1.507188 & 1.160663 \\ \hline
        1.33352 &- &- &- & 0.106017 & 1.298617 & 1.467469 & 1.146151 \\ \hline
        1.77828 &- &- &- & 0.018336 & 0.530154 & 1.391381 & 1.119483 \\ \hline
        3.16228 &- &- &- &- & 0.023197 & 0.536229 & 0.964849 \\ \hline
        5.62341 &- &- &- &- & 3.1e-05 & 0.005921 & 0.056159 \\ \hline
        10.0 &- &- &- &- &- &- & 2.4e-05 \\ \hline
        17.78279 &- &- &- &- &- &- &- \\ \hline
    \end{tabular}
    \label{tab:const_cv_e}
\end{table}

\subsection{Su-Olson problem with a Gaussian source}\label{sec:su_gauss_thin}
Returning to the linearized Su-Olson problem, we consider a Gaussian source. Here the source is defined by \wgb{Eq.~\eqref{eq:gaussian_source}} where the uncollided solution is taken from \cite{bennett2022benchmarks} for the full transport solution or Eq.\eqref{eq:uncol_gauss_s2} for the $S_2$ solution. We set $x_0=0.5$ and the source duration is still $t_0 = 10$. In \cite{movingmesh}, smooth Gaussian sources allowed for geometric convergence of the solution at all times. We expect the same result in this application.

Since there are no discontinuities induced by nonsmoothness in the source, we are able to employ a far simpler mesh function than what was used for the thin square source problems. We only guess the edge of the problem domain and span the given space evenly with stationary edges. The moving mesh was not used in this case because earlier tests in \cite{movingmesh} revealed the mesh to be non-useful in smooth problems. The uncollided solution however, was employed. We include Gaussian sources though they do not prove to be challenging enough problems to require the full application of our method because we can achieve very accurate solutions.


\begin{figure}
    \centering
    \begin{subfigure}[b]{0.48\textwidth}
    \includegraphics[width=\textwidth]{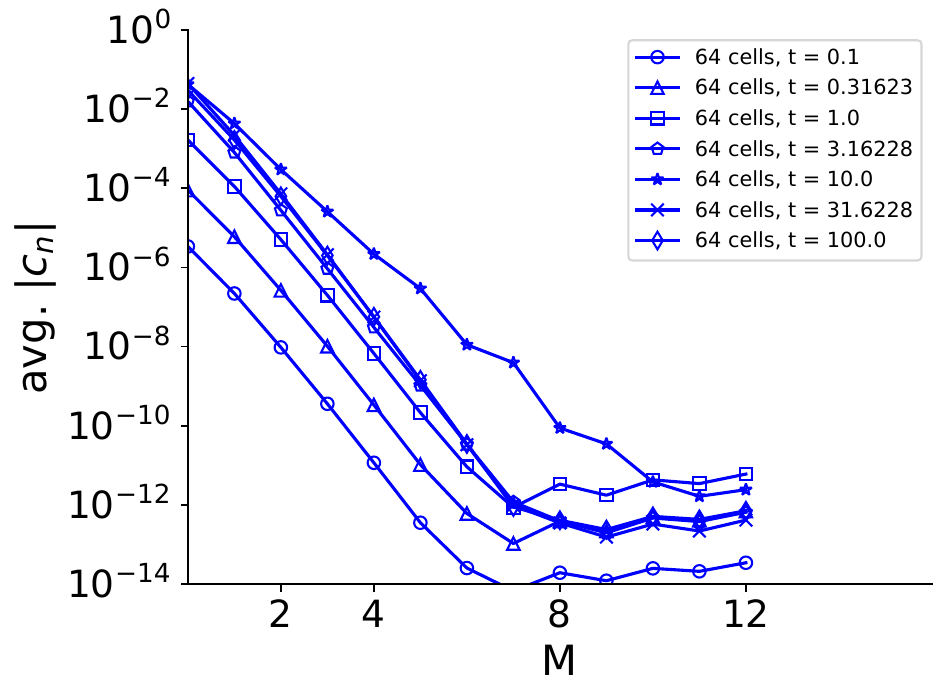}
    \caption{Radiation energy density, $\lphi$}
    \label{subfig:su_gaussian_phi}
    \end{subfigure}
    \centering
    \begin{subfigure}[b]{0.48\textwidth}
        \includegraphics[width=\textwidth]{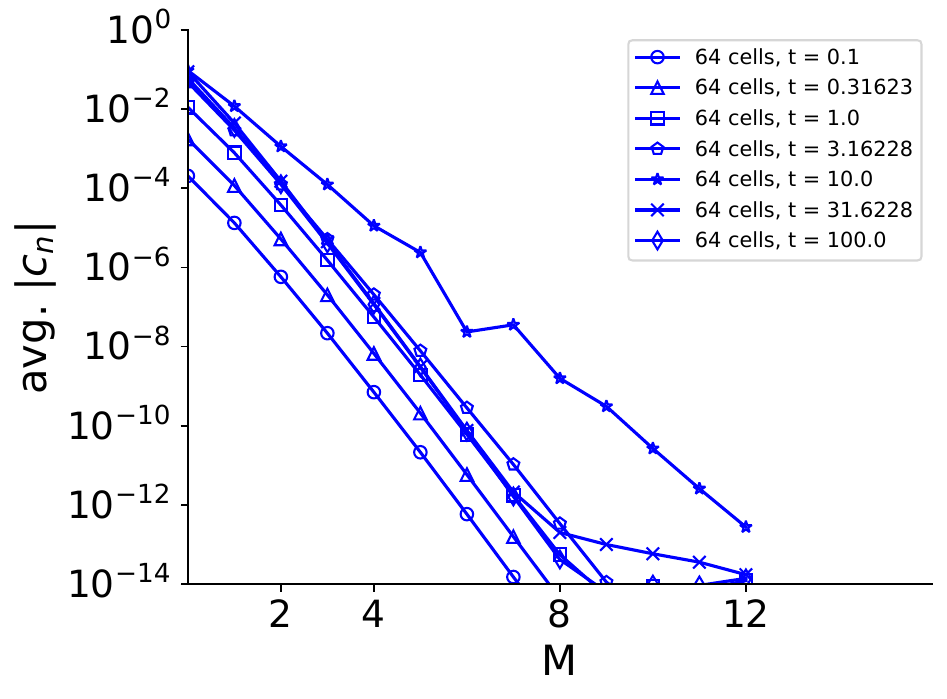}
        \caption{Material energy density, $\Le$}
        \label{subfig:su_gaussian_e}
    \end{subfigure}
      \caption{Log-linear scaled average value of the solution expansion coefficients (found by Eqs.~\eqref{eq:coeffs_phi}) for the optically thin ($\sigma_a=1$ cm$^{-1}$) Su-Olson Gaussian source problem where $x_0=0.5$, $t_0=10$. The quadrature order for all results is $S_{16}$. All results were calculated with a static mesh and uncollided source treatment.}
    \label{fig:su_gaussian_convergence}
\end{figure}


\begin{figure}
    \centering
    \begin{subfigure}[b]{0.48\textwidth}
    \includegraphics[width=\textwidth]{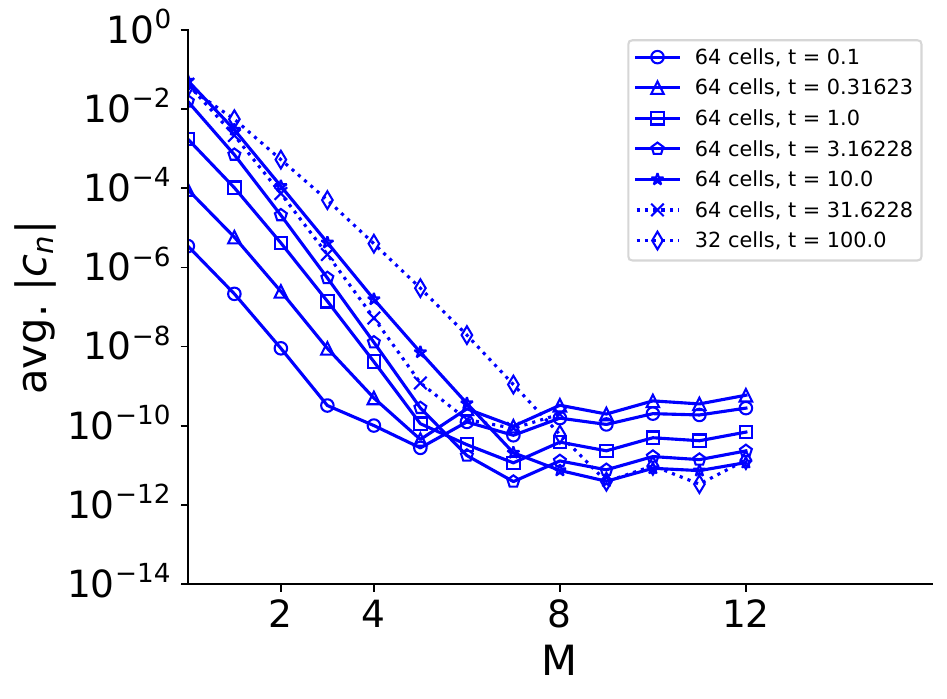}
    \caption{Radiation energy density, $\lphi$}
    \label{subfig:su_gaussian_s2_phi}
    \end{subfigure}
    \centering
    \begin{subfigure}[b]{0.48\textwidth}
        \includegraphics[width=\textwidth]{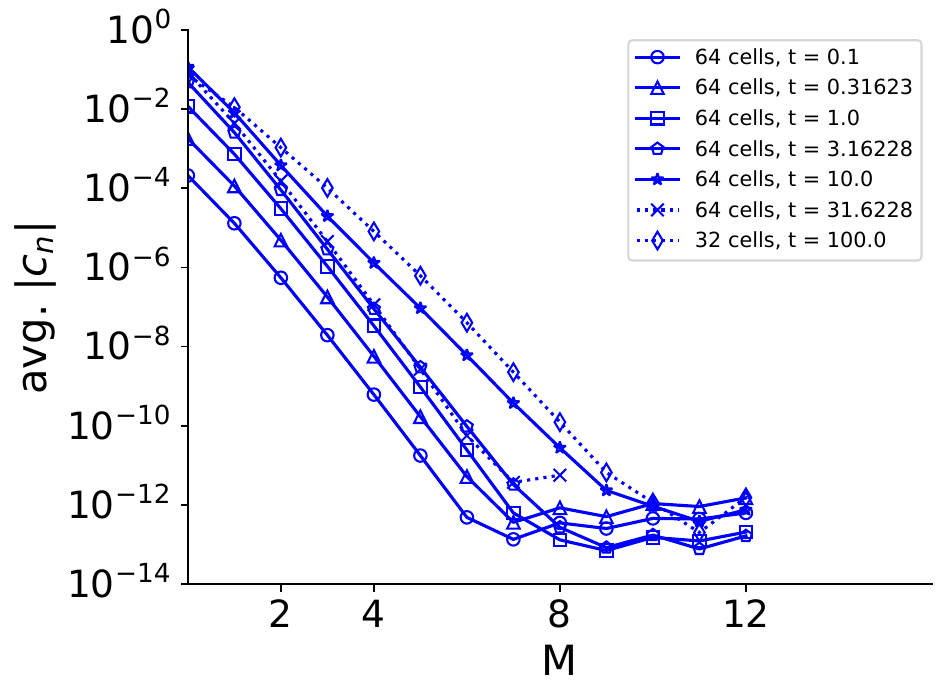}
        \caption{Material energy density, $\Le$}
        \label{subfig:su_gaussian_s2_e}
    \end{subfigure}
      \caption{Log-linear scaled average value of the solution expansion coefficients (found by Eqs.~\eqref{eq:coeffs_phi}) for the optically thin ($\sigma_a=1$ cm$^{-1}$) $S_2$ Su-Olson Gaussian source problem where $x_0=0.5$, $t_0=10$. All results were calculated with a moving mesh and uncollided source treatment except for the $t=31.6228$ and $t=100$ cases where a standard source treatment was used. The dashed lines represent solutions found with a moving mesh.}
    \label{fig:su_gaussian_s2_convergence}
\end{figure}


\begin{figure}
    \centering
    \begin{subfigure}[b]{0.3\textwidth}
    \includegraphics[width=\textwidth]{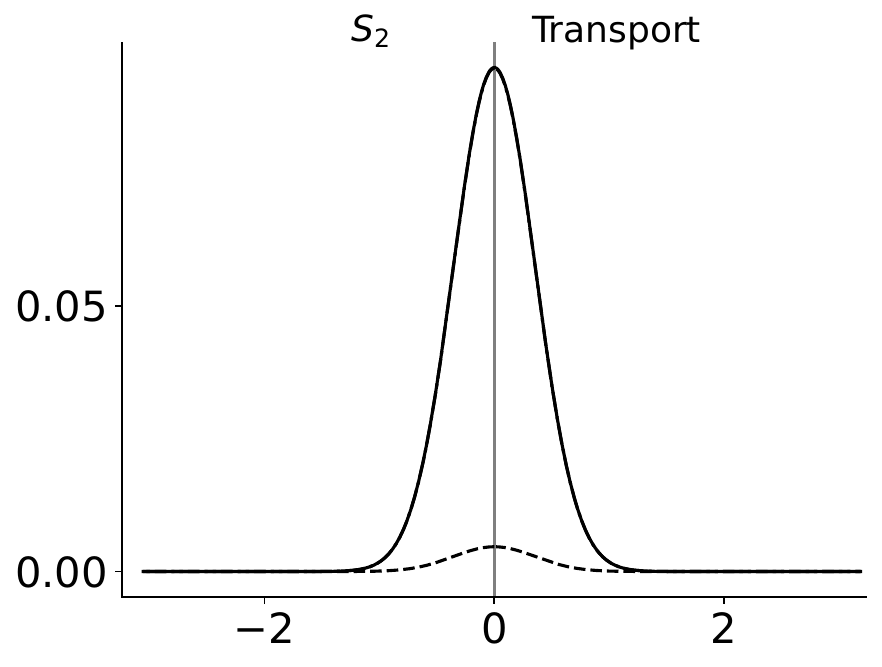}
    \caption{$t=0.1$}
    \end{subfigure}
    \centering
    \begin{subfigure}[b]{0.3\textwidth}
        \includegraphics[width=\textwidth]{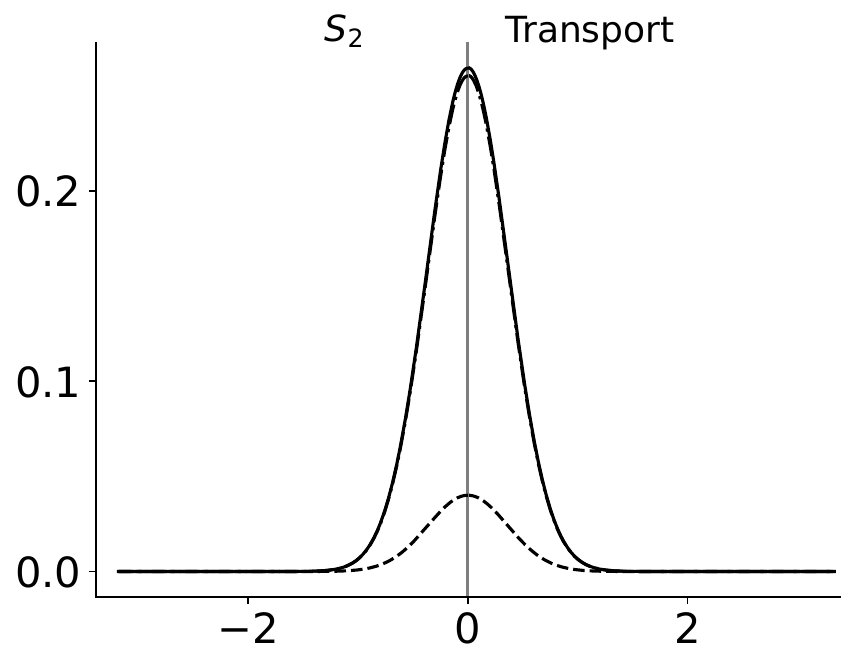}
        \caption{$t=0.31623$}
    \end{subfigure}
    \centering
    \begin{subfigure}[b]{0.3\textwidth}
        \includegraphics[width=\textwidth]{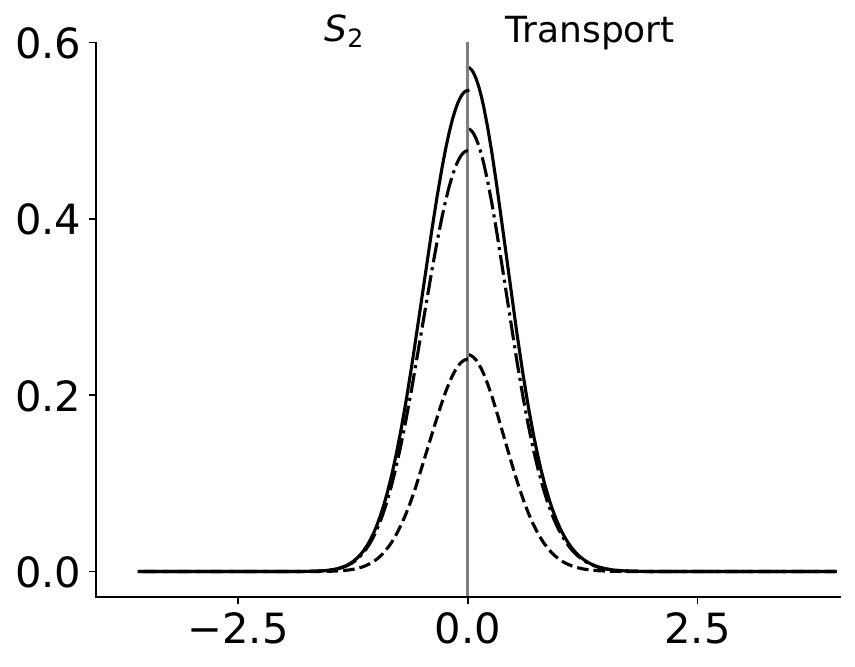}
        \caption{$t=1$}
    \end{subfigure}
    \begin{subfigure}[b]{0.3\textwidth}
        \includegraphics[width=\textwidth]{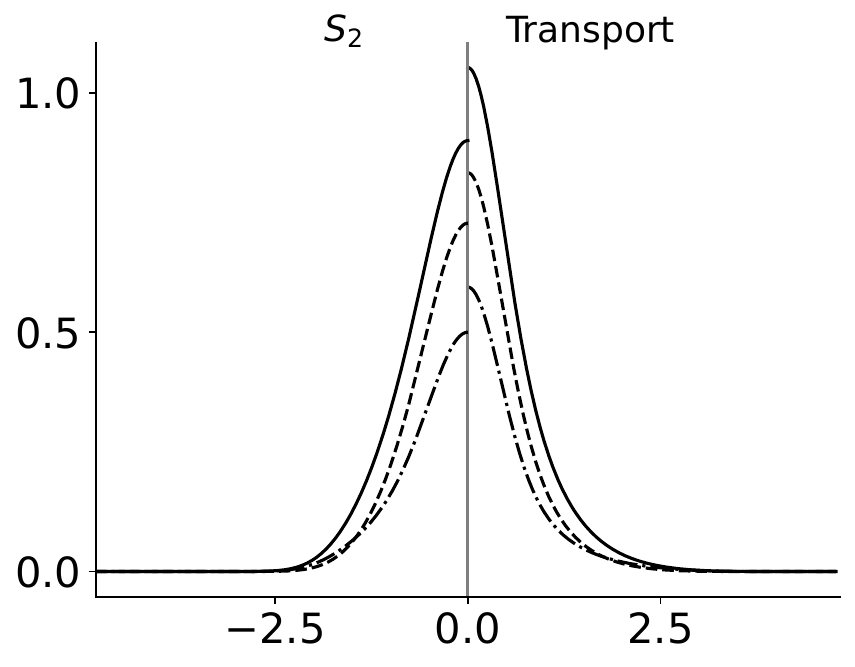}
        \caption{$t=3.16228$}
    \end{subfigure}
    \begin{subfigure}[b]{0.3\textwidth}
        \includegraphics[width=\textwidth]{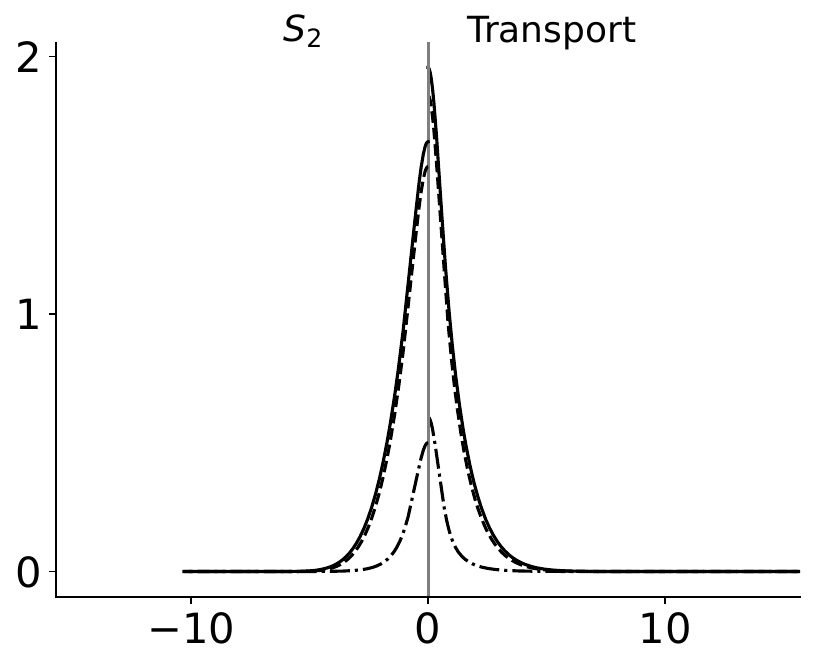}
        \caption{$t=10$}
    \end{subfigure}
    \begin{subfigure}[b]{0.3\textwidth}
        \includegraphics[width=\textwidth]{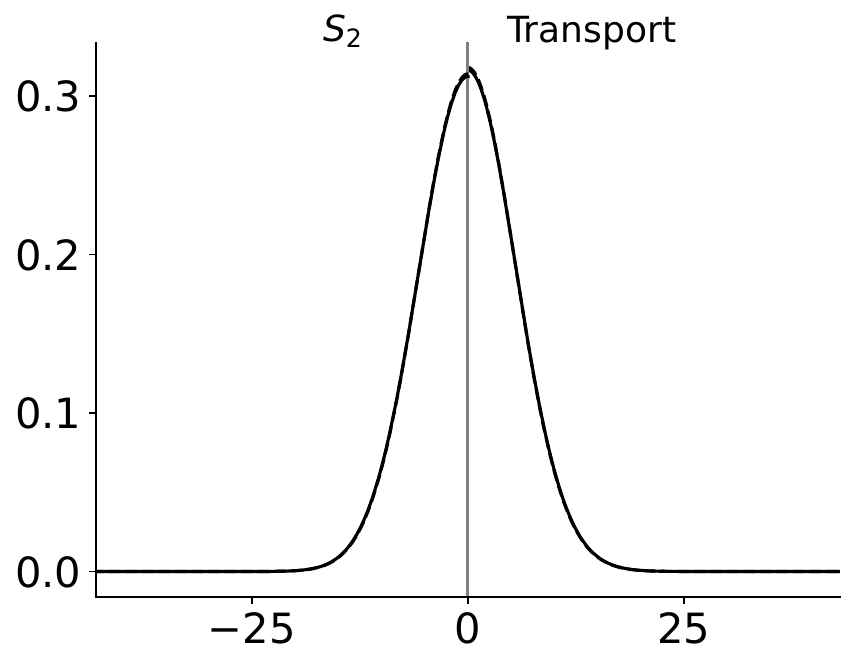}
        \caption{$t=100$}
        \label{}
    \end{subfigure}
      \caption{$S_2$ (left of $x=0$) and full transport (right of $x=0$) solutions for the optically thin Su-Olson Gaussian source problem with $x_0=0.5$, $t_0=10$. Solid lines are scalar flux, $\lphi$, \wgb{dash-dotted lines are the uncollided scalar flux, $\lphi_u$,} and dashed are material energy density, $\Le$.}
    \label{fig:su_gauss_solutions} 
\end{figure}

While $S_{256}$ was used for the full transport solutions for the thin square sources, we only use $S_{64}$ \wgb{for} the Gaussian sources. This choice is informed by tests run in \cite{movingmesh} that showed that far fewer quadrature points are required to resolve the angular error. 
\begin{table}[!ht]
    \centering
     \caption{Transport (top) and $S_2$ (bottom) results for the scalar flux, $\lphi$, for the thin Gaussian source Su-Olson problem with $x_0=0.5$, $t_0=10$}
     
    \begin{tabular}{|l|l|l|l|l|l|l|l|}
    \hline
        $\boldsymbol{x/t}$ & 0.1 & 0.31623 & 1.0 & 3.16228 & 10.0 & 31.6228 & 100.0 \\ \hline
        0.01 & 0.094869 & 0.264712 & 0.571441 & 1.053107 & 1.956705 & 0.610025 & 0.316341 \\ \hline
        0.1 & 0.091217 & 0.255181 & 0.556247 & 1.033176 & 1.933017 & 0.609634 & 0.31629 \\ \hline
        0.17783 & 0.083721 & 0.235541 & 0.524582 & 0.991429 & 1.883176 & 0.608783 & 0.316178 \\ \hline
        0.31623 & 0.063836 & 0.182853 & 0.436813 & 0.873994 & 1.741091 & 0.6061 & 0.315826 \\ \hline
        0.45 & 0.042514 & 0.125118 & 0.334025 & 0.732182 & 1.564765 & 0.602104 & 0.315298 \\ \hline
        0.5 & 0.035215 & 0.104949 & 0.29568 & 0.67763 & 1.49511 & 0.600263 & 0.315054 \\ \hline
        0.56234 & 0.027081 & 0.082142 & 0.250049 & 0.611096 & 1.408369 & 0.597705 & 0.314714 \\ \hline
        0.75 & 0.010198 & 0.033042 & 0.137011 & 0.433926 & 1.163525 & 0.588305 & 0.313452 \\ \hline
        1.0 & 0.001799 & 0.006564 & 0.050023 & 0.267122 & 0.898484 & 0.572018 & 0.311224 \\ \hline
        1.33352 & 8.2e-05 & 0.000371 & 0.008965 & 0.139587 & 0.64098 & 0.544297 & 0.3073 \\ \hline
        1.77828 &- & 2e-06 & 0.000387 & 0.05756 & 0.40731 & 0.498676 & 0.300446 \\ \hline
        3.16228 &- &- &- & 0.00137 & 0.087854 & 0.327865 & 0.268818 \\ \hline
        5.62341 &- &- &- &- & 0.003363 & 0.096556 & 0.189497 \\ \hline
        10.0 &- &- &- &- &- & 0.003523 & 0.063953 \\ \hline
        17.78279 &- &- &- &- &- &- & 0.002424 \\ \hline
        \hline
        \hline
        $\boldsymbol{x/t}$ & 0.1 & 0.31623 & 1.0 & 3.16228 & 10.0 & 31.6228 & 100.0 \\ \hline
        0.01 & 0.00467 & 0.040073 & 0.240699 & 0.727949 & 1.572225 & 0.594815 & 0.314048 \\ \hline
        0.1 & 0.00449 & 0.03858 & 0.23398 & 0.717893 & 1.560265 & 0.594488 & 0.313999 \\ \hline
        0.17783 & 0.004119 & 0.035509 & 0.219946 & 0.696571 & 1.534799 & 0.593775 & 0.313893 \\ \hline
        0.31623 & 0.003137 & 0.027317 & 0.180837 & 0.6345 & 1.45974 & 0.591526 & 0.313557 \\ \hline
        0.45 & 0.002085 & 0.018437 & 0.134764 & 0.554569 & 1.360669 & 0.58817 & 0.313054 \\ \hline
        0.5 & 0.001726 & 0.015365 & 0.117566 & 0.522032 & 1.319396 & 0.586621 & 0.312821 \\ \hline
        0.56234 & 0.001326 & 0.011918 & 0.09716 & 0.480715 & 1.266036 & 0.584468 & 0.312496 \\ \hline
        0.75 & 0.000497 & 0.004632 & 0.047669 & 0.359391 & 1.101318 & 0.576529 & 0.311293 \\ \hline
        1.0 & 8.7e-05 & 0.000863 & 0.013277 & 0.223358 & 0.893342 & 0.562683 & 0.309166 \\ \hline
        1.33352 & 3e-06 & 4.3e-05 & 0.001308 & 0.100634 & 0.657392 & 0.538848 & 0.305419 \\ \hline
        1.77828 &- &- & 1.8e-05 & 0.022908 & 0.417809 & 0.498866 & 0.298865 \\ \hline
        3.16228 &- &- &- &- & 0.068504 & 0.34004 & 0.268475 \\ \hline
        5.62341 &- &- &- &- & 0.000121 & 0.098551 & 0.191126 \\ \hline
        10.0 &- &- &- &- &- & 0.001487 & 0.064775 \\ \hline
        17.78279 &- &- &- &- &- &- & 0.001958 \\ \hline
        $\boldsymbol{RMSE}$ & 9.105e-10 & 8.561e-10 & 2.604e-09 & 7.839e-09 & 1.321e-08 & 6.808e-09 & 6.899e-05 \\ \hline
    \end{tabular}
    \label{tab:su_gauss_phi}
\end{table}

\begin{table}[!ht]
    \centering
    \caption{Transport (top) and $S_2$ (bottom) results for the material energy density, $\Le$, for the thin Gaussian source Su-Olson problem  with $x_0=0.5$, $t_0=10$}

    \begin{tabular}{|l|l|l|l|l|l|l|l|}
    \hline
        $\boldsymbol{x/t}$ & 0.1 & 0.31623 & 1.0 & 3.16228 & 10.0 & 31.6228 & 100.0 \\ \hline
        0.01 & 0.00467 & 0.040089 & 0.245869 & 0.833299 & 1.847977 & 0.623026 & 0.318058 \\ \hline
        0.1 & 0.00449 & 0.038593 & 0.238404 & 0.815919 & 1.824555 & 0.622608 & 0.318006 \\ \hline
        0.17783 & 0.004119 & 0.035517 & 0.222913 & 0.779578 & 1.775289 & 0.621695 & 0.317892 \\ \hline
        0.31623 & 0.003137 & 0.027313 & 0.18051 & 0.677838 & 1.63499 & 0.61882 & 0.317534 \\ \hline
        0.45 & 0.002085 & 0.018426 & 0.132108 & 0.556161 & 1.461245 & 0.614538 & 0.316998 \\ \hline
        0.5 & 0.001726 & 0.015355 & 0.114504 & 0.509791 & 1.392751 & 0.612566 & 0.316749 \\ \hline
        0.56234 & 0.001326 & 0.011908 & 0.093967 & 0.453643 & 1.307592 & 0.609828 & 0.316404 \\ \hline
        0.75 & 0.000497 & 0.004629 & 0.04582 & 0.307085 & 1.068295 & 0.59977 & 0.315121 \\ \hline
        1.0 & 8.7e-05 & 0.000865 & 0.013544 & 0.175481 & 0.812011 & 0.58237 & 0.312856 \\ \hline
        1.33352 & 3e-06 & 4.4e-05 & 0.001743 & 0.082634 & 0.567439 & 0.552834 & 0.308867 \\ \hline
        1.77828 &- &- & 5e-05 & 0.029302 & 0.35102 & 0.504445 & 0.301903 \\ \hline
        3.16228 &- &- &- & 0.000297 & 0.069582 & 0.325807 & 0.269788 \\ \hline
        5.62341 &- &- &- &- & 0.002245 & 0.091971 & 0.189454 \\ \hline
        10.0 &- &- &- &- &- & 0.003086 & 0.063216 \\ \hline
        17.78279 &- &- &- &- &- &- & 0.002324 \\ \hline
        \hline
        \hline
        $\boldsymbol{x/t}$ & 0.1 & 0.31623 & 1.0 & 3.16228 & 10.0 & 31.6228 & 100.0 \\ \hline
        0.01 & 0.00467 & 0.040073 & 0.240699 & 0.727949 & 1.572225 & 0.594815 & 0.314048 \\ \hline
        0.1 & 0.00449 & 0.03858 & 0.23398 & 0.717893 & 1.560265 & 0.594488 & 0.313999 \\ \hline
        0.17783 & 0.004119 & 0.035509 & 0.219946 & 0.696571 & 1.534799 & 0.593775 & 0.313893 \\ \hline
        0.31623 & 0.003137 & 0.027317 & 0.180837 & 0.6345 & 1.45974 & 0.591526 & 0.313557 \\ \hline
        0.45 & 0.002085 & 0.018437 & 0.134764 & 0.554569 & 1.360669 & 0.58817 & 0.313054 \\ \hline
        0.5 & 0.001726 & 0.015365 & 0.117566 & 0.522032 & 1.319396 & 0.586621 & 0.312821 \\ \hline
        0.56234 & 0.001326 & 0.011918 & 0.09716 & 0.480715 & 1.266036 & 0.584468 & 0.312496 \\ \hline
        0.75 & 0.000497 & 0.004632 & 0.047669 & 0.359391 & 1.101318 & 0.576529 & 0.311293 \\ \hline
        1.0 & 8.7e-05 & 0.000863 & 0.013277 & 0.223358 & 0.893342 & 0.562683 & 0.309166 \\ \hline
        1.33352 & 3e-06 & 4.3e-05 & 0.001308 & 0.100634 & 0.657392 & 0.538848 & 0.305419 \\ \hline
        1.77828 &- &- & 1.8e-05 & 0.022908 & 0.417809 & 0.498866 & 0.298865 \\ \hline
        3.16228 &- &- &- &- & 0.068504 & 0.34004 & 0.268475 \\ \hline
        5.62341 &- &- &- &- & 0.000121 & 0.098551 & 0.191126 \\ \hline
        10.0 &- &- &- &- &- & 0.001487 & 0.064775 \\ \hline
        17.78279 &- &- &- &- &- &- & 0.001958 \\ \hline
        $\boldsymbol{RMSE}$ & 9.105e-10 & 8.561e-10 & 2.604e-09 & 7.839e-09 & 1.321e-08 & 6.808e-09 & 6.899e-05 \\ \hline
    \end{tabular}
    \label{tab_su_gauss_e}
\end{table}

With the temperature defined by Eq~\eqref{eq:eos_su} we present solutions in Tables \ref{tab:su_gauss_phi} and \ref{tab_su_gauss_e} with convergence results shown in Figures \ref{fig:su_gaussian_convergence} and \ref{fig:su_gaussian_s2_convergence}. Solutions are shown in Figure \ref{fig:su_gauss_solutions}. As illustrated in the aforementioned convergence plots, the problem converges geometrically even with a standard static mesh. We quickly note that for $t=31.6228$ and $t=100$ in the $S_2$ results in Figure \ref{fig:su_gaussian_s2_convergence}, a moving mesh was employed. In this case, the mesh moved with a constant speed from the initial width to the specified final width. This was not done out of necessity, but rather to ascertain \wgb{whether} the moving mesh was useful in these problems. 

On the difference between the full and $S_2$ solutions, we point out that it can be seen in Figure \ref{fig:su_gauss_solutions} that the $S_2$ solution is better \wgbt{compared with the square sources at estimating the solution at early times. This is because the solution is mostly uncollided at early times and the $S_2$ and full transport uncollided solutions for a Gaussian source are similar. For intermediate times, the full transport solution has more angular dependence and the $S_2$ solution is not as accurate. At later times, collisions smooth the angular flux of the transport solution and reduce the angular dependence, bringing the system closer to the diffusion approximation and $S_2$ and transport again agree.}

\subsection{Constant $\cv$ problem with a Gaussian source}
\begin{figure}
    \centering
    \begin{subfigure}[b]{0.3\textwidth}
    \includegraphics[width=\textwidth]{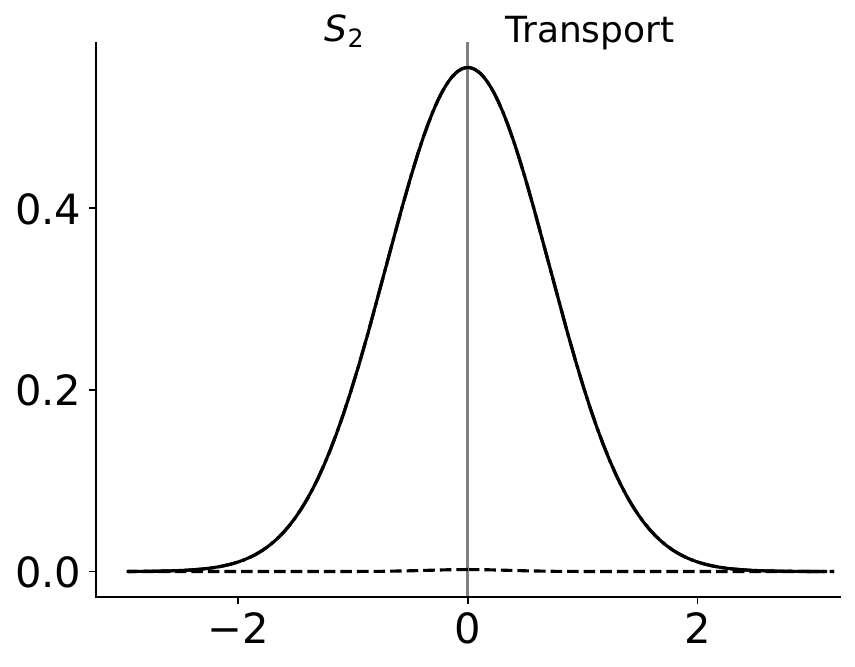}
    \caption{$t=0.1$}
    \end{subfigure}
    \centering
    \begin{subfigure}[b]{0.3\textwidth}
        \includegraphics[width=\textwidth]{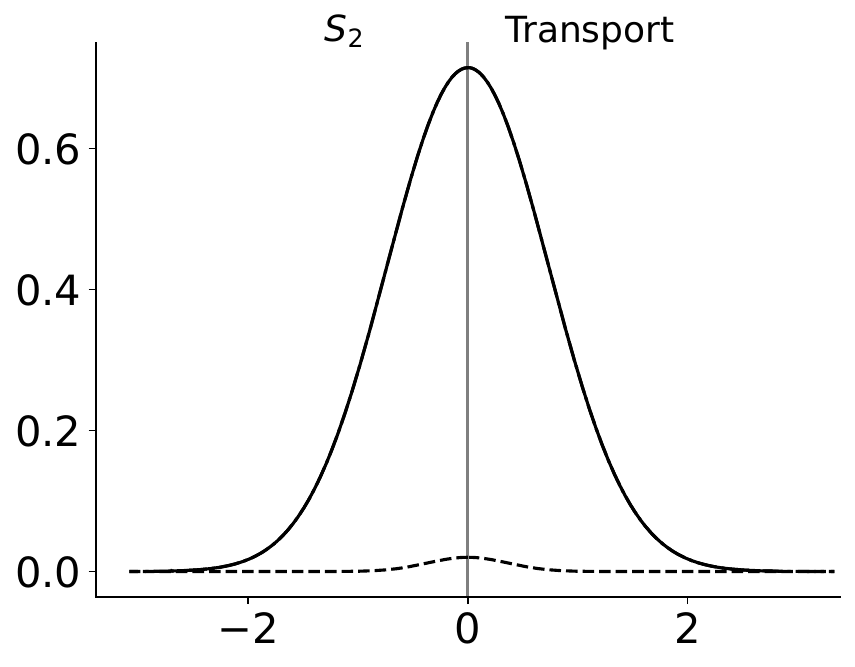}
        \caption{$t=0.31623$}
    \end{subfigure}
    \centering
    \begin{subfigure}[b]{0.3\textwidth}
        \includegraphics[width=\textwidth]{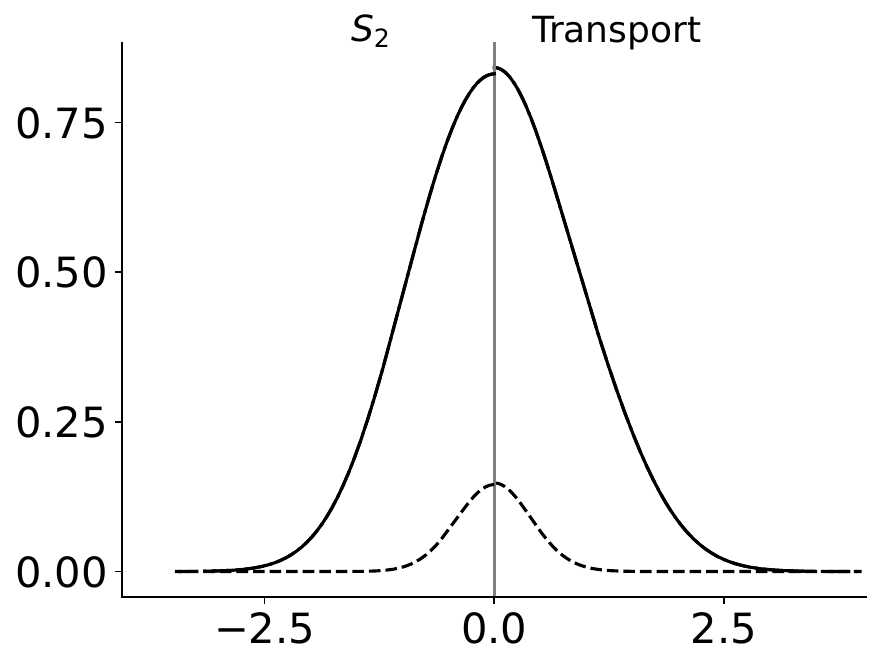}
        \caption{$t=1$}
    \end{subfigure}
    \begin{subfigure}[b]{0.3\textwidth}
        \includegraphics[width=\textwidth]{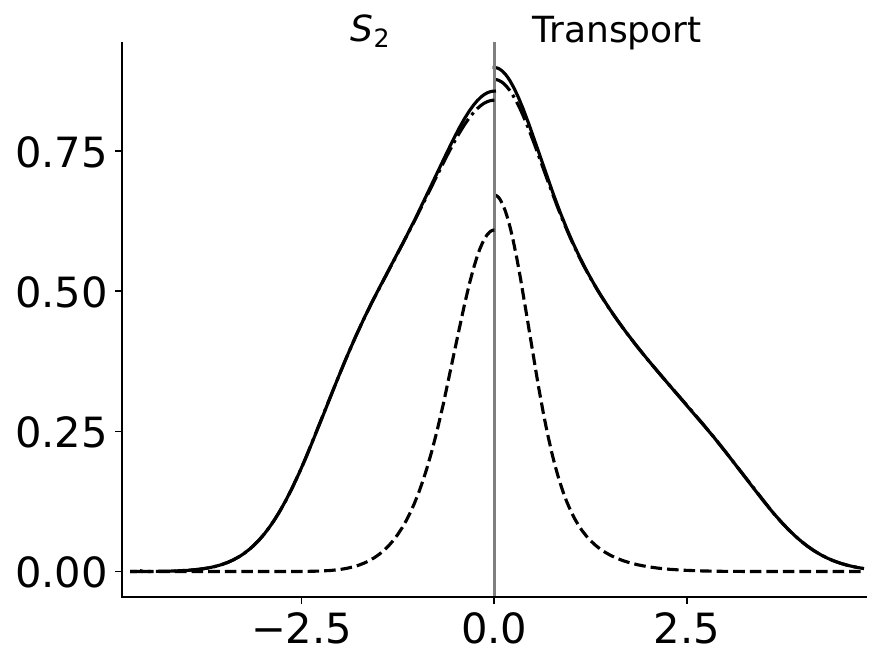}
        \caption{$t=3.16228$}
    \end{subfigure}
    \begin{subfigure}[b]{0.3\textwidth}
        \includegraphics[width=\textwidth]{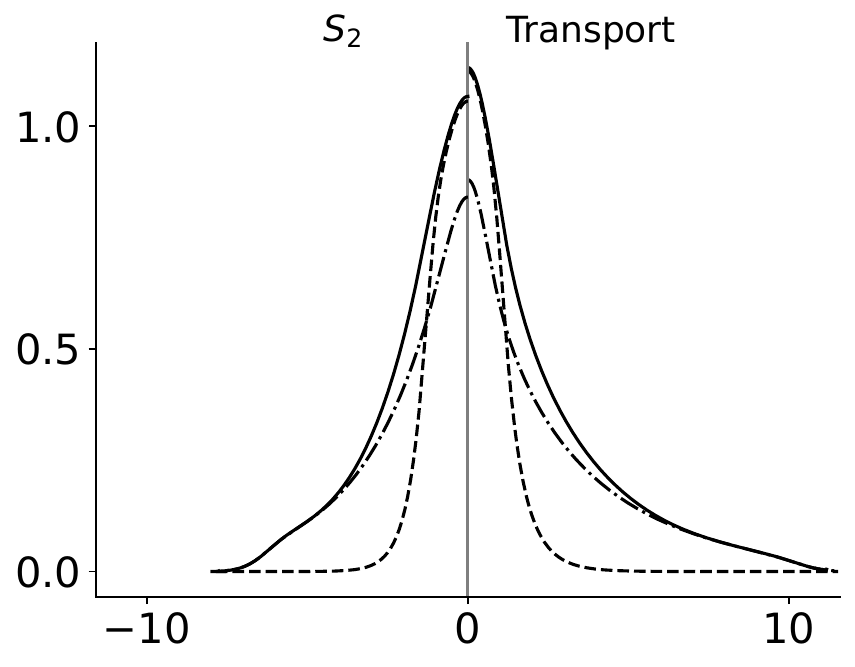}
        \caption{$t=10$}
    \end{subfigure}
    \begin{subfigure}[b]{0.3\textwidth}
        \includegraphics[width=\textwidth]{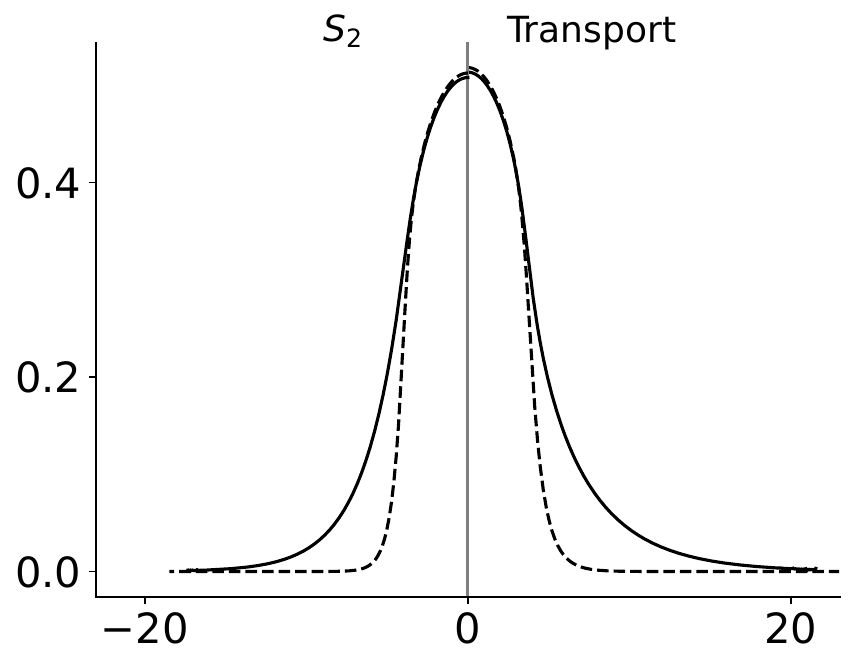}
        \caption{$t=100$}
    \end{subfigure}
      \caption{$S_2$ (left of $x=0$) and full transport (right of $x=0$) solutions for the optically thin constant $\cv$ Gaussian source problem with $x_0=0.5$, $t_0=10$. Solid lines are radiation temperature $\lphi^{1/4}$,\wgb{dash-dotted lines are the uncollided radiation temperature, $\lphi_u^{1/4},$} and dashed are temperature, $\lT$.}
    \label{fig:nl_gaussian_solutions} 
\end{figure}

 We include a problem with the same source and parameters as the last section but with a constant specific heat so that the temperature to material energy density conversion is given by Eq.~\eqref{eq:eos:nonlin}. We specify the dimensional specific heat to be $C_\mathrm{v0} = 0.03 \:\mathrm{GJ}\cdot\mathrm{cm}^{-3}\cdot\mathrm{keV}^{-1}$. Solutions are shown in Figure \ref{fig:nl_gaussian_solutions}. While we have less certainty in forecasting the behavior of these nonlinear results, we still expect geometric convergence since there are no sources of nonsmoothness. 

\begin{figure}
    \centering
    \begin{subfigure}[b]{0.48\textwidth}
    \includegraphics[width=\textwidth]{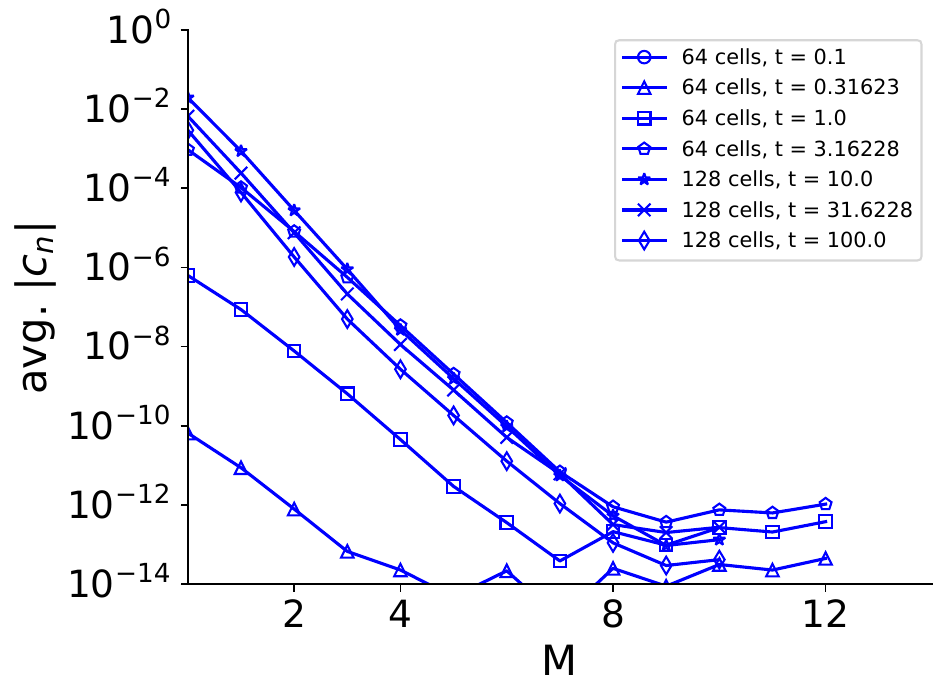}
    \caption{Radiation energy density, $\lphi$}
    \label{subfig:nl_gaussian_phi}
    \end{subfigure}
    \centering
    \begin{subfigure}[b]{0.48\textwidth}
        \includegraphics[width=\textwidth]{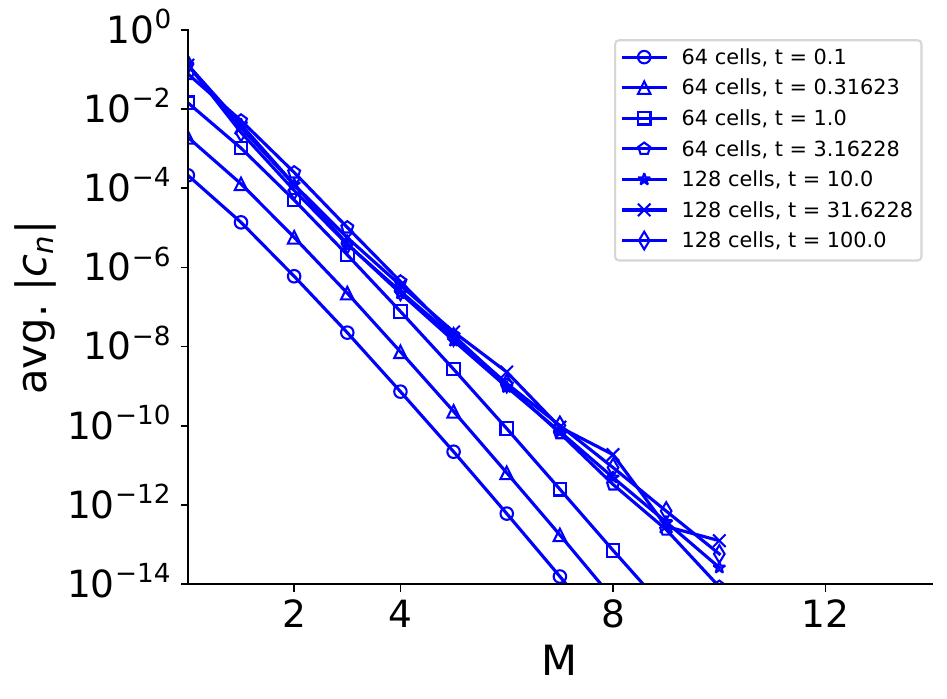}
        \caption{Material energy density, $\Le$}
        \label{subfig:nl_gaussian_e}
    \end{subfigure}
      \caption{Log-linear scaled average value of the solution expansion coefficients (found by Eqs.~\eqref{eq:coeffs_phi}) for the optically thin ($\sigma_a=1$ cm$^{-1}$) constant $\cv$ Gaussian source problem where $x_0=0.5$, $t_0=10$. The quadrature order for all results is $S_{16}$. All results were calculated with a static mesh and uncollided source treatment.}
    \label{fig:const_cv_gaussian_thin_conv}
\end{figure}

\begin{figure}
    \centering
    \begin{subfigure}[b]{0.48\textwidth}
    \includegraphics[width=\textwidth]{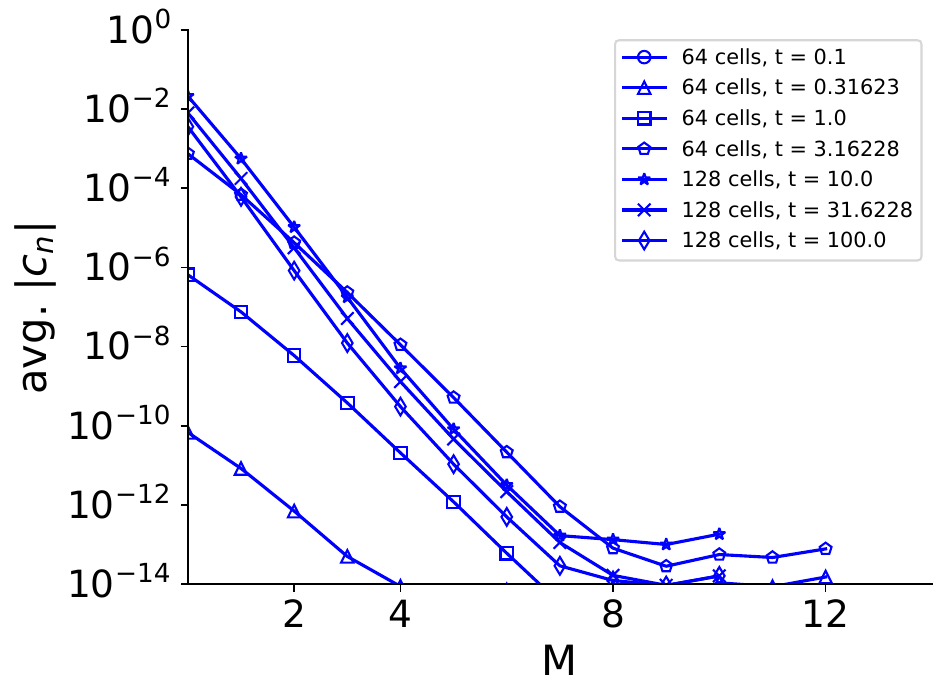}
    \caption{Radiation energy density, $\lphi$}
    \label{subfig:nl_gaussian_s2_phi}
    \end{subfigure}
    \centering
    \begin{subfigure}[b]{0.48\textwidth}
        \includegraphics[width=\textwidth]{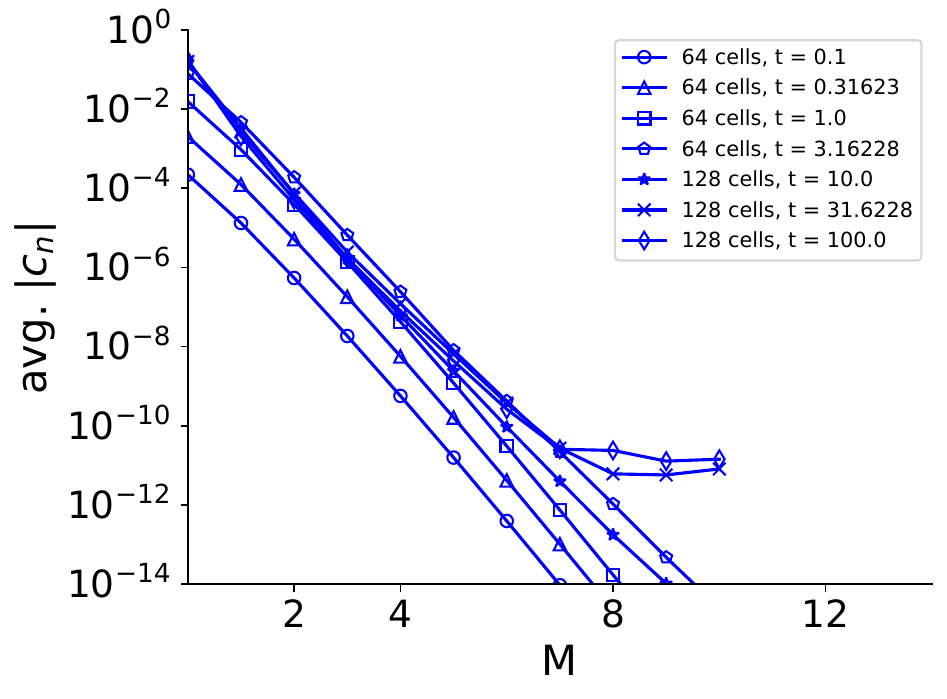}
        \caption{Material energy density, $\Le$}
        \label{subfig:nl_gaussian_s2_e}
    \end{subfigure}
      \caption{Log-linear scaled average value of the solution expansion coefficients (found by Eqs.~\eqref{eq:coeffs_phi}) for the optically thin ($\sigma_a=1$ cm$^{-1}$) $S_2$ constant $\cv$ Gaussian source problem where $x_0=0.5$, $t_0=10$. All results were calculated with a static mesh and uncollided source treatment.}
    \label{fig:const_cv_gaussian_thin_s2_conv}
\end{figure}

Like the Gaussian source in the linearized system, we specified a static mesh that evenly spans some estimated width of the actual solution domain. The convergence results in Figures \ref{fig:const_cv_gaussian_thin_conv} and \ref{fig:const_cv_gaussian_thin_s2_conv} show that this was sufficient to achieve geometric convergence at all chosen times. We also see the phenomenon first observed in Section \ref{sec:nl_square_thin} of extremely fast convergence for the scalar flux at early times. This is once again the result of the scalar flux being most uncollided at these times, making solving for the collided portion simpler. 

\wgbt{Like the linearized Gaussian,} during the time the source is on\wgbt{,} \wgbt{the discrepancy} between the $S_2$ and transport solutions \wgbt{grows.} This discrepancy \wgbt{diminishes} after the source is turned off \wgbt{and scattering reduces the angular dependence of the transport solution}. Similar to the nonlinear square source, the solution is not in equilibrium by $t=100$. This leads us to draw the conclusion that equilibrium is not as impacted by the nonsmoothness of the source as it is by the functional form of the specific heat, since the Su-Olson problem with a nonsmooth source goes more quickly into equilibrium than this constant $\cv$ smooth source problem. We present the solutions in Tables \ref{tab:thin_gauss_phi} and \ref{tab:thin_gauss_e}.



\begin{table}[!ht]
    \centering
    \caption{Transport (top) and $S_2$ (bottom) results for the scalar flux, $\lphi$, for the thin Gaussian source constant $\cv$ problem with $x_0=0.5$, $t_0=10$, and $C_\mathrm{v0} = 0.03 \:\mathrm{GJ}\cdot\mathrm{cm}^{-3}\cdot\mathrm{keV}^{-1}$}
    
    \begin{tabular}{|l|l|l|l|l|l|l|l|}
    \hline
        $\boldsymbol{x/t}$ & 0.1 & 0.31623 & 1.0 & 3.16228 & 10.0 & 31.6228 & 100.0 \\ \hline
        0.01 & 0.094715 & 0.260632 & 0.501868 & 0.653649 & 1.639873 & 0.218959 & 0.069516 \\ \hline
        0.1 & 0.091069 & 0.251252 & 0.488575 & 0.636913 & 1.614763 & 0.21862 & 0.069466 \\ \hline
        0.17783 & 0.083585 & 0.231922 & 0.460879 & 0.602409 & 1.561824 & 0.217881 & 0.069356 \\ \hline
        0.31623 & 0.063732 & 0.180063 & 0.384158 & 0.509408 & 1.41001 & 0.215547 & 0.069011 \\ \hline
        0.45 & 0.042445 & 0.123229 & 0.294366 & 0.405067 & 1.219261 & 0.212063 & 0.068495 \\ \hline
        0.5 & 0.035158 & 0.103372 & 0.260874 & 0.367198 & 1.142975 & 0.210455 & 0.068256 \\ \hline
        0.56234 & 0.027037 & 0.080916 & 0.221011 & 0.322683 & 1.047037 & 0.208219 & 0.067923 \\ \hline
        0.75 & 0.010181 & 0.032561 & 0.122095 & 0.213179 & 0.768814 & 0.19997 & 0.066689 \\ \hline
        1.0 & 0.001796 & 0.006473 & 0.045352 & 0.123297 & 0.461235 & 0.185567 & 0.064514 \\ \hline
        1.33352 & 8.2e-05 & 0.000367 & 0.008352 & 0.064418 & 0.223392 & 0.160718 & 0.060695 \\ \hline
        1.77828 &- & 2e-06 & 0.000371 & 0.029558 & 0.097211 & 0.118847 & 0.054065 \\ \hline
        3.16228 &- &- &- & 0.001165 & 0.011091 & 0.014263 & 0.024308 \\ \hline
        5.62341 &- &- &- &- & 0.00034 & 0.00057 & 0.000634 \\ \hline
        10.0 &- &- &- &- &- & 4e-06 & 3e-06 \\ \hline
        17.78279 &- &- &- &- &- &- &- \\ \hline
        \hline
        \hline
        \hline
        \textbf{x/t} & 0.1 & 0.31623 & 1.0 & 3.16228 & 10.0 & 31.6228 & 100.0 \\ \hline
        0.01 & 0.094714 & 0.260323 & 0.477228 & 0.539628 & 1.295862 & 0.197266 & 0.066556 \\ \hline
        0.1 & 0.091067 & 0.250998 & 0.467159 & 0.53082 & 1.283232 & 0.197001 & 0.066511 \\ \hline
        0.17783 & 0.083584 & 0.231775 & 0.445829 & 0.512305 & 1.256308 & 0.196423 & 0.066411 \\ \hline
        0.31623 & 0.063733 & 0.180133 & 0.384035 & 0.459709 & 1.176709 & 0.194598 & 0.066097 \\ \hline
        0.45 & 0.042446 & 0.123414 & 0.305857 & 0.394987 & 1.071024 & 0.191874 & 0.065626 \\ \hline
        0.5 & 0.035159 & 0.103564 & 0.274806 & 0.369646 & 1.026763 & 0.190617 & 0.065409 \\ \hline
        0.56234 & 0.027038 & 0.081091 & 0.236322 & 0.338318 & 0.969319 & 0.188869 & 0.065106 \\ \hline
        0.75 & 0.010181 & 0.032613 & 0.132763 & 0.251849 & 0.790231 & 0.18242 & 0.063983 \\ \hline
        1.0 & 0.001796 & 0.006438 & 0.045179 & 0.164176 & 0.560306 & 0.171164 & 0.062005 \\ \hline
        1.33352 & 8.1e-05 & 0.000354 & 0.005814 & 0.08995 & 0.308784 & 0.151783 & 0.058534 \\ \hline
        1.77828 &- & 1e-06 & 0.000114 & 0.03286 & 0.12204 & 0.119388 & 0.052522 \\ \hline
        3.16228 &- &- & -0.0 & 2e-06 & 0.006114 & 0.019949 & 0.02605 \\ \hline
        5.62341 &- &- &- &- & 4.6e-05 & 0.000264 & 0.000564 \\ \hline
        10.0 &- &- &- &- &- &- &- \\ \hline
        17.78279 &- &- &- &- &- &- &- \\ \hline
    \end{tabular}
    \label{tab:thin_gauss_phi}
\end{table}

\begin{table}[!ht]
    \centering
    \caption{Transport (top) and $S_2$ (bottom) results for the material energy density, $\Le$, for the thin Gaussian source constant $\cv$ problem with $x_0=0.5$, $t_0=10$, and $C_\mathrm{v0} = 0.03 \:\mathrm{GJ}\cdot\mathrm{cm}^{-3}\cdot\mathrm{keV}^{-1}$}
    
    \begin{tabular}{|l|l|l|l|l|l|l|l|}
    \hline
        $\boldsymbol{x/t}$ & 0.1 & 0.31623 & 1.0 & 3.16228 & 10.0 & 31.6228 & 100.0 \\ \hline
        0.01 & 0.004825 & 0.044224 & 0.323377 & 1.4684 & 2.455993 & 1.518526 & 1.133064 \\ \hline
        0.1 & 0.004638 & 0.042571 & 0.313295 & 1.438115 & 2.446023 & 1.51794 & 1.132861 \\ \hline
        0.17783 & 0.004255 & 0.039172 & 0.292399 & 1.373348 & 2.424567 & 1.516657 & 1.132417 \\ \hline
        0.31623 & 0.003241 & 0.030112 & 0.235379 & 1.183116 & 2.359381 & 1.512586 & 1.131013 \\ \hline
        0.45 & 0.002154 & 0.020302 & 0.170708 & 0.944125 & 2.267966 & 1.506441 & 1.128901 \\ \hline
        0.5 & 0.001783 & 0.016913 & 0.147332 & 0.851718 & 2.227558 & 1.503576 & 1.12792 \\ \hline
        0.56234 & 0.001369 & 0.013111 & 0.120189 & 0.740141 & 2.172608 & 1.499564 & 1.126549 \\ \hline
        0.75 & 0.000513 & 0.005089 & 0.057328 & 0.458276 & 1.96946 & 1.484442 & 1.121422 \\ \hline
        1.0 & 8.9e-05 & 0.000949 & 0.016339 & 0.231272 & 1.546952 & 1.456735 & 1.112196 \\ \hline
        1.33352 & 4e-06 & 4.8e-05 & 0.002002 & 0.097311 & 0.889883 & 1.4042 & 1.095386 \\ \hline
        1.77828 &- &- & 5.5e-05 & 0.032461 & 0.400128 & 1.294119 & 1.064084 \\ \hline
        3.16228 &- &- &- & 0.000322 & 0.042204 & 0.349552 & 0.857807 \\ \hline
        5.62341 &- &- &- &- & 0.001039 & 0.014856 & 0.054666 \\ \hline
        10.0 &- &- &- &- &- & 8.8e-05 & 0.000336 \\ \hline
        17.78279 &- &- &- &- &- &- &- \\ \hline
        \hline
        \hline
        \hline
        \textbf{x/t} & 0.1 & 0.31623 & 1.0 & 3.16228 & 10.0 & 31.6228 & 100.0 \\ \hline
        0.01 & 0.004825 & 0.044206 & 0.317475 & 1.332131 & 2.309445 & 1.479441 & 1.120796 \\ \hline
        0.1 & 0.004638 & 0.042557 & 0.308253 & 1.31107 & 2.303385 & 1.478948 & 1.120606 \\ \hline
        0.17783 & 0.004255 & 0.039164 & 0.289032 & 1.265902 & 2.29029 & 1.47787 & 1.120189 \\ \hline
        0.31623 & 0.003241 & 0.030116 & 0.235788 & 1.131312 & 2.250065 & 1.474452 & 1.118869 \\ \hline
        0.45 & 0.002154 & 0.020312 & 0.173764 & 0.954656 & 2.192602 & 1.469301 & 1.116886 \\ \hline
        0.5 & 0.001783 & 0.016923 & 0.150839 & 0.882882 & 2.166882 & 1.466903 & 1.115965 \\ \hline
        0.56234 & 0.001369 & 0.013121 & 0.123828 & 0.79271 & 2.13172 & 1.463548 & 1.114678 \\ \hline
        0.75 & 0.000513 & 0.005092 & 0.059398 & 0.540414 & 2.00314 & 1.45095 & 1.109868 \\ \hline
        1.0 & 8.9e-05 & 0.000947 & 0.016015 & 0.293197 & 1.73921 & 1.428067 & 1.101228 \\ \hline
        1.33352 & 4e-06 & 4.7e-05 & 0.001515 & 0.113254 & 1.125017 & 1.385564 & 1.085536 \\ \hline
        1.77828 &- &- & 2e-05 & 0.022889 & 0.453017 & 1.302486 & 1.05652 \\ \hline
        3.16228 &- &- &- &- & 0.021263 & 0.375028 & 0.880385 \\ \hline
        5.62341 &- &- &- &- & 3.7e-05 & 0.00417 & 0.03231 \\ \hline
        10.0 &- &- &- &- &- & 1e-06 & 1.4e-05 \\ \hline
        17.78279 &- &- &- &- &- &- &- \\ \hline
    \end{tabular}
    \label{tab:thin_gauss_e}
\end{table}

\FloatBarrier

\section{Optically thick results}\label{sec:thick_results}
Here we include results for problems that we consider optically thick. By this, we mean that the source width is far greater than a mean free path. We accomplish this by specifying, $\sigma_a=800 \:\:\mathrm{cm}^{-1}$ and $z_0 < 1$ cm ($z_0$ is the dimensional $x_0$). In this section, we include \wgb{$S_2$ and transport} results for linearized Gaussian and square sources as well as a constant $\cv$ Gaussian source problem. We do not include a constant $\cv$ square source, since our method could not resolve the nonlinear, non\wgb{-}equilibrium, and very sharp wave that the square source induces.

Since in an optically thick problem results are of interest only after many mean free times, we give results for $\tau \approx 0.01, 0.1,$ and $1$ ns. By these times, the uncollided source is negligible and any discontinuous wavefronts have decayed. For these reasons, the problems in this section do not employ an uncollided source or a moving mesh. \wgb{On the solution plots in this section, we show the diffusion solution for the energy density (for the linearized problems) and the temperature (for the nonlinear problems). These are included to demonstrate the qualitative difference between transport and diffusion for thick problems and were calculated with a numerical \wgbt{non-flux limited,} non-equilibrium diffusion solver \wgbt{\cite{humbird2017adjoint}}.}

\subsection{Su-Olson problem with a square source}
To keep the dimensional spatial domain manageable, we set $l = \frac{1}{800} $ in Eqs.~\eqref{eq:transport_nondim} and \eqref{eq:material_nondim}. This makes the dimensional and nondimensional domains the same, but stiffens the system. We are again using a square source (Eq.~\eqref{eq:square_source}) with $x_0=0.5$ and \wgb{$t_0 = 0.0125$}. We forgo the use \wgb{of} an uncollided source since the evaluation times of interest are long after the uncollided solution has decayed to zero.

For the mesh in this problem, only a static mesh was necessary for satisfactory convergence. We use a initialization outlined in  Section \ref{subsec:su_square_thin} but that initial width $\delta x$ is not set to a small number, but to a guess of the solution width at the evaluation time and the edges never move. Essentially, the mesh is the same as the initial mesh for the thin square source but covering the \wgb{entire} domain. The Gauss-Legendre spacing of the edges has the effect of concentrating static edges around the source edge which  makes it more likely that the region where the wavefront will be is resolved. 

The initial guess for the solution width was important and was refined with each run increasing the number of spatial divisions. Since negative solutions are possible in our DG formulation and more likely to occur when there is a sharp wavefront, the temperature was calculated with $\lT = \mathrm{sign}(\Le)|\Le|^{1/4}$.

\begin{figure}
    \centering
    \begin{subfigure}[b]{0.3\textwidth}
    \includegraphics[width=\textwidth]{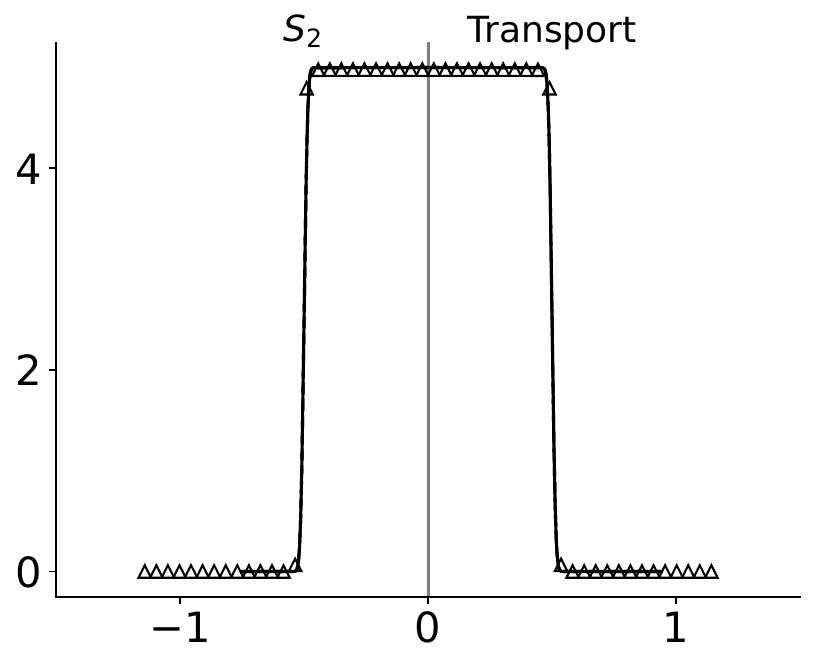}
    \caption{$t=0.3$}
    \end{subfigure}
    \centering
    \begin{subfigure}[b]{0.3\textwidth}
        \includegraphics[width=\textwidth]{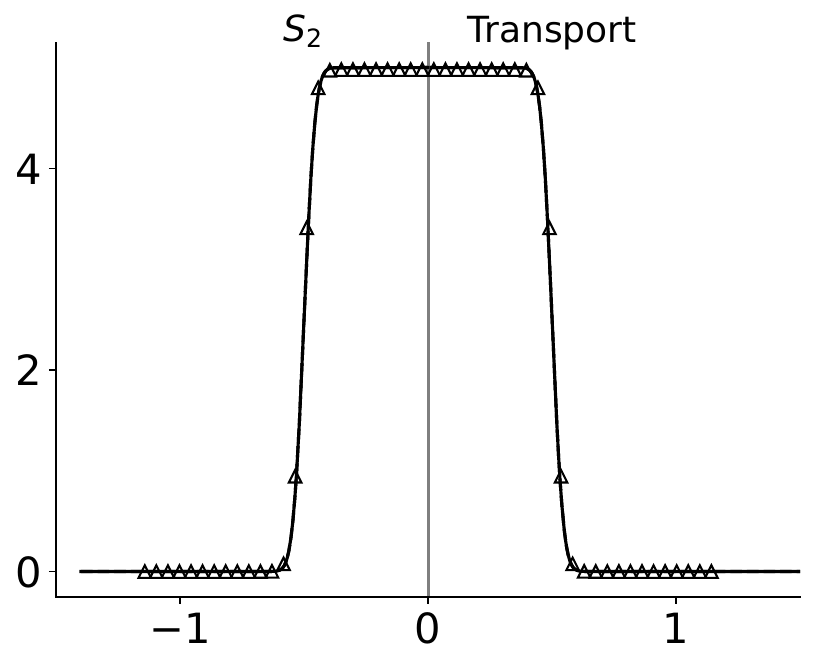}
        \caption{$t=3$}
    \end{subfigure}
    \centering
    \begin{subfigure}[b]{0.3\textwidth}
        \includegraphics[width=\textwidth]{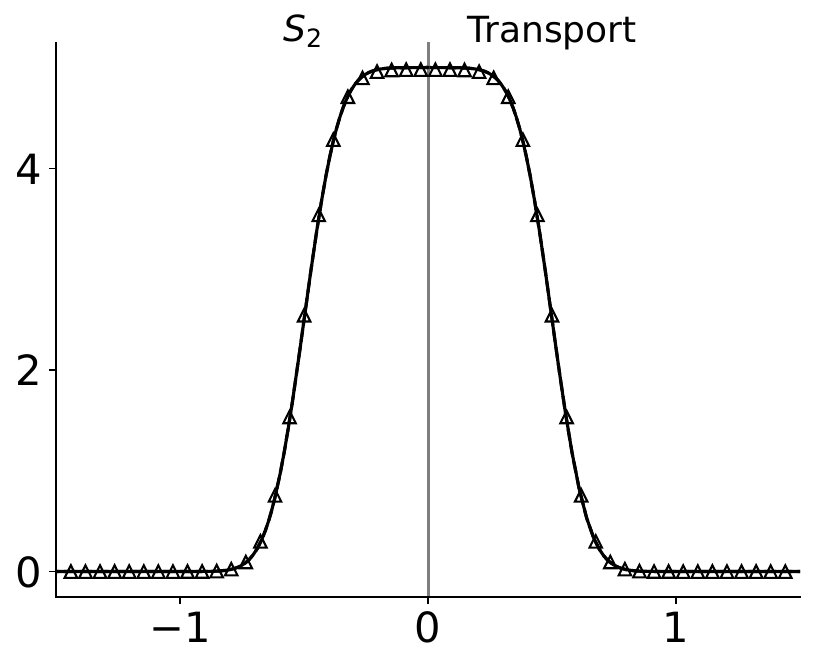}
        \caption{$t=30$}
    \end{subfigure}
      \caption{$S_2$ (left of $x=0$) and full transport (right of $x=0$) solutions for the optically thick Su-Olson square source problem with $x_0=0.5$, \wgb{$t_0=0.0125$}. Solid lines are scalar flux, $\lphi$, and dashed are material energy density, $\Le$. \wgb{Triangles are the diffusion solution for the energy density.} On the scale of this figure, the solid and dashed lines are coincident.}
    \label{fig:su_square_thick_solutions} 
\end{figure}

\begin{figure}
    \centering
    \begin{subfigure}[b]{0.48\textwidth}
    \includegraphics[width=\textwidth]{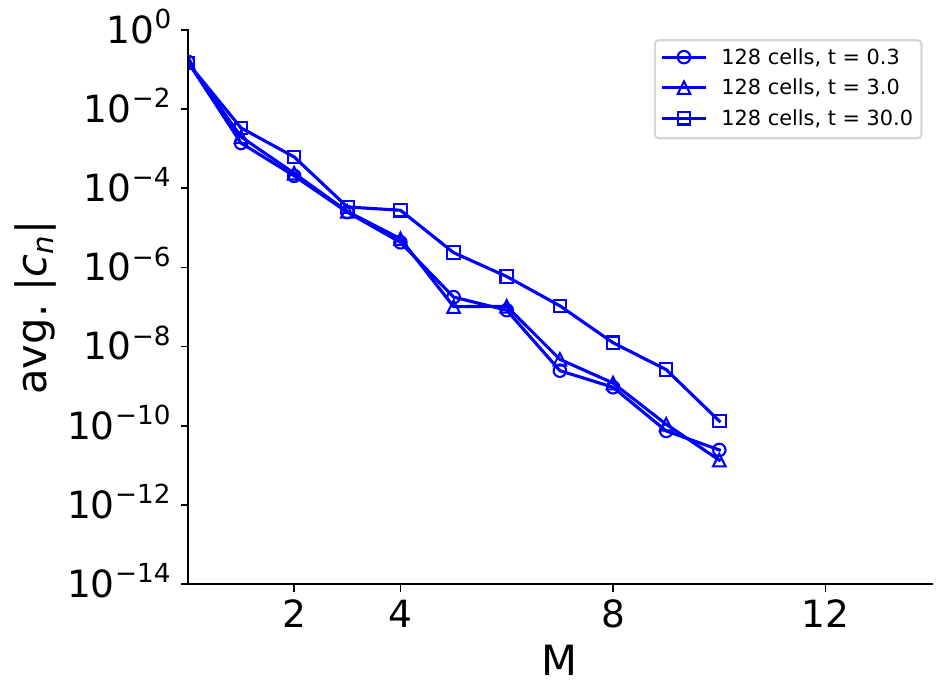}
    \caption{Radiation energy density, $\lphi$}
    \label{subfig:su_square_thick_phi}
    \end{subfigure}
    \centering
    \begin{subfigure}[b]{0.48\textwidth}
        \includegraphics[width=\textwidth]{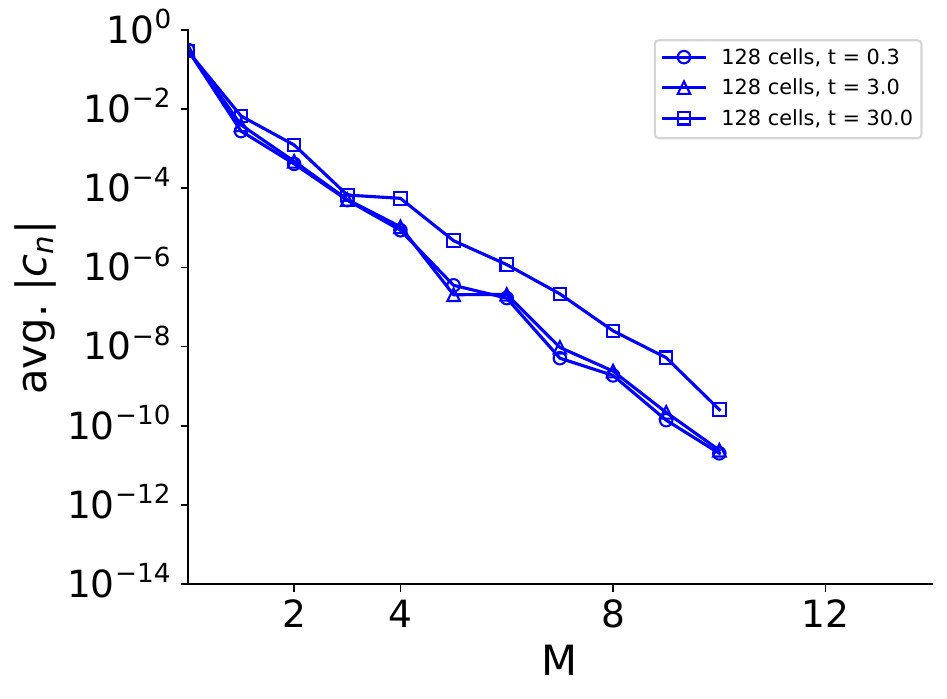}
        \caption{Material energy density, $\Le$}
        \label{subfig:su_square_thick_e}
    \end{subfigure}
      \caption{Log-linear scaled average value of the solution expansion coefficients (found by Eqs.~\eqref{eq:coeffs_phi}) for the optically thick ($\sigma_a=800$ cm$^{-1}$) Su-Olson square source problem where $x_0=0.5$, \wgb{$t_0=0.0125$}. The quadrature order for all results is $S_{16}$. All results were calculated with a static mesh and standard source treatment.}
    \label{fig:su_square_thick_convergence}
\end{figure}

\begin{figure}
    \centering
    \begin{subfigure}[b]{0.48\textwidth}
    \includegraphics[width=\textwidth]{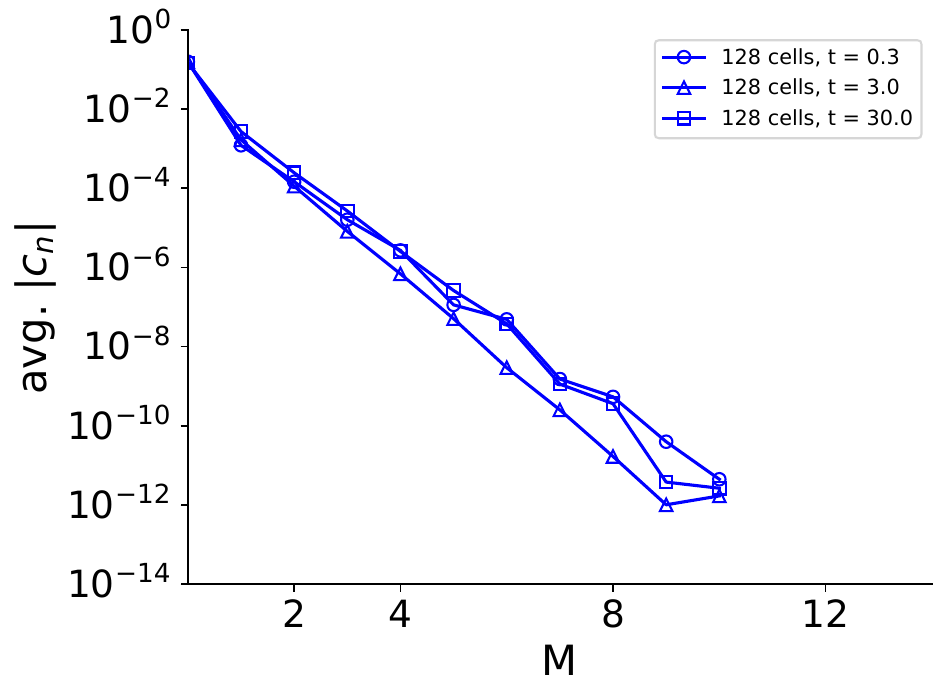}
    \caption{Radiation energy density, $\lphi$}
    \label{subfig:su_square_thick_s2_phi}
    \end{subfigure}
    \centering
    \begin{subfigure}[b]{0.48\textwidth}
        \includegraphics[width=\textwidth]{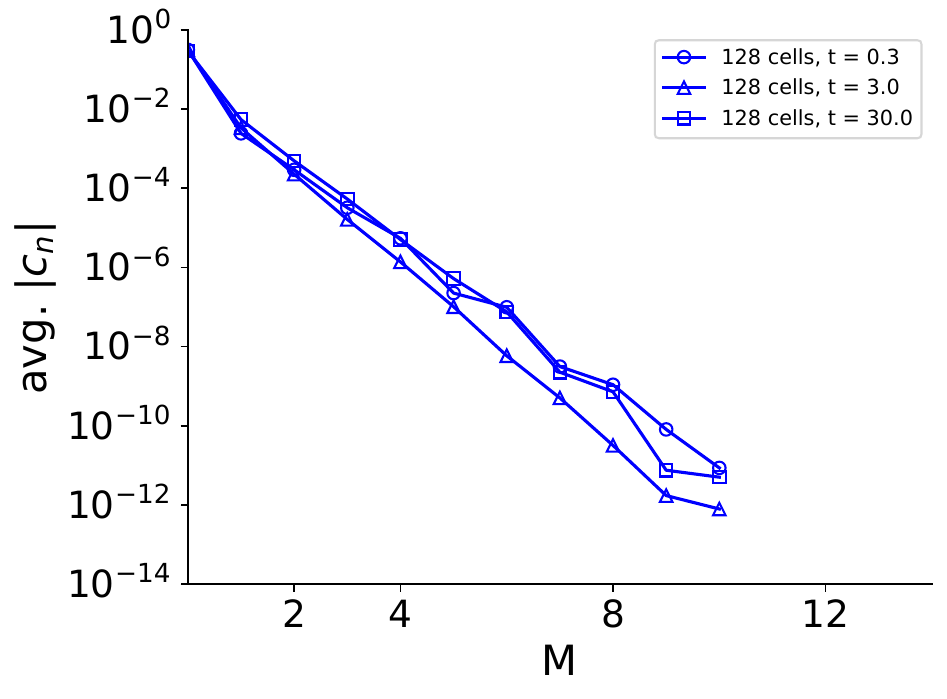}
        \caption{Material energy density, $\Le$}
        \label{subfig:su_square_thick_s2_e}
    \end{subfigure}
      \caption{Log-linear scaled average value of the solution expansion coefficients (found by Eqs.~\eqref{eq:coeffs_phi}) for the optically thick ($\sigma_a=800$ cm$^{-1}$) $S_2$ Su-Olson square source problem where $x_0=0.5$, \wgb{$t_0 = 0.0125$}. All results were calculated with a static mesh and standard source treatment.}
    \label{fig:su_square_thick_s2_convergence}
\end{figure}

The solution plots for this problem (Figure \ref{fig:su_square_thick_solutions}) show that for the chosen times, the solution is in local equilibrium. Unlike the selected thin square source solutions where a discontinuous wave travelling at the wavespeed determines the speed the solution travels, here a wave resembling a nonlinear heat waves moves outwards while the source is on. Geometric convergence, shown in Figures \ref{fig:su_square_thick_convergence} and \ref{fig:su_square_thick_s2_convergence}, is only possible since the nonsmooth portions of the scalar flux have decayed to zero and the solution is in equilibrium, which has the effect of smoothing the leading edge of the wavefront. 

We also take not\wgb{e} of the similarity between the transport and $S_2$ solutions, apparent in the solution plots and in Tables \ref{tab:su_thick_square_phi} and \ref{tab:su_thick_square_e}. This is expected from the optically thick results, since we saw \wgb{the} the two solutions converge at long times \wgb{when} \wgb{they have both come into} equilibrium. \wgb{The solution plots also show \wgbt{qualitative agreement between transport/$S_2$ and diffusion.}}  

\begin{table}[!ht]
    \centering
     \caption{Transport (top) and $S_2$ (bottom) results for the scalar flux, $\lphi$,  for the thick square source Su-Olson problem with $x_0=0.5$,  \wgb{$t_0 = 0.0125$}}
     
    \begin{tabular}{|l|l|l|l|}
    \hline
        $\boldsymbol{x/t}$ & 0.3 & 3.0 & 30.0 \\ \hline
       - & 4.999998 & 4.999998 & 4.999957 \\ \hline
        0.0579 & 4.999998 & 4.999998 & 4.999802 \\ \hline
        0.1158 & 4.999998 & 4.999998 & 4.998522 \\ \hline
        0.1737 & 4.999998 & 4.999998 & 4.991207 \\ \hline
        0.2316 & 4.999998 & 4.999998 & 4.959099 \\ \hline
        0.2895 & 4.999998 & 4.999998 & 4.850729 \\ \hline
        0.3474 & 4.999998 & 4.999957 & 4.569407 \\ \hline
        0.4053 & 4.999998 & 4.981594 & 4.007694 \\ \hline
        0.4632 & 4.997551 & 4.256673 & 3.144975 \\ \hline
        0.5211 & 0.141706 & 1.375262 & 2.125719 \\ \hline
        0.5789 &- & 0.063798 & 1.200785 \\ \hline
        0.6368 &- & 0.000274 & 0.552639 \\ \hline
        0.6947 &- &- & 0.203932 \\ \hline
        0.7526 &- &- & 0.05963 \\ \hline
        0.8105 &- &- & 0.013701 \\ \hline
        0.8684 &- &- & 0.002459 \\ \hline
        0.9263 &- &- & 0.000343 \\ \hline
        0.9842 &- &- & 3.7e-05 \\ \hline
        1.0421 &- &- & 3e-06 \\ \hline
        1.1 &- &- &- \\ \hline
        \hline
        \hline
        \hline
        $\boldsymbol{x/t}$ & 0.3 & 3.0 & 30.0 \\ \hline
       - & 4.999999 & 5.0 & 4.999961 \\ \hline
        0.0579 & 4.999999 & 5.0 & 4.999807 \\ \hline
        0.1158 & 4.999999 & 5.0 & 4.998528 \\ \hline
        0.1737 & 4.999999 & 5.0 & 4.991218 \\ \hline
        0.2316 & 4.999999 & 5.0 & 4.959116 \\ \hline
        0.2895 & 4.999999 & 4.999999 & 4.850739 \\ \hline
        0.3474 & 4.999999 & 4.999961 & 4.56939 \\ \hline
        0.4053 & 4.999999 & 4.981698 & 4.007652 \\ \hline
        0.4632 & 4.997908 & 4.256289 & 3.144948 \\ \hline
        0.5211 & 0.141029 & 1.375703 & 2.125739 \\ \hline
        0.5789 &- & 0.063676 & 1.200832 \\ \hline
        0.6368 &- & 0.000265 & 0.552668 \\ \hline
        0.6947 &- &- & 0.203932 \\ \hline
        0.7526 &- &- & 0.059617 \\ \hline
        0.8105 &- &- & 0.013692 \\ \hline
        0.8684 &- &- & 0.002455 \\ \hline
        0.9263 &- &- & 0.000342 \\ \hline
        0.9842 &- &- & 3.7e-05 \\ \hline
        1.0421 &- &- & 3e-06 \\ \hline
        1.1 &- &- &- \\ \hline
    \end{tabular}
    \label{tab:su_thick_square_phi}
\end{table}

\begin{table}[!ht]
    \centering
     \caption{Transport (top) and $S_2$ (bottom) results for the material energy density, $\Le$,  for the thick square source Su-Olson problem with $x_0=0.5$, \wgb{$t_0 = 0.0125$}. Convergence results for these answers are plotted in Figures \ref{fig:su_square_thick_convergence} and \ref{fig:su_square_thick_s2_convergence}.}
    \begin{tabular}{|l|l|l|l|}
    \hline
       - & 0.3 & 3.0 & 30.0 \\ \hline
       - & 4.999998 & 4.999998 & 4.999957 \\ \hline
        0.0579 & 4.999998 & 4.999998 & 4.999802 \\ \hline
        0.1158 & 4.999998 & 4.999998 & 4.998522 \\ \hline
        0.1737 & 4.999998 & 4.999998 & 4.991208 \\ \hline
        0.2316 & 4.999998 & 4.999998 & 4.959105 \\ \hline
        0.2895 & 4.999998 & 4.999998 & 4.850742 \\ \hline
        0.3474 & 4.999998 & 4.999957 & 4.56943 \\ \hline
        0.4053 & 4.999998 & 4.981624 & 4.007719 \\ \hline
        0.4632 & 4.997609 & 4.256925 & 3.144988 \\ \hline
        0.5211 & 0.140405 & 1.375054 & 2.125712 \\ \hline
        0.5789 &- & 0.063722 & 1.200762 \\ \hline
        0.6368 &- & 0.000273 & 0.552615 \\ \hline
        0.6947 &- &- & 0.203916 \\ \hline
        0.7526 &- &- & 0.059622 \\ \hline
        0.8105 &- &- & 0.013699 \\ \hline
        0.8684 &- &- & 0.002458 \\ \hline
        0.9263 &- &- & 0.000343 \\ \hline
        0.9842 &- &- & 3.7e-05 \\ \hline
        1.0421 &- &- & 3e-06 \\ \hline
        1.1 &- &- &- \\ \hline
        \hline
        \hline
        \hline
       - & 0.3 & 3.0 & 30.0 \\ \hline
       - & 4.999999 & 5.0 & 4.999961 \\ \hline
        0.0579 & 4.999999 & 5.0 & 4.999807 \\ \hline
        0.1158 & 4.999999 & 5.0 & 4.998529 \\ \hline
        0.1737 & 4.999999 & 5.0 & 4.99122 \\ \hline
        0.2316 & 4.999999 & 5.0 & 4.959121 \\ \hline
        0.2895 & 4.999999 & 4.999999 & 4.850752 \\ \hline
        0.3474 & 4.999999 & 4.999961 & 4.569413 \\ \hline
        0.4053 & 4.999999 & 4.981729 & 4.007677 \\ \hline
        0.4632 & 4.997962 & 4.256541 & 3.144961 \\ \hline
        0.5211 & 0.139713 & 1.375495 & 2.125732 \\ \hline
        0.5789 &- & 0.063599 & 1.200809 \\ \hline
        0.6368 &- & 0.000264 & 0.552644 \\ \hline
        0.6947 &- &- & 0.203916 \\ \hline
        0.7526 &- &- & 0.05961 \\ \hline
        0.8105 &- &- & 0.013689 \\ \hline
        0.8684 &- &- & 0.002455 \\ \hline
        0.9263 &- &- & 0.000342 \\ \hline
        0.9842 &- &- & 3.6e-05 \\ \hline
        1.0421 &- &- & 3e-06 \\ \hline
        1.1 &- &- &- \\ \hline
    \end{tabular}
    \label{tab:su_thick_square_e}
\end{table}

\subsection{Su-Olson problem with a Gaussian source}\label{sec:su_thick_gauss}
In order to provide consistent examples across optically thin and thick regimes and as a trial run for the constant $\cv$ thick Gaussian \wgb{source} of the next section, we include here a\wgb{n} optically thick Gaussian source. For this and the nonlinear Gaussian \wgb{source}, we specify the length parameter $x_0=0.375$. The source duration, $t_0$, is still \wgb{$0.0125$} and $l=\frac{1}{800}$. Once again, the source is given by Eq.~\eqref{eq:gaussian_source} and we do not use an uncollided source. Like the thin Gaussian sources, a moving mesh is not necessary. 

\begin{figure}
    \centering
    \begin{subfigure}[b]{0.3\textwidth}
    \includegraphics[width=\textwidth]{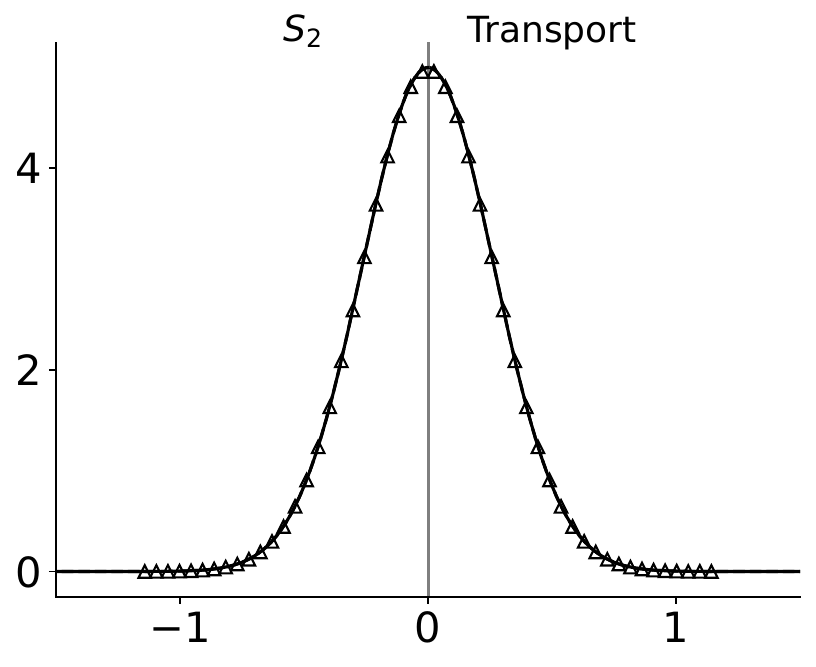}
    \caption{$t=0.3$}
    \end{subfigure}
    \centering
    \begin{subfigure}[b]{0.3\textwidth}
        \includegraphics[width=\textwidth]{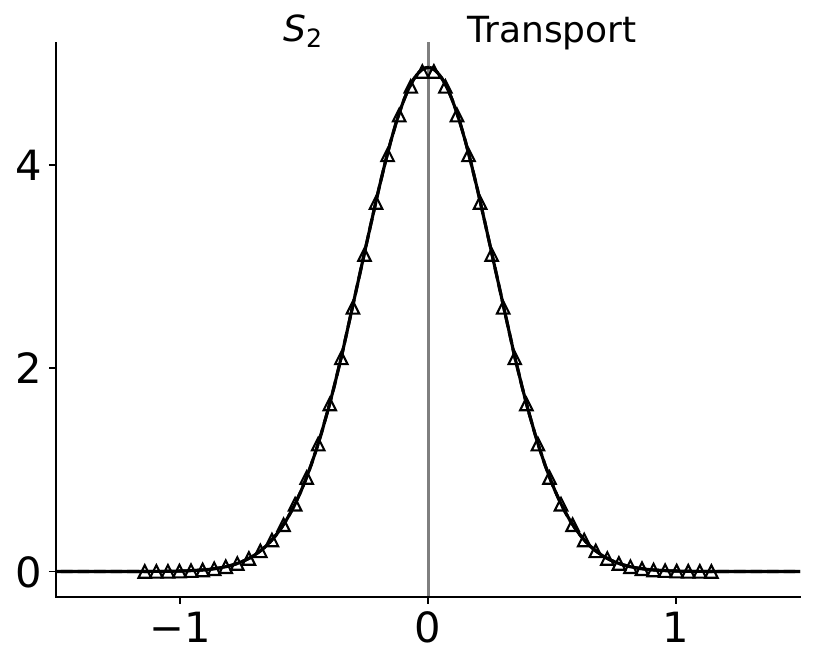}
        \caption{$t=3$}
    \end{subfigure}
    \centering
    \begin{subfigure}[b]{0.3\textwidth}
        \includegraphics[width=\textwidth]{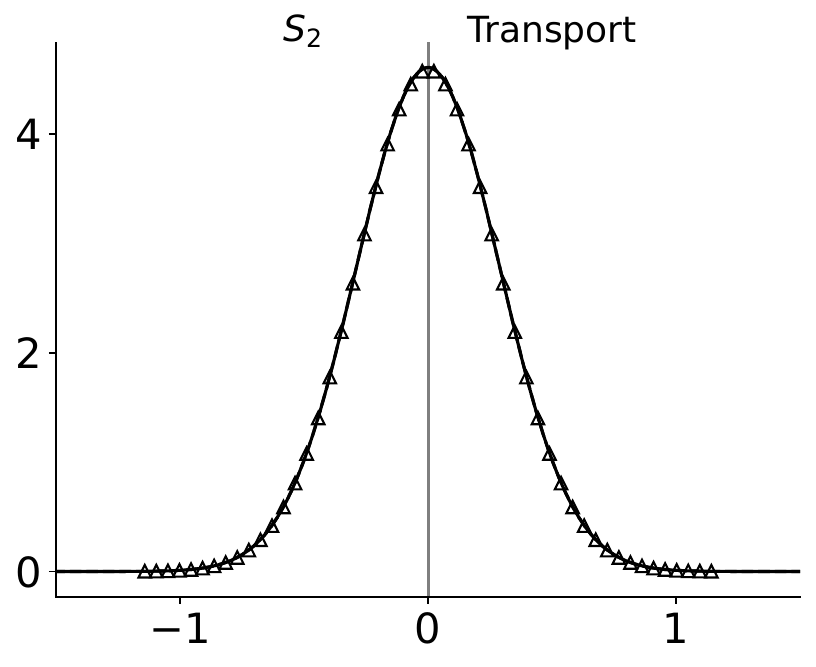}
        \caption{$t=30$}
    \end{subfigure}
      \caption{$S_2$ (left of $x=0$) and full transport (right of $x=0$) solutions for the optically thick Su-Olson Gaussian source problem with $x_0=0.375$, \wgb{$t_0 =0.0125$}. Solid lines are scalar flux, $\lphi$, and dashed are material energy density, $\Le$. \wgb{Triangles are the diffusion solution for the energy density.} On the scale of this figure, the solid and dashed lines are coincident.}
    \label{fig:su_gaussian_thick_solutions} 
\end{figure}

\begin{figure}
    \centering
    \begin{subfigure}[b]{0.48\textwidth}
    \includegraphics[width=\textwidth]{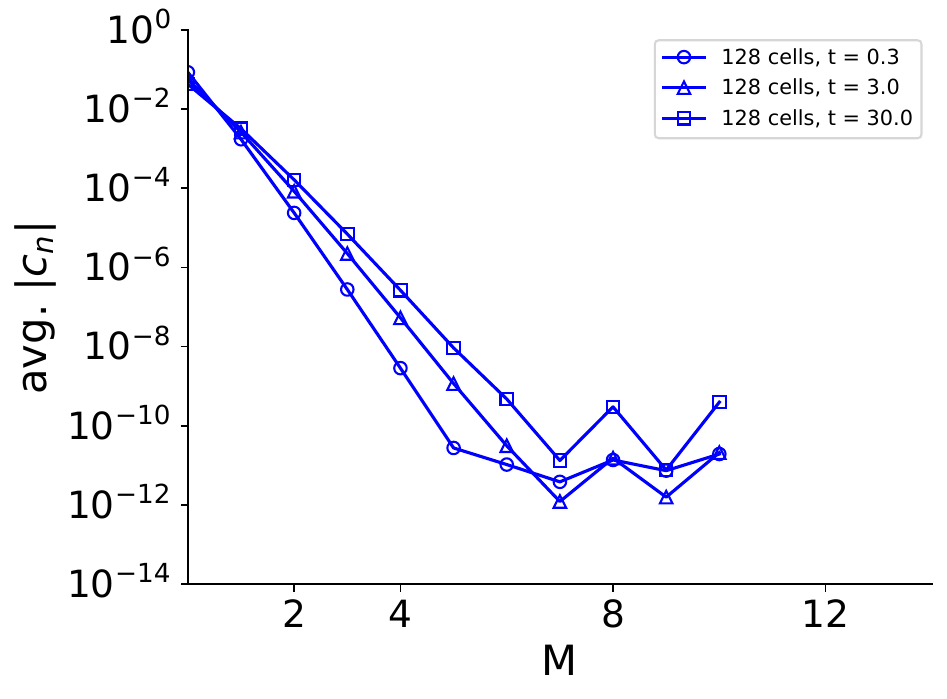}
    \caption{Radiation energy density, $\lphi$}
    \label{subfig:su_gauss_thick_phi}
    \end{subfigure}
    \centering
    \begin{subfigure}[b]{0.48\textwidth}
        \includegraphics[width=\textwidth]{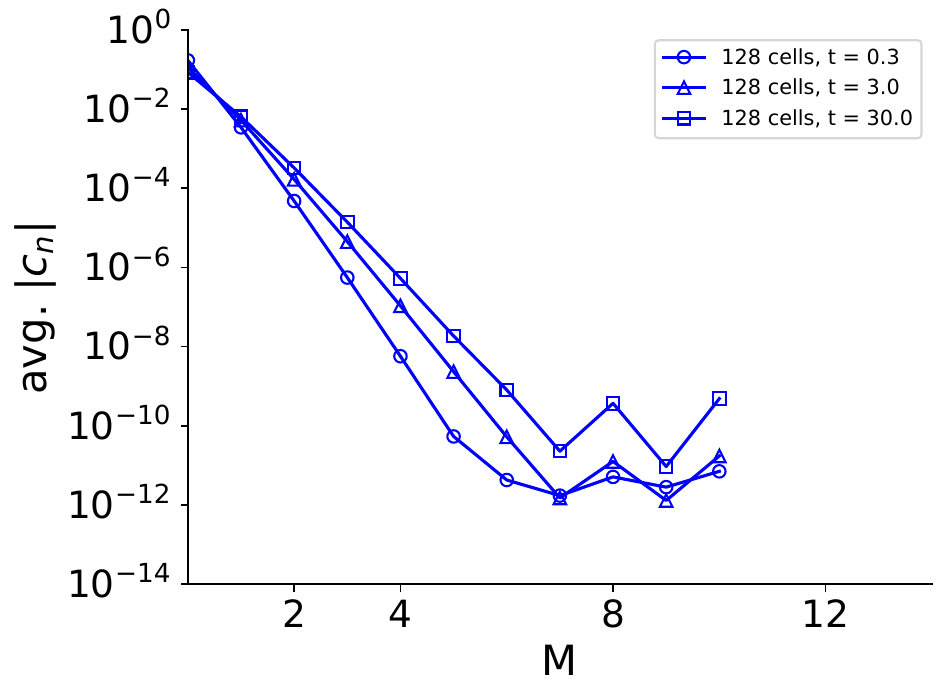}
        \caption{Material energy density, $\Le$}
        \label{subfig:su_gauss_thick_e}
    \end{subfigure}
      \caption{Log-linear scaled average value of the solution expansion coefficients (found by Eqs.~\eqref{eq:coeffs_phi}) for the optically thick ($\sigma_a=800$ cm$^{-1}$) Su-Olson Gaussian source problem where $x_0=0.375$, \wgb{$t_0=0.0125$}. The quadrature order for all results is $S_{16}$. All results were calculated with a static mesh and standard source treatment.}
    \label{fig:su_gaussian_thick_convergence}
\end{figure}

\begin{figure}
    \centering
    \begin{subfigure}[b]{0.48\textwidth}
    \includegraphics[width=\textwidth]{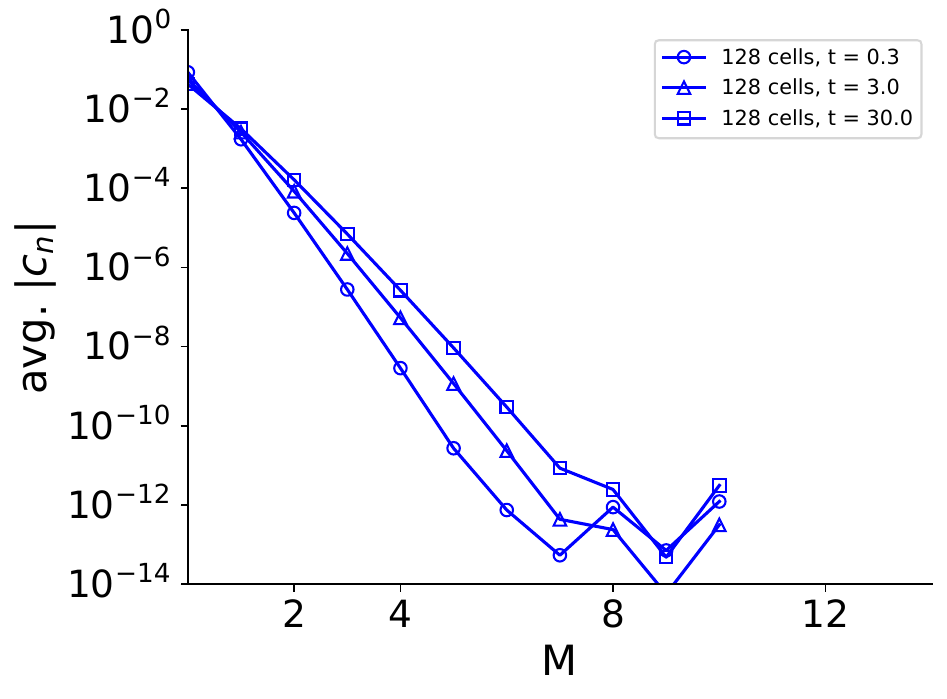}
    \caption{Radiation energy density, $\lphi$}
    \label{subfig:su_gauss_thick_s2_phi}
    \end{subfigure}
    \centering
    \begin{subfigure}[b]{0.48\textwidth}
        \includegraphics[width=\textwidth]{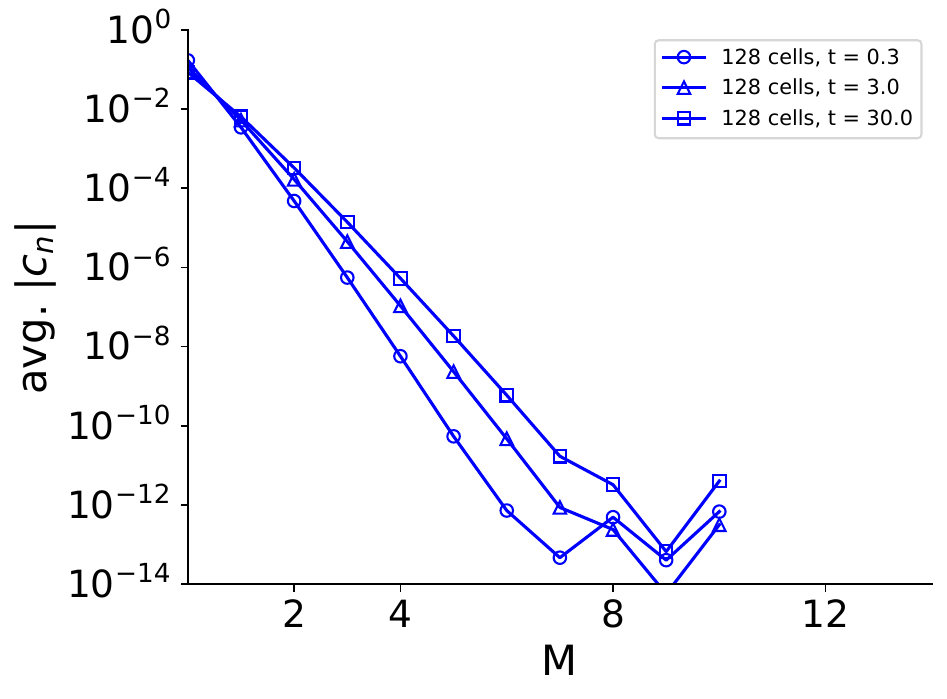}
        \caption{Material energy density, $\Le$}
        \label{subfig:su_gauss_thick_s2_e}
    \end{subfigure}
      \caption{Log-linear scaled average value of the solution expansion coefficients (found by Eqs.~\eqref{eq:coeffs_phi}) for the optically thick ($\sigma_a=800$ cm$^{-1}$) $S_2$ Su-Olson square Gaussian problem where $x_0=0.375$, $t_0=10$. All results were calculated with a static mesh and standard source treatment.}
    \label{fig:su_gaussian_thick_s2_convergence}
\end{figure}

Figure \ref{fig:su_gaussian_thick_solutions} shows that like the optically thick square source, the solutions are in equilibrium during the selected time window. There is however, no wavefront but the solution maintains Gaussian characteristics. Like the square source, the transport, \wgbt{diffusion,} and $S_2$ solutions are very similar. Geometric convergence of our standard DG method is shown in Figures \ref{fig:su_gaussian_thick_convergence} and \ref{fig:su_gaussian_thick_s2_convergence}. We note that in this problem and the Su-Olson, square source $128$ spatial cells were required to achieve the desired rate of convergence, though the solution was smooth. This was due to the small length scales.

\begin{table}[!ht]
    \centering
     \caption{Transport (top) and $S_2$ (bottom) results for the scalar flux, $\lphi$, for the thick Gaussian source Su-Olson problem with $x_0=0.375$,  \wgb{$t_0=0.0125$}} 
     
    \begin{tabular}{|l|l|l|l|}
    \hline
        $\boldsymbol{x/t}$ & 0.3 & 3.0 & 30.0 \\ \hline
       - & 4.99565 & 4.956221 & 4.607285 \\ \hline
        0.0842 & 4.750452 & 4.71669 & 4.414224 \\ \hline
        0.1684 & 4.084736 & 4.065342 & 3.882236 \\ \hline
        0.2526 & 3.175988 & 3.173439 & 3.13421 \\ \hline
        0.3368 & 2.232954 & 2.243553 & 2.322698 \\ \hline
        0.4211 & 1.418754 & 1.435689 & 1.579266 \\ \hline
        0.5053 & 0.815508 & 0.832457 & 0.986082 \\ \hline
        0.5895 & 0.423872 & 0.437156 & 0.565183 \\ \hline
        0.6737 & 0.199217 & 0.207914 & 0.297361 \\ \hline
        0.7579 & 0.084665 & 0.089558 & 0.143615 \\ \hline
        0.8421 & 0.032536 & 0.034938 & 0.063669 \\ \hline
        0.9263 & 0.011306 & 0.012344 & 0.025911 \\ \hline
        1.0105 & 0.003552 & 0.00395 & 0.009679 \\ \hline
        1.0947 & 0.001009 & 0.001144 & 0.003319 \\ \hline
        1.1789 & 0.000259 & 0.0003 & 0.001044 \\ \hline
        1.2632 & 6e-05 & 7.1e-05 & 0.000301 \\ \hline
        1.3474 & 1.2e-05 & 1.5e-05 & 7.9e-05 \\ \hline
        1.4316 & 2e-06 & 2e-06 & 1.9e-05 \\ \hline
        1.5158 &- &- & 4e-06 \\ \hline
        1.6 &- &- &- \\ \hline
        \hline
        \hline
        \hline
        $\boldsymbol{x/t}$ & 0.3 & 3.0 & 30.0 \\ \hline
       - & 4.995653 & 4.956229 & 4.607284 \\ \hline
        0.0842 & 4.750456 & 4.716697 & 4.414223 \\ \hline
        0.1684 & 4.084739 & 4.065349 & 3.882238 \\ \hline
        0.2526 & 3.175991 & 3.173444 & 3.134214 \\ \hline
        0.3368 & 2.232956 & 2.243557 & 2.322703 \\ \hline
        0.4211 & 1.418755 & 1.435691 & 1.57927 \\ \hline
        0.5053 & 0.815508 & 0.832459 & 0.986085 \\ \hline
        0.5895 & 0.423872 & 0.437157 & 0.565185 \\ \hline
        0.6737 & 0.199218 & 0.207914 & 0.297362 \\ \hline
        0.7579 & 0.084665 & 0.089558 & 0.143614 \\ \hline
        0.8421 & 0.032536 & 0.034938 & 0.063669 \\ \hline
        0.9263 & 0.011306 & 0.012344 & 0.02591 \\ \hline
        1.0105 & 0.003552 & 0.00395 & 0.009679 \\ \hline
        1.0947 & 0.001009 & 0.001144 & 0.003319 \\ \hline
        1.1789 & 0.000259 & 0.0003 & 0.001044 \\ \hline
        1.2632 & 6e-05 & 7.1e-05 & 0.000301 \\ \hline
        1.3474 & 1.2e-05 & 1.5e-05 & 7.9e-05 \\ \hline
        1.4316 & 2e-06 & 2e-06 & 1.9e-05 \\ \hline
        1.5158 &- &- & 4e-06 \\ \hline
        1.6 &- &- &- \\ \hline
    \end{tabular}
\end{table}

\begin{table}[!ht]
    \centering
     \caption{Transport (top) and $S_2$ (bottom) results for the material energy density, $\Le$,  for the thick Gaussian source Su-Olson problem with $x_0=0.375$, \wgb{$t_0=0.0125$}} 
     
    \begin{tabular}{|l|l|l|l|}
    \hline
        $\boldsymbol{x/t}$ & 0.3 & 3.0 & 30.0 \\ \hline
       - & 4.995668 & 4.956239 & 4.6073 \\ \hline
        0.0842 & 4.750468 & 4.716705 & 4.414236 \\ \hline
        0.1684 & 4.084745 & 4.065351 & 3.882244 \\ \hline
        0.2526 & 3.17599 & 3.17344 & 3.134212 \\ \hline
        0.3368 & 2.232949 & 2.243548 & 2.322695 \\ \hline
        0.4211 & 1.418746 & 1.435681 & 1.57926 \\ \hline
        0.5053 & 0.8155 & 0.832449 & 0.986075 \\ \hline
        0.5895 & 0.423866 & 0.43715 & 0.565178 \\ \hline
        0.6737 & 0.199213 & 0.20791 & 0.297357 \\ \hline
        0.7579 & 0.084663 & 0.089556 & 0.143612 \\ \hline
        0.8421 & 0.032535 & 0.034937 & 0.063668 \\ \hline
        0.9263 & 0.011305 & 0.012343 & 0.02591 \\ \hline
        1.0105 & 0.003552 & 0.003949 & 0.009679 \\ \hline
        1.0947 & 0.001009 & 0.001144 & 0.003319 \\ \hline
        1.1789 & 0.000259 & 0.0003 & 0.001044 \\ \hline
        1.2632 & 6e-05 & 7.1e-05 & 0.000301 \\ \hline
        1.3474 & 1.2e-05 & 1.5e-05 & 7.9e-05 \\ \hline
        1.4316 & 2e-06 & 2e-06 & 1.9e-05 \\ \hline
        1.5158 &- &- & 4e-06 \\ \hline
        1.6 &- &- &- \\ \hline
        \hline
        \hline
        \hline
        $\boldsymbol{x/t}$& 0.3 & 3.0 & 30.0 \\ \hline
       - & 4.995672 & 4.956247 & 4.607298 \\ \hline
        0.0842 & 4.750472 & 4.716712 & 4.414236 \\ \hline
        0.1684 & 4.084748 & 4.065358 & 3.882246 \\ \hline
        0.2526 & 3.175992 & 3.173446 & 3.134216 \\ \hline
        0.3368 & 2.232951 & 2.243552 & 2.3227 \\ \hline
        0.4211 & 1.418747 & 1.435684 & 1.579264 \\ \hline
        0.5053 & 0.8155 & 0.832451 & 0.986078 \\ \hline
        0.5895 & 0.423866 & 0.437151 & 0.565179 \\ \hline
        0.6737 & 0.199214 & 0.20791 & 0.297357 \\ \hline
        0.7579 & 0.084663 & 0.089556 & 0.143611 \\ \hline
        0.8421 & 0.032535 & 0.034937 & 0.063667 \\ \hline
        0.9263 & 0.011305 & 0.012343 & 0.025909 \\ \hline
        1.0105 & 0.003552 & 0.003949 & 0.009679 \\ \hline
        1.0947 & 0.001009 & 0.001144 & 0.003319 \\ \hline
        1.1789 & 0.000259 & 0.0003 & 0.001044 \\ \hline
        1.2632 & 6e-05 & 7.1e-05 & 0.000301 \\ \hline
        1.3474 & 1.2e-05 & 1.5e-05 & 7.9e-05 \\ \hline
        1.4316 & 2e-06 & 2e-06 & 1.9e-05 \\ \hline
        1.5158 &- &- & 4e-06 \\ \hline
        1.6 &- &- &- \\ \hline
    \end{tabular}
\end{table}


\subsection{Constant $\cv$ Gaussian problem}
Finally, we provide results for the constant $\cv$ optically thick problem with a Gaussian source. Like the linear version of this problem, $x_0=0.375$, \wgb{$t_0=0.0125$}, and $l = \frac{1}{800}$. We choose our constant opacity to be the same as we used for the optically thin case, \wgb{and the constant heat capacity} $C_\mathrm{v0}=0.03 \wgb{\:\mathrm{GJ}\cdot\mathrm{cm}^{-3}\cdot\mathrm{keV}^{-1}}$ \wgb{(which is the same as for the optically thin problems)}. The source is again given by Eq.~\eqref{eq:gaussian_source} and the uncollided solution is not used. \wgb{As with} the linear thick Gaussian \wgb{source problem}, a static mesh is employed. 

\begin{figure}
    \centering
    \begin{subfigure}[b]{0.3\textwidth}
    \includegraphics[width=\textwidth]{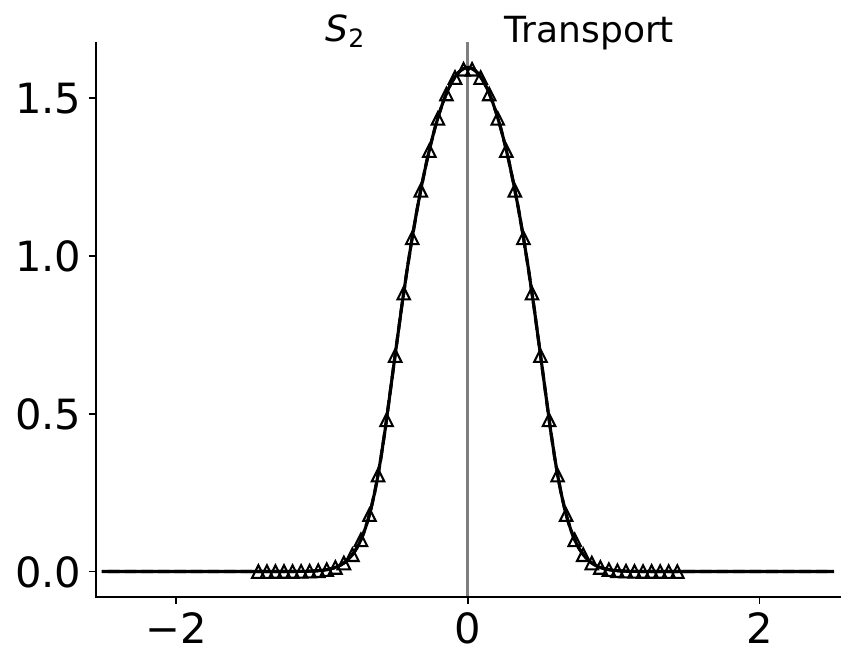}
    \caption{$t=0.3$}
    \end{subfigure}
    \centering
    \begin{subfigure}[b]{0.3\textwidth}
        \includegraphics[width=\textwidth]{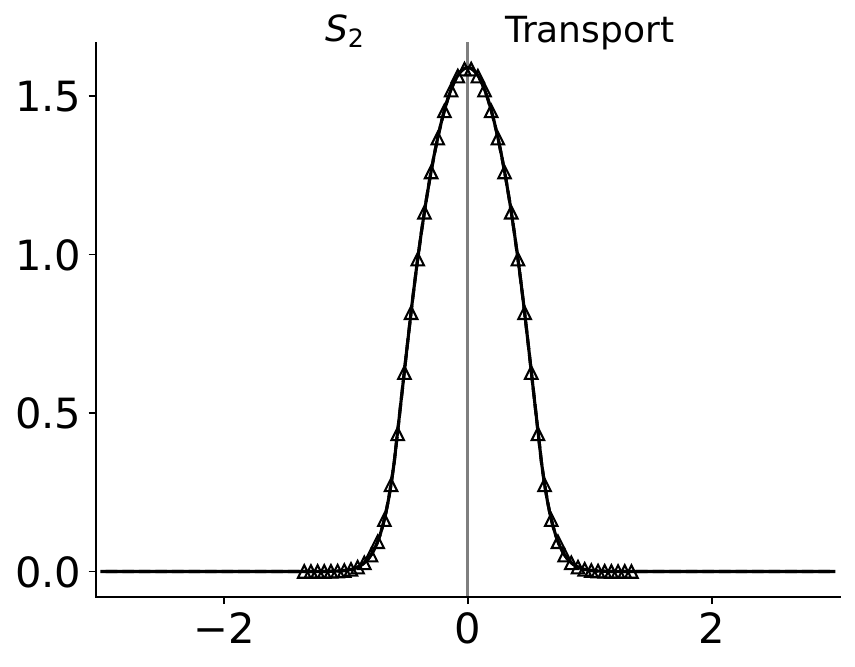}
        \caption{$t=3$}
    \end{subfigure}
    \centering
    \begin{subfigure}[b]{0.3\textwidth}
        \includegraphics[width=\textwidth]{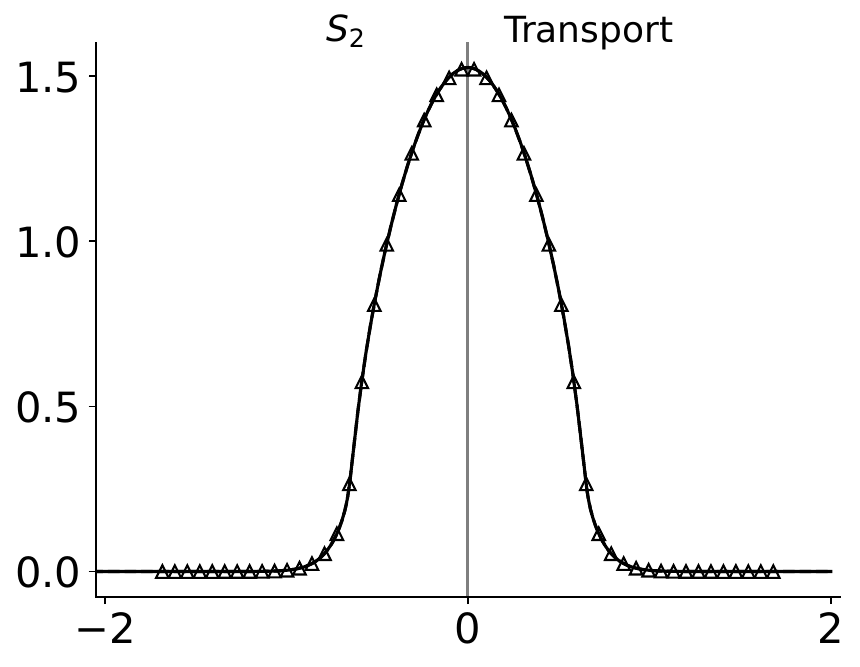}
        \caption{$t=30$}
    \end{subfigure}
      \caption{$S_2$ (left of $x=0$) and full transport (right of $x=0$) solutions for the optically thick constant $\cv$ Gaussian source problem with $x_0=0.375$, \wgb{$t_0=0.0125$}. Solid lines are radiation temperature $\lphi^{1/4}$, and dashed are temperature, $\lT$. \wgb{Triangles are the  diffusion solution for the temperature.} On the scale of this figure, the solid and dashed lines are coincident.}
    \label{fig:nl_gaussian_thick_solutions} 
\end{figure}
\begin{figure}
    \centering
    \begin{subfigure}[b]{0.48\textwidth}
    \includegraphics[width=\textwidth]{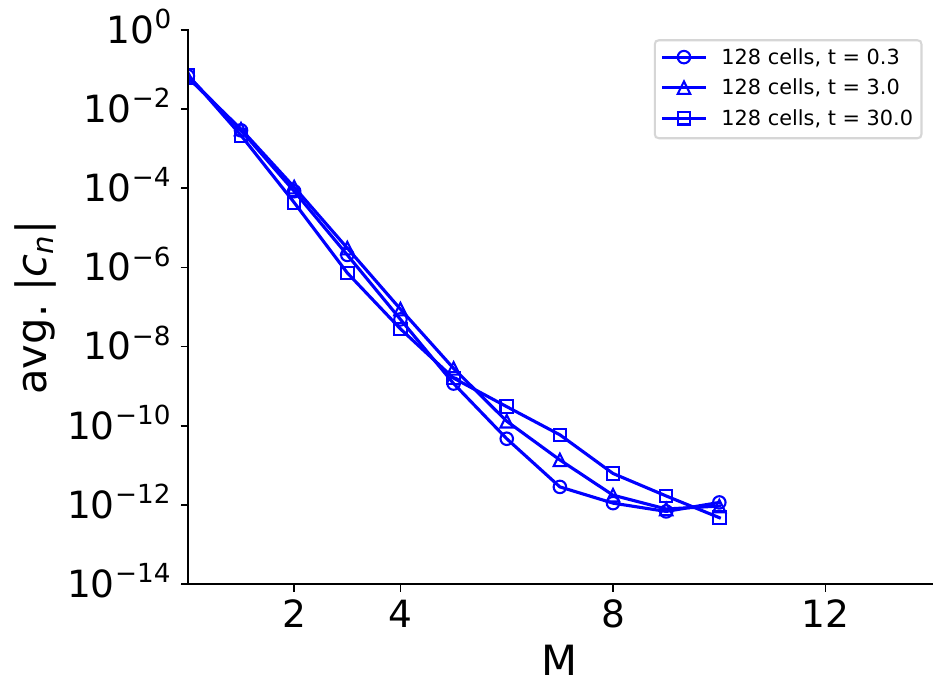}
    \caption{Radiation energy density, $\lphi$}
    \label{subfig:nl_gauss_thick_phi}
    \end{subfigure}
    \centering
    \begin{subfigure}[b]{0.48\textwidth}
        \includegraphics[width=\textwidth]{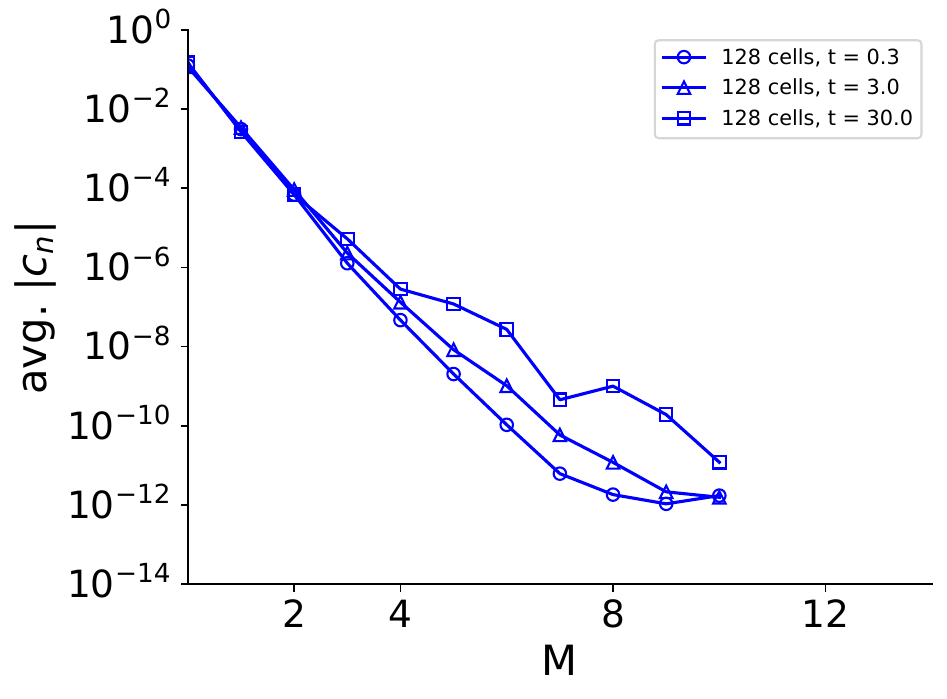}
        \caption{Material energy density, $\Le$}
        \label{subfig:nl_gauss_thick_e}
    \end{subfigure}
      \caption{Log-linear scaled average value of the solution expansion coefficients (found by Eqs.~\eqref{eq:coeffs_phi}) for the optically thick ($\sigma_a=800$ cm$^{-1}$) constant $\cv$ Gaussian source problem where $x_0=0.375$, \wgb{$t_0=0.0125$}. The quadrature order for all results is $S_{16}$. All results were calculated with a static mesh and standard source treatment.}
    \label{fig:nl_gaussian_thick_convergence}
\end{figure}


\begin{figure}
    \centering
    \begin{subfigure}[b]{0.48\textwidth}
    \includegraphics[width=\textwidth]{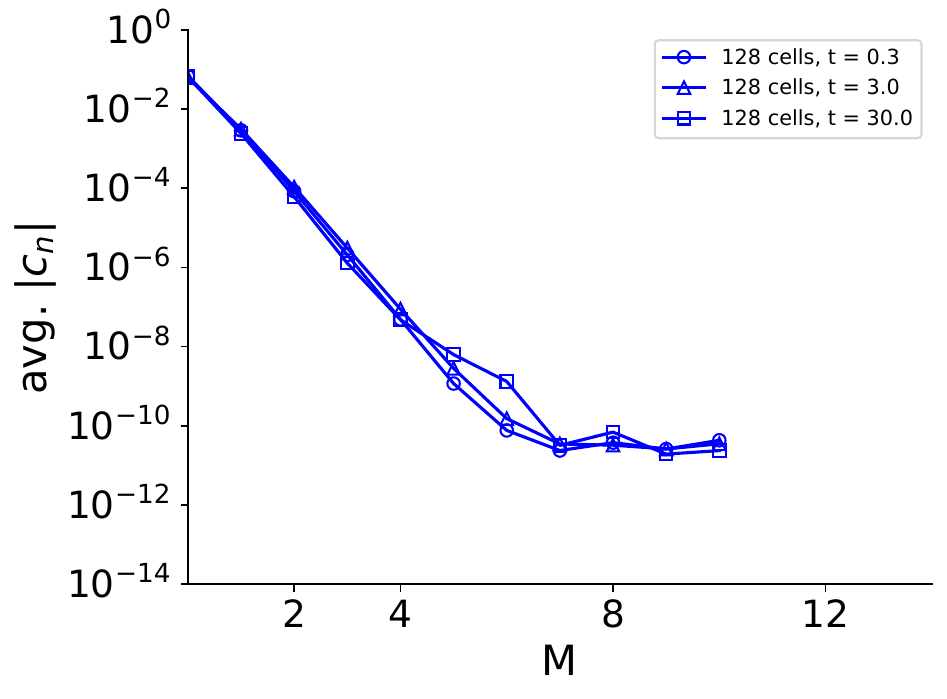}
    \caption{Radiation energy density, $\lphi$}
    \label{subfig:nl_gauss_thick_s2_phi}
    \end{subfigure}
    \centering
    \begin{subfigure}[b]{0.48\textwidth}
        \includegraphics[width=\textwidth]{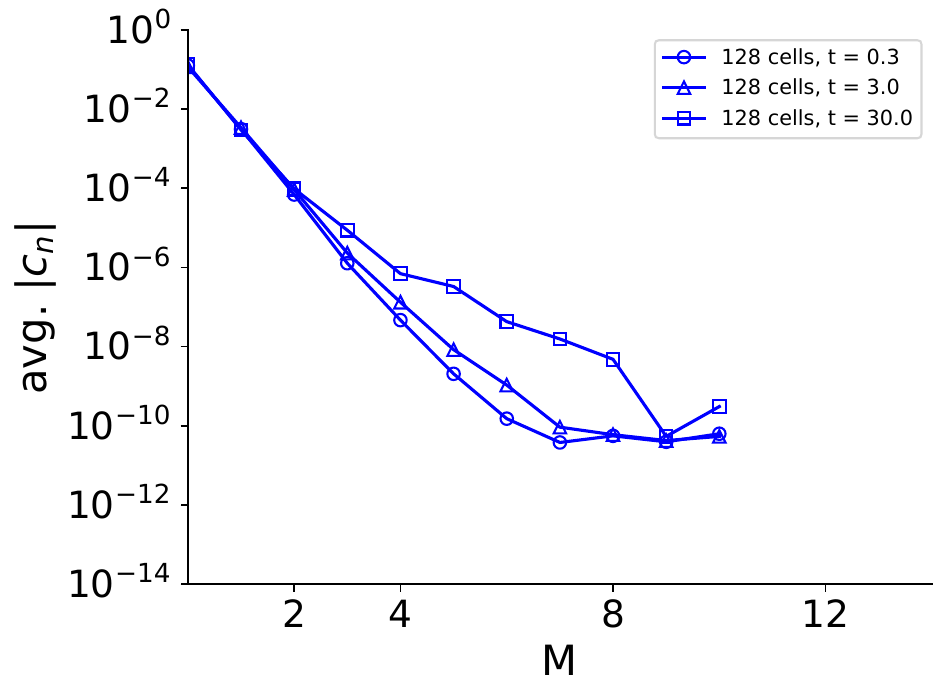}
        \caption{Material energy density, $\Le$}
        \label{subfig:nl_gauss_thick_s2_e}
    \end{subfigure}
      \caption{Log-linear scaled average value of the solution expansion coefficients (found by Eqs.~\eqref{eq:coeffs_phi}) for the optically thick ($\sigma_a=800$ cm$^{-1}$) $S_2$ constant $\cv$ Gaussian problem where $x_0=0.375$, \wgb{$t_0 =0.0125$}. All results were calculated with a static mesh and standard source treatment.}
    \label{fig:nl_gaussian_s2_thick_convergence}
\end{figure}


\begin{table}[!ht]
    \centering
    \caption{Transport (top) and $S_2$ (bottom) results for the scalar flux, $\lphi$, for the thick Gaussian source constant $\cv$ problem with $x_0=0.375$, \wgb{$t_0=0.0125$}, and $C_\mathrm{v0} = 0.03 \:\mathrm{GJ}\cdot\mathrm{cm}^{-3}\cdot\mathrm{keV}^{-1}$}
    
    \begin{tabular}{|l|l|l|l|}
    \hline
        $\boldsymbol{x/t}$ & 0.3 & 3.0 & 30.0 \\ \hline
       - & 6.494763 & 6.374101 & 5.426524 \\ \hline
        0.0789 & 6.11596 & 6.012864 & 5.186732 \\ \hline
        0.1579 & 5.085009 & 5.026915 & 4.518059 \\ \hline
        0.2368 & 3.682628 & 3.677866 & 3.56146 \\ \hline
        0.3158 & 2.248727 & 2.285336 & 2.499432 \\ \hline
        0.3947 & 1.081952 & 1.134228 & 1.513531 \\ \hline
        0.4737 & 0.352326 & 0.389958 & 0.733651 \\ \hline
        0.5526 & 0.060476 & 0.069031 & 0.230211 \\ \hline
        0.6316 & 0.00502 & 0.005255 & 0.01702 \\ \hline
        0.7105 & 0.000253 & 0.000255 & 0.000274 \\ \hline
        0.7895 & 8e-06 & 8e-06 & 8e-06 \\ \hline
        0.8684 &- &- &- \\ \hline
        \hline
        \hline
        \hline
        $\boldsymbol{x/t}$ & 0.3 & 3.0 & 30.0 \\ \hline
       - & 6.494762 & 6.374098 & 5.42651 \\ \hline
        0.0789 & 6.115958 & 6.012861 & 5.18672 \\ \hline
        0.1579 & 5.085008 & 5.026914 & 4.518052 \\ \hline
        0.2368 & 3.682627 & 3.677866 & 3.561461 \\ \hline
        0.3158 & 2.248727 & 2.285337 & 2.499437 \\ \hline
        0.3947 & 1.081952 & 1.134229 & 1.513539 \\ \hline
        0.4737 & 0.352326 & 0.389959 & 0.733658 \\ \hline
        0.5526 & 0.060476 & 0.069032 & 0.230216 \\ \hline
        0.6316 & 0.00502 & 0.005255 & 0.017021 \\ \hline
        0.7105 & 0.000253 & 0.000255 & 0.000274 \\ \hline
        0.7895 & 8e-06 & 8e-06 & 8e-06 \\ \hline
        0.8684 &- &- &- \\ \hline
    \end{tabular}
    \label{tab:nl_thick_gauss_phi}
\end{table}

\begin{table}[!ht]
    \centering
    \caption{Transport (top) and $S_2$ (bottom) results for the material energy density, $\Le$, for the thick Gaussian source constant $\cv$ problem with $x_0=0.375$, \wgb{$t_0=0.0125$}, and $C_\mathrm{v0} = 0.03 \:\mathrm{GJ}\cdot\mathrm{cm}^{-3}\cdot\mathrm{keV}^{-1}$}
    
    \centering
    \begin{tabular}{|l|l|l|l|}
    \hline
        \textbf{x/t} & 0.3 & 3.0 & 30.0 \\ \hline
       - & 3.490028 & 3.473704 & 3.33671 \\ \hline
        0.0737 & 3.444608 & 3.429805 & 3.303993 \\ \hline
        0.1474 & 3.309364 & 3.299069 & 3.20642 \\ \hline
        0.2211 & 3.086882 & 3.083965 & 3.04554 \\ \hline
        0.2947 & 2.780372 & 2.787696 & 2.82358 \\ \hline
        0.3684 & 2.389727 & 2.410536 & 2.540796 \\ \hline
        0.4421 & 1.913502 & 1.950669 & 2.194284 \\ \hline
        0.5158 & 1.363511 & 1.408883 & 1.769126 \\ \hline
        0.5895 & 0.82657 & 0.846776 & 1.213573 \\ \hline
        0.6632 & 0.436854 & 0.439279 & 0.485835 \\ \hline
        0.7368 & 0.210529 & 0.210694 & 0.212429 \\ \hline
        0.8105 & 0.093598 & 0.093606 & 0.093685 \\ \hline
        0.8842 & 0.038508 & 0.038508 & 0.038511 \\ \hline
        0.9579 & 0.014665 & 0.014665 & 0.014665 \\ \hline
        1.0316 & 0.005169 & 0.005169 & 0.005169 \\ \hline
        1.1053 & 0.001686 & 0.001686 & 0.001686 \\ \hline
        1.1789 & 0.00051 & 0.00051 & 0.00051 \\ \hline
        1.2526 & 0.000142 & 0.000142 & 0.000142 \\ \hline
        1.3263 & 3.6e-05 & 3.6e-05 & 3.6e-05 \\ \hline
        1.4 & 8e-06 & 8e-06 & 8e-06 \\ \hline
        \hline
        \hline
        \hline
        \textbf{x/t} & 0.3 & 3.0 & 30.0 \\ \hline
       - & 3.490028 & 3.473704 & 3.336708 \\ \hline
        0.0737 & 3.444608 & 3.429805 & 3.303991 \\ \hline
        0.1474 & 3.309364 & 3.299069 & 3.206419 \\ \hline
        0.2211 & 3.086882 & 3.083965 & 3.04554 \\ \hline
        0.2947 & 2.780372 & 2.787696 & 2.823581 \\ \hline
        0.3684 & 2.389727 & 2.410537 & 2.540799 \\ \hline
        0.4421 & 1.913502 & 1.95067 & 2.194288 \\ \hline
        0.5158 & 1.363511 & 1.408884 & 1.769132 \\ \hline
        0.5895 & 0.82657 & 0.846774 & 1.213586 \\ \hline
        0.6632 & 0.436853 & 0.439278 & 0.485758 \\ \hline
        0.7368 & 0.210529 & 0.210693 & 0.212427 \\ \hline
        0.8105 & 0.093598 & 0.093606 & 0.093684 \\ \hline
        0.8842 & 0.038508 & 0.038508 & 0.038511 \\ \hline
        0.9579 & 0.014665 & 0.014665 & 0.014665 \\ \hline
        1.0316 & 0.005169 & 0.005169 & 0.005169 \\ \hline
        1.1053 & 0.001686 & 0.001686 & 0.001686 \\ \hline
        1.1789 & 0.00051 & 0.00051 & 0.00051 \\ \hline
        1.2526 & 0.000142 & 0.000142 & 0.000142 \\ \hline
        1.3263 & 3.6e-05 & 3.6e-05 & 3.6e-05 \\ \hline
        1.4 & 8e-06 & 8e-06 & 8e-06 \\ \hline
    \end{tabular}
    \label{tab:nl_thick_gauss_e}
\end{table}

Although the Gaussian \wgb{profile} is slightly misshapen in the solution plots (Figure \ref{fig:nl_gaussian_thick_solutions}) when compared to the linearized Gaussian \wgb{source problem results}, the solution is smooth like the linearized problem and spectral convergence is observed in Figures \ref{fig:nl_gaussian_thick_convergence} and \ref{fig:nl_gaussian_s2_thick_convergence}. The scalar flux and material energy density are given in Tables \ref{tab:nl_thick_gauss_phi} and \ref{tab:nl_thick_gauss_e}. \wgb{As with the linearized Gaussian, the diffusion approximation is qualitatively correct.} 

\FloatBarrier

\section{Conclusions}
We have presented benchmark solutions to \wgb{constant opacity} time dependent radiative transfer problems with two functional forms for the specific heat in optically thin and thick media. These solutions will be useful to researchers who seek a nonlinear radiative transfer benchmark for verification purposes, desire more digits of accuracy for the Su-Olson type problem, or who intend to resolve a mean free path in the optically thick limit with a transport code. Although discontinuous sources are inconvenient for DG methods and required a complex mesh treatment to find accurate results\wgb{,} we have provided solutions for square sources since they are simple to simulate in numerical codes and are already implemented in codes that run the Su-Olson problem. Researchers who implement the DG friendly Gaussian source that we have defined will be able to converge to even more accurate results than for the square source. 



To support our claim of benchmark quality results, we presented the convergence of the magnitude of coefficients in the solution expansion for each result. While this does not necessarily guarantee the correctness of the system being solved, it does provide confidence that the solution is converged to a particular \wgb{tolerance}. We further increased confidence in our solutions by running $S_2$ benchmarks for the linearized problems to high accuracy and comparing to the original published results in \cite{SU19971035}. Furthermore, the nonlinear problems were checked for systematic errors with a $S_n$ code. \wgb{Future work will include non-constant opacity problems to provide better coverage for radiative transfer codes.}

\section*{Acknowledgement}
\noindent This work was supported by a Los Alamos National Laboratory
under contract \#599008, “Verification for Radiation Hydrodynamics Simulations”.
\FloatBarrier



\appendix
\renewcommand{\thesubsection}{\thesection.\Roman{subsection}}

\section{Uncollided solutions to the $S_2$ transport equation}\label{sec:s2_uncol}
This section contains solutions to Eq.~\eqref{eq:uncol_s2_greens} for a square and a Gaussian source, which are used for the uncollided source treatment in $S_2$ problems. 
\subsection{Gaussian source}
With Eq.~\eqref{eq:gaussian_source} as $S$ in Eq.~\eqref{eq:uncol_s2_greens}, the integral evaluates to, 
\begin{multline}\label{eq:uncol_gauss_s2}
    \lphi_u^{gs}(x,t) =   \frac{1}{4} \sqrt{3 \pi } x_0  e^{\frac{3 x_0 ^2}{4}-\sqrt{3} x}
   \wgb{\left(\text{erf}\left(\frac{1}{\sqrt{3}x_0}\left(t_0-t\right)+\frac{x}{x_0}-\frac{\sqrt{3}}{2}x_0\right)+ \right.} \\ 
   \wgb{\left.e^{2 \sqrt{3} x} \left(\text{erf}\left(\frac{t}{\sqrt{3} x_0 }+\frac{x}{x_0 } + \frac{\sqrt{3} x_0
   }{2}\right)-\text{erf}\left(\frac{1}{\sqrt{3}x_0}\left(t-t_0\right)+\frac{x}{x_0}+\frac{\sqrt{3}}{2}x_0\right)\right)+\text{erf}\left(\frac{t}{\sqrt{3}x_0} - \frac{x}{x_0}+ \frac{\sqrt{3}}{2}x_0\right)\right)}
\end{multline}

\subsection{Square source}
To evaluate the integral, Eq.~\eqref{eq:uncol_s2_greens} for a square source, Eq.~\eqref{eq:square_source}, each possible case of integration limits allowed by the step functions in the source must be considered. This will result in a piecewise function for the uncollided scalar flux. If we define, 
\begin{equation}
    F(t,\tau) =-\frac{1}{2} \exp(-t+\tau),
\end{equation}
then, if $t\leq t_0$,
\begin{multline}\label{eq:uncol_square_s2}
    \phi^{ss}_u(x,t) = \\ \begin{cases}
    F\bigg{|}_{\mathrm{max}(0,\tau_a)}^{\mathrm{max}(0,\tau_b)}   & (|x|>x_0)\\
    \\
     \begin{cases}
       2 F \bigg{|}_{0}^{\mathrm{min}(t,t_0)} &  (t +\sqrt{3}|x| \leq \sqrt{3}x_0) \\
        \\
        F\bigg{|}_{\tau_a}^{\tau_c} + 2 F\bigg{|}_{\tau_c}^{\mathrm{min}(t,t_0)} & \left(t + \sqrt{3}\,|x| \geq \sqrt{3}\,x_0\right) \: \& \: \left(t-\sqrt{3}(|x|+x_0) >0\right) \\
        \\
            F\bigg{|}_{0}^{\mathrm{max}(0,\tau_c)} + 2 F\bigg{|}_{\mathrm{max}(0,\tau_c)}^{\mathrm{min}(t,t_0)} & \left(t + \sqrt{3}\,|x| \geq \sqrt{3}\,x_0\right) \: \& \: \left(t-\sqrt{3}(|x|+x_0) \leq 0\right)
    \end{cases} & (|x|\leq x_0)
    \end{cases}
\end{multline}

or, if $t>t_0$,
\begin{equation}\label{eq:square_s2_uncol_2}
    \phi^{ss}_u(x,t) = \begin{cases}
      F\bigg{|}_{\mathrm{min}\left(\tau_e, t_0\right)}^{\mathrm{min}\left(\tau_d, t_0\right)} & x_0 - \frac{\sqrt{3}}{3}(t-t_0) \leq x \leq x_0 + \frac{\sqrt{3}}{3}(t-t_0) \\
      \\
      F\bigg{|}_{\mathrm{min}\left(\tau_e, t_0\right)}^{\mathrm{min}\left(\tau_d, t_0\right)} & x > x_0 + \frac{\sqrt{3}}{3}(t-t_0) \\
      \\
      \begin{cases}
      2 F\bigg{|}_0^{t_0} & t - \frac{3}{\sqrt{3}}(x_0-x) \leq 0 \\
      \\
      2F\bigg{|}^{t_0}_{\tau_d} + F\bigg{|}_0^{\tau_d} &  t - \frac{3}{\sqrt{3}}(x_0-x) > 0 \:\mathrm{and}\: t - \frac{3}{\sqrt{3}}(x_0-x) < t_0
      \end{cases} & x < x_0 - \frac{\sqrt{3}}{3}(t-t_0)
    \end{cases}
\end{equation}

Where
\begin{equation}
    \tau_a = t-\sqrt{3}(|x|+x_0),
\end{equation}
\begin{equation}
    \tau_b  = t-\sqrt{3}(|x|-x_0),
\end{equation}
\begin{equation}
    \tau_c  = t-\sqrt{3}(x_0-|x|),
\end{equation}
\begin{equation}
    \tau_d = \left|t-\frac{3}{\sqrt{3}}(x-x_0)\right|,
\end{equation}
and
\begin{equation}
    \tau_e = \left|t-\frac{3}{\sqrt{3}}(x+x_0)\right|.
\end{equation}


\FloatBarrier

\bibliography{references.bib}

\end{document}